\documentclass[11pt,a4paper]{article}
\usepackage{amsfonts}
\usepackage{times,latexsym}
\usepackage{url}
\usepackage[T1]{fontenc}
\usepackage{graphicx}
\usepackage{tabularray}
\usepackage{afterpage}

\usepackage{booktabs,multirow,array}
\usepackage[dvipsnames,table]{xcolor}
\usepackage{adjustbox}
\usepackage{amsmath}
\usepackage{tikz}
\usepackage{multicol}
\usetikzlibrary{arrows,shapes,positioning,shadows,trees}
\usepackage{longtable,pdflscape,booktabs}
\usepackage[section]{placeins}
\usepackage{amsmath}

\FloatBarrier
\tikzset{
  basic/.style  = {draw, rounded corners=6pt, text width=5cm, drop shadow, font=\sffamily, rectangle},
  root/.style   = {basic, rounded corners=6pt, thin, align=center,
                   fill=orange!20},
  level 2 red/.style = {basic, rounded corners=6pt, thin,align=center, fill=purple!50,
                   text width=9.5em},
level 2 green/.style = {basic, rounded corners=6pt, thin,align=center, fill=teal!50,
                   text width=8em},
  level 3 red/.style = {basic,rounded corners=6pt, thin, align=center, fill=red!40,
                 text width=11em},
  level 3 green/.style = {basic,rounded corners=6pt, thin, align=center, fill=green!40,
                 text width=11em},
  level 4 red/.style = {basic,rounded corners=6pt, thin, align=center, fill=pink!60,
                 text width=11em},
  level 4 green/.style = {basic,rounded corners=6pt, thin, align=center, fill=green!10,
                 text width=11em}
}

\definecolor{light-gray}{HTML}{E5E4E2}

\PassOptionsToPackage{table}{xcolor}

\usepackage[acceptedwitha,hyperref,]{tacl2021v1}

\usepackage{xspace,mfirstuc,tabulary}
\usepackage{subcaption}

\title{Securing Large Language Models: Threats, Vulnerabilities and Responsible Practices}

\author{
Sara Abdali\Thanks{Corresponding author}$^\diamond$ \and
Richard Anarfi\Thanks{These authors contributed equally.}$^\dagger$ \and
CJ Barberan$^\dagger$ \and
Jia He$^\dagger$ \and
Erfan Shayegani$^\dagger$
\\  
$^\diamond$Microsoft, Redmond, WA, USA \\
\texttt{saraabdali@microsoft.com}
\\ 
$^\dagger$Microsoft, Boston, MA, USA \\
\texttt{ranarfi,cjbarberan,hejia@microsoft}
\\
University of California, Riverside (UCR), Riverside, CA, USA \\
\texttt{sshay004@ucr.edu}
}

\date{}

\begin{document}
\pagenumbering{arabic}
\maketitle
\begin{abstract}
Large language models (LLMs) have significantly transformed the landscape of Natural Language Processing (NLP). Their impact extends across a diverse spectrum of tasks, revolutionizing how we approach language understanding and generations. Nevertheless, alongside their remarkable utility, LLMs introduce critical security and risk considerations. These challenges warrant careful examination to ensure responsible deployment and safeguard against potential vulnerabilities. This research paper thoroughly investigates security and privacy concerns related to LLMs from five thematic perspectives: security and privacy concerns, vulnerabilities against adversarial attacks, potential harms caused by misuses of LLMs, mitigation strategies to address these challenges while identifying limitations of current strategies. Lastly, the paper recommends promising avenues for future research to enhance the security and risk management of LLMs.\footnote{This study represents independent research conducted by the authors and does not necessarily represent the views or opinions of any organizations.}

\end{abstract}

\section{Introduction}

Recently, Large Language Models (LLMs) have initiated a significant paradigm shift in the areas of Natural Language Processing (NLP) including Natural Language Generation (NLG). LLMs are generally characterized by a large number of parameters -- usually ranging from millions to trillions -- and are constructed using deep neural networks, mainly transformer architectures~\cite{Vaswani2017AttentionIA,Lin2021ASO}. They undergo pre-training on massive amounts of text data, often collected from the web, and leverage self-supervised~\cite{Jiao2023LogicLLMES}, semi-supervised~\cite{Shi2023RethinkingSL}, or Reinforcement Learning (RL)~\cite{Ouyang2022TrainingLM,Gulcehre2023ReinforcedS} methods for pre-training and fine-tuning. Researchers continue to explore ways to fine-tune and adapt these LLMs for specific applications, making them indispensable tools in the NLP community and beyond.

\par LLMs have demonstrated remarkable abilities to generate coherent, human-like text, often for a given textual input, also referred to as a prompt~\cite{Zhao2023ASO}. 
\par For example, LLMs can assist users in communicating effectively, providing consistent and context-aware responses. They enable users to access information quickly, summarizing large volumes of text or answering complex queries. Moreover, LLMs contribute to producing varied and inventive content, whether it's generating poetry, stories, or code snippets. Beyond individual use cases, they also play a crucial role in education, research, and innovation across diverse fields, including science, art, and literature.
\par More specifically, LLMs achieve impressive results on various NLP tasks, such as text generation~\cite{Senadeera2022ControlledTG}, question answering~\cite{Zaib2021BERTCoQACBC,Bhat2023InvestigatingAO}, sentiment analysis~\cite{Batra2021BERTBasedSA,Kheiri2023SentimentGPTEG}, as well as augmenting human abilities by improving Human-Computer Interactions (HCI)~\cite{Oppenlaender2023MappingTC,hamalainen2023evaluating}.

\par While LLMs have demonstrated significant performance improvements across various tasks, they also present critical challenges with respect to security, privacy, and ethical protocols. 
\par For instance, LLMs are generally pre-trained on a huge volume of text from the web, which could potentially contain sensitive, personal, or confidential information. This poses a risk of leakage or misuse by adversaries.~\cite{Weidinger2021EthicalAS}. They could be also used for generating biased, toxic, harmful and discriminatory content~\cite{kuchnik2023validating}, infringing intellectual property rights~\cite{peng2023are,stokel2022ai}, bypassing corporate security protocols~\cite{Shayegani2023SurveyOV,Mozes2023UseOL}, or other malicious purposes, such as generating cyber-security attacks~\cite{Han2023FedMLSecurityAB} and spreading misinformation and propaganda~\cite{Vykopal2023DisinformationCO,Mozes2023UseOL}.
\par To foster responsible and ethical use of LLMs, it is essential to develop methods and frameworks that assess, improve, and govern LLMs in accordance with the principles of fairness, accountability, transparency, and explainability. This requires a comprehensive and interdisciplinary investigation of LLMs from a security, ethical and risk mitigation perspective.
\par While there are several existing studies~\cite{Ganguli2022RedTL,Huang2023ASO,Sun2023SafetyAO,Deshpande2023AnthropomorphizationOA,Wang2023DecodingTrustAC} that examine the security and risks of LLMs, the fast-paced advancement and innovation in this domain calls for a rigorous and systematic analysis.

\par With the goal of raising awareness and promoting responsible practices, we explore the potential threats and vulnerabilities associated with LLMs and categorize them into model-based, training-time and inference-time vulnerabilities. 
\par Additionally, we explore solutions and best practices to ensure their safe and responsible use. Our approach involves a rigorous investigation and evaluation of security and risk mitigation aspects related to LLMs. By doing so, we aim to highlight gaps and limitations in existing research and propose future directions. In summary, the key aspects of our research are:
\begin{itemize}
    \item \textbf{Security Risks:} We identify security concerns arising from LLM usage, including information leakage, memorization and security holes in codes generated by LLMs.  
    \item \textbf{Vulnerabilities and Risks of Adversarial Attacks}: We discuss LLMs' susceptibility to adversarial attacks including model-based, training-based and inference-based attacks. 
    \item \textbf{Risks of Misuse:} We analyze the risks and misuses associated with LLMs including bias, discrimination, and misinformation.
 \item \textbf{Risk Mitigation Strategies:} Our comprehensive assessment covers mitigation strategies like red teaming, model editing, watermarking, and AI-generated text detection techniques including discussions on limitations and trade-offs.
 \item \textbf{Future Research Directions:} We explore new research avenues aimed at addressing security and risk issues related to LLMs.
\end{itemize}
\par The rest of the paper is organized as follows:
\par As a preliminary step, we provide a glossary of the main terms that are frequently used in this work in the section \ref{sec:background} to promote the the readability of this paper and eliminate unnecessary repetition.  Then, in the section 
 \ref{sec:vulns}, we introduce some of the major vulnerabilities of LLMs by classifying them into three main categories: Model-based, training-based and inference-based attacks and their respective countermeasures. In addition, in section \ref{sec:security}, we investigate the security issues that emerge with LLM usage. We elaborate further risks and misuses of LLMs in section \ref{fig:risk&misuse}. Afterwards, in section \ref{sec:mitigation} we discuss mitigation strategies to reduce such risks, followed by section~\ref{sec:newops} where we propose new avenues of research. Finally, in section~\ref{sec:cons} we conclude. An overview of the paper's content is illustrated in figure \ref{fig:crownjewel}.
  \section{Background}
\label{sec:background}
In order to enhance the clarity of this paper and avoid redundancy, we provide a glossary of the key terms that are frequently used in this work. We present table~\ref{tab:glossary} that shows the terms, their concise definitions, and the paper sections where they are elaborated in depth. We encourage the readers to refer to the glossary whenever they encounter a term that is unfamiliar or unclear to them. The glossary is intended to serve as a quick reference guide and not as a comprehensive explanation of the concepts.

\onecolumn
\begin{center}
\small
\centering
\begin{longtable}{p{0.14\textwidth} p{0.67\textwidth} p{.068\textwidth}} 
\hline
\rowcolor{orange!20}\centering\textbf{Terminology}& \centering\textbf{Description} &\textbf{Section}\\
\hline
\endfirsthead
\multicolumn{3}{c}{} \\
\hline
\rowcolor{orange!20}\centering\textbf{Terminology} & \centering\textbf{Description} &\textbf{Section} \\
\hline
\endhead
\hline
\multicolumn{3}{c}{Continued on the next page} \\
\endfoot
\hline
\endlastfoot
\rowcolor{orange!2}
\centering Unlearning & A data pre-processing step for retraining or fine-tuning LLMs. It explicitly removes data points identified as vulnerable to leakage from dataset and retrains or fine-tunes LLMs on the processed dataset.~\cite{cao2015towards, chakraborty-etal-2024-textual}& \ref{sec:PII}\\

\hline
\hline
\rowcolor{orange!2}
  \centering Memorization &Memorization is the phenomenon of LLMs retaining and reproducing information from their training data. Memorization can be beneficial for tasks that require factual or linguistic knowledge, but also problematic for privacy, security, and quality reasons~\cite{Hartmann2023SoKMI,Zhong2023MemoryBankEL}.& \ref{sec:mem} \\
\hline
\hline
\rowcolor{orange!2}
\centering Association & association is the ability of LLMs to form connections between different pieces of information, such as words, entities, concepts, or events. Association in LLMs can enable various applications, such as knowledge retrieval, text summarization, and question answering. However, association in LLMs can also pose challenges, such as privacy leakage, hallucination, and bias~\cite{Shao2023QuantifyingAC,Du2023QuantifyingAA,chen-ding-2023-probing}& \ref{sec:mem}\\
\hline
\hline
\rowcolor{orange!2}

\centering Auditing & Auditing is to perform an examination to understand the implications and consequences of LLM memorization. For example, in auditing verbatim memorization, the examination would include a setup to generate arbitrary generated strings in order to detect if the LLM can provide an output of the said arbitrary generated string with high probability.~\cite{hartmann2023sok} &  \ref{sec:mem} \\
\hline
\hline
\rowcolor{orange!2}
\centering Emergent Misalignment&refers to unintended, unpredictable behavior that arises during fine-tuning, leading models to exhibit unsafe or unaligned responses across multiple dimensions.~\cite{betley2025emergent}&\ref{sec:code}\\
\hline
\hline
\rowcolor{orange!2}
\centering Adversarial Attack&An adversarial attack is a method that leverages the vulnerabilities or shortcomings of an LLM to induce erroneous or deceptive outputs. Adversarial attacks can be utilized for malicious purposes, such as creating misinformation, circumventing security protocols, or undermining the reliability of the model~\cite{bachu2024unfairalignmentexaminingsafety, Shayegani2023SurveyOV}.&\ref{sec:vulns}\\
\hline
\hline
\rowcolor{orange!2}
\centering Attack Success
Rate (ASR) &An attack success rate is a measure of the efficacy of an adversarial attack on a machine learning model. It is computed as the fraction of successful attacks over the total number of attacks. A successful attack is one that makes the model produce an erroneous prediction or output . An attack success rate can change depending on the type of attack, the type of model, the type of task, and the degree of perturbation~\cite{Wu2021PerformanceEO}.&\ref{sec:vulns}\\
\hline
\hline
\rowcolor{orange!2}
\centering Model Extraction Attack&A form of adversarial attack that leverages a large number of queries and their corresponding responses to extract the knowledge or parameters of an LLM. The extracted information can then be used to train a reduced parameter model that approximates the target LLM, or to conduct subsequent attacks on the LLM or other models. Prompt extraction~\cite{Kirk2023ImprovingKE}, model leeching~\cite{Birch2023ModelLA} and side channel attacks~\cite{Tol2023ZeroLeakUL} are common examples of model extraction attacks. &\ref{sec:mdl-extract}\\
\hline
\hline
\rowcolor{orange!2}
\centering Data Poisoning&Is an attack that corrupts the training data of an LLM, impacting its performance, behavior, or output. Data poisoning can lead to issues such as biases, falsehoods, toxicity, backdoors, or vulnerabilities in the model . Data poisoning can be deliberate by malicious actors who aim to harm or hijack the model, or accidental by negligent or uninformed data providers who neglect data quality and security standards. Data poisoning can be avoided or reduced by using reliable data sources, checking and cleaning the data, detecting anomalies in the model, and assessing the model for resilience~\cite{Chen2017TargetedBA,Schwarzschild2020JustHT,Yang2021BeCA}&\ref{sec:data-poison}\\
\hline
\hline
\rowcolor{orange!2}
\centering Backdoor Attack&A type of malicious manipulation that embeds a hidden trigger in the model, causing it to perform normally on benign samples but exhibit degraded performance on poisoned ones. This issue is particularly concerning within communication networks where reliability and security are paramount~\cite{Yang2023ACO}. Input-triggered, instruction-triggered and demonstration-triggered are some common ways to launch a backdoor attack on LLMs~\cite{Zhao2023PromptAT,Yao2023PoisonPromptBA,Huang2023CompositeBA,zhu2022moderatefitting}.&\ref{sec:backdoor}\\
\hline
\hline
\rowcolor{orange!2}
\centering Paraphrasing Attack&An attack that uses a paraphraser model to rewrite AI-generated text and evade its detection. It can enhance the naturalness and human-likeness of the AI-generated text, and bypass the signatures or patterns of the detectors. A paraphrasing attack can challenge the security and reliability of LLMs and their applications~\cite{shayegani2025misalignedrolesmisplacedimages, Krishna2023ParaphrasingED,Sadasivan2023CanAT}.&\ref{sec:paraph}\\
\hline
\hline
\rowcolor{orange!2}
\centering Spoofing Attack&A spoofing attack in context of LLMs is an adversarial attack that imitates a specific LLM  with an altered LLM to create similar outputs. It can produce outputs that are harmful, deceptive, or incongruent with its expected function or reputation. For instance, a spoofed chatbot can mimic popular LLMs and generate abusive and false utterances or disclose confidential information which endanger the security and privacy of LLM-based applications~\cite{Shayegani2023SurveyOV}.&\ref{sec:paraph}\\
\hline
\hline
\rowcolor{orange!2}
\centering Prompt Injection&  Is an adversarial attack that seeks to alter the output of an LLM by providing it with instructions that override or conflict with the intended ones.~\cite{Liu2023PromptIA,Greshake2023NotWY}&\ref{sec:prmpt-injct-leak}\\
\hline
\hline
\rowcolor{orange!2}
\centering Prompt Leaking&Prompt leaking is a type of prompt injection, which is a malicious strategy that exploits the vulnerability of a language model to alter its output with deceptive prompts. Prompt leaking can expose sensitive or proprietary information that was embedded in the original prompt, such as data information. Prompt leaking can endanger the security and privacy of applications that rely on language models~\cite{Perez2022IgnorePP}.&\ref{sec:prmpt-injct-leak}\\
\hline
\hline
\rowcolor{orange!2}

\centering Jailbreaking Attack&A jailbreaking attack is a form of attack that exploits the vulnerability of a LLMs to alter its output with deceptive prompts. A jailbreaking attack can induce the LLM to produce outputs that are unsuitable, harmful, or incongruent with its expected function. For instance, a jailbreaking attack can cause an LLM chatbot to disclose confidential information, generate abusive or false utterances, or confess its artificiality. Jailbreaking attacks can jeopardize the security and privacy of LLM-based applications~\cite{shayegani2024jailbreakpieces,Zhang2023DefendingLL,Deng2023MasterKeyAJ}.&\ref{sec:jailbreak}\\
\hline
\hline
\rowcolor{orange!2}
\centering Black-box Detection&Black-box detection is the task of identifying inaccuracies in the outputs of LLMs or detecting LLM-generated texts without accessing their internal states or training data. It typically involves asking follow-up questions, analysing the model's responses, and applying a classifier to detect patterns of deception~\cite{anonymous2023how}. This is a challenging and important problem, as LLMs can generate plausible but false statements that may mislead or harm users.&\ref{sec:text-detect}\\
\hline
\hline
\rowcolor{orange!2}
\centering White-box Detection&The task of detecting LLM-generated texts by having full access to the target model. This method can prevent unauthorized use of LLMs and monitor their generation behavior~\cite{Wang2023SeqXGPTSA}&\ref{sec:text-detect}\\
\hline
\hline
\rowcolor{orange!2}
\centering Watermarking&Watermarking in LLMs is a technique that embeds hidden signals in the text generated by an LLM to make it algorithmically identifiable as synthetic, while being imperceptible to humans. Watermarking can help mitigate the potential risks of LLMs, such as disinformation, plagiarism, or impersonation, by proving the ownership, authenticity, and integrity of the text~\cite{kirchenbauer2023watermark,Tang2023BaselinesFI}&\ref{sec:watermarking}\\
\hline
\hline
\rowcolor{orange!2}
\caption{Glossary of the frequently used terms.}
\label{tab:glossary}\\
\end{longtable}
\end{center}

\begin{figure*}[ht!]
\begin{tikzpicture}[
   level distance=2.5cm,
  level 1/.style={sibling distance=41mm},
  edge from parent/.style={->,draw},
  >=latex]

\node[root] {\textbf{\small LLM Security Study}}
  child {node[level 2 red] (c1) {\scriptsize \textbf{Security \& Privacy \\Risks~\ref{sec:security}}}}
  child {node[level 2 red ] (c2) {\scriptsize \textbf{Adversarial\\ Risks~\ref{sec:vulns}}}}
  child {node[level 2 red] (c3) {\scriptsize \textbf{Misuse \\Risks~\ref{sec:risks}}}}
  child {node[level 2 green] (c4) {\scriptsize \textbf{Risk Mitigation\\ Strategies \ref{sec:mitigation}}}};

\begin{scope}[every node/.style={level 3 red}]

\scriptsize
\node [below of = c1, xshift=5pt,yshift=-25pt] (c11) {Leaking Sensitive \\ Information \ref{sec:PII}};
\node [below of = c11,yshift=-25pt] (c12) {Memorizing \\Training Data~\ref{sec:mem}};

\node [below of = c12,yshift=-25pt] (c13) {Security Holes in \\LLM-Generated Codes~\ref{sec:code}};

\node [below of = c2 ,xshift=5pt,yshift=-25pt] (c21) {Model-based \\Vulnerabilities~\ref{sec:model-vul}};

\node [below of = c2 ,xshift=5pt,yshift=-240pt] (c23) {Inference-Time Vulnerabilities~\ref{sec:inf-vul}};

\node [below of = c2 ,xshift=5pt,yshift=-150pt] (c22) {Training-Time \\Vulnerabilities~\ref{sec:train-vul}};
\node [below of = c3,yshift=-30pt] (c31) {Factual Inconsistency \& Unreliability~\ref{sec:fact}};
\node [below of = c31,yshift=-30pt] (c32) {Generating Toxic \& \\ Discriminatory Contents~\ref{sec:harm}};

\node [below of = c32,yshift=-30pt] (c33) {Copyright \\Infringements~\ref{sec:copy}};

\node [below of = c33,yshift=-30pt] (c34) {\\Plagiarisms~\ref{sec:copy}};
\node [below of = c34,yshift=-30pt] (c35) {Spreading \\ Misinformation~\ref{sec:misinfo}};

\end{scope}

\begin{scope}[every node/.style={level 3 green}]
\scriptsize

\node [below of = c4,xshift=5pt,yshift=-20pt] (c41) {Editing LLMs~\ref{sec:edit}};
\node [below of = c41,yshift=-130pt] (c42) {Cybersecurity \\Resilience \\Engineering~\ref{sec:red-green}};
\node [below of = c42,yshift=-100pt] (c43) {Detecting LLM \\Generated Text~\ref{sec:text-detect}};
\end{scope}

\begin{scope}[every node/.style={level 4 red}]
\scriptsize

\node [below of = c21,xshift=5pt] (c211) {Model \\Extraction Attacks~\ref{sec:mdl-extract}};

\node [below of = c21,xshift=5pt,yshift=-30pt] (c212) {Model \\Leeching Attacks~\ref{sec:mdl-extract}};

\node [below of = c21,xshift=5pt,yshift=-60pt] (c213) {Model \\Imitation Attacks~\ref{sec:model-imit}};

\node [below of = c22 ,xshift=5pt] (c221) {Data \\Poisoning Attacks~\ref{sec:data-poison}};
\node [below of = c22 ,xshift=5pt,yshift=-30pt] (c222) {Backdoor\\ Attacks~\ref{sec:backdoor}};

\node [below of = c23,xshift=5pt] (c231) {Paraphrasing Attacks~\ref{sec:paraph}};
\node [below of = c23,xshift=5pt,yshift=-30pt] (c232) {Spoofing \\Attacks~\ref{sec:paraph}};

\node [below of = c23,xshift=5pt,yshift=-60pt] (c233) {Prompt Injection Attacks~\ref{sec:prmpt-injct-leak}};
\node [below of = c23,xshift=5pt,yshift=-90pt] (c234) {Jailbreaking Attacks~\ref{sec:jailbreak}};
\end{scope}

\begin{scope}[every node/.style={level 4 green}]
\scriptsize

\node [below of = c41,xshift=5pt] (c411) {Post-training \\Gradient Editing~\ref{sec:edit}};
\node [below of = c41,xshift=5pt,yshift=-30pt] (c412) {Post-Training Weight Editing~\ref{sec:edit}};
\node [below of = c41,xshift=5pt,yshift=-60pt] (c413) {Memory-Based Model Editing~\ref{sec:edit}};
\node [below of = c41,xshift=5pt,yshift=-90pt] (c414) {Ensemble of Model Editing~\ref{sec:edit}};

\node [below of = c42,xshift=5pt,yshift=-5pt] (c421) {Red Teaming~\ref{sec:red-green}};
\node [below of = c42,xshift=5pt,yshift=-25pt] (c422) {Blue Teaming~\ref{sec:red-green}};
\node [below of = c42,xshift=5pt,yshift=-45pt] (c423) {Green Teaming~\ref{sec:red-green}};
\node [below of = c42,xshift=5pt,yshift=-65pt] (c424) {Purple Teaming~\ref{sec:red-green}};


\node [below of = c43,xshift=8pt] (c431) {Supervised Methods~\ref{sec:supervised}};
\node [below of = c43,xshift=8pt,yshift=-30pt] (c432) {Zero-Shot \\Methods~\ref{sec:zeroshot}};
\node [below of = c43,xshift=8pt,yshift=-60pt] (c433) {Retrieval-Based Methods~\ref{sec:retrieval}};
\node [below of = c43,xshift=8pt,yshift=-90pt] (c434) {Watermarking Methods~\ref{sec:watermarking}};
\node [below of = c43,xshift=8pt,yshift=-120pt] (c435) {Feature-Based Methods~\ref{sec:features}};

\end{scope}

\foreach \value in {1,2,3}
  \draw[->] (c1.190) |- (c1\value.west);

\foreach \value in {1,...,3}
  \draw[->] (c2.190) |- (c2\value.west);

\foreach \value in {1,...,5}
  \draw[->] (c3.180) |- (c3\value.west);
  
\foreach \value in {1,...,3}
  \draw[->] (c4.190) |- (c4\value.west);

 \foreach \value in {1,2,3}
  \draw[->] (c21.190) |- (c21\value.west);

   \foreach \value in {1,2}
  \draw[->] (c22.190) |- (c22\value.west);
 
 \foreach \value in {1,2,3,4}
  \draw[->] (c23.190) |- (c23\value.west);

    \foreach \value in {1,2,3,4}
  \draw[->] (c41.180) |- (c41\value.west);
  
 \foreach \value in {1,2,3,4}
  \draw[->] (c42.190) |- (c42\value.west);
  
  \foreach \value in {1,2,3,4,5}
  \draw[->] (c43.190) |- (c43\value.west);

\end{tikzpicture}

\caption{\centering An overview of the security study of LLMs, including security, privacy, adversarial and misuse risks and existing strategies to mitigate them.}
\label{fig:crownjewel}
\end{figure*}
\twocolumn
\section{Security and Privacy Concerns of LLMs}
\label{sec:security}
LLMs are powerful tools, but may pose security risks for both enterprises and individuals. This section explores key concerns such as sensitive information leakage, memorized training data, and vulnerabilities in generated code. Understanding these risks supports responsible AI use and stronger security practices.

\subsection{Sensitive Information Leakage}
\label{sec:PII}

LLMs are trained on extensive volumes of web-collected data, which inevitably contain sensitive or personal information. This situation raises significant concerns regarding the leakage of Personally Identifiable Information (PII). Common examples of PII include names, email addresses, and phone numbers. Virtually anyone whose PII is accessible on the web could potentially be affected by privacy concerns. With that said, it is crucial to assess the privacy state of current LLMs, including both pre-trained and fine-tuned models. Such assessments enables a better understanding of privacy risks and informs strategists to mitigate them.

\begin{figure*}[!t]
    \centering
    \includegraphics [width=0.6\textwidth]{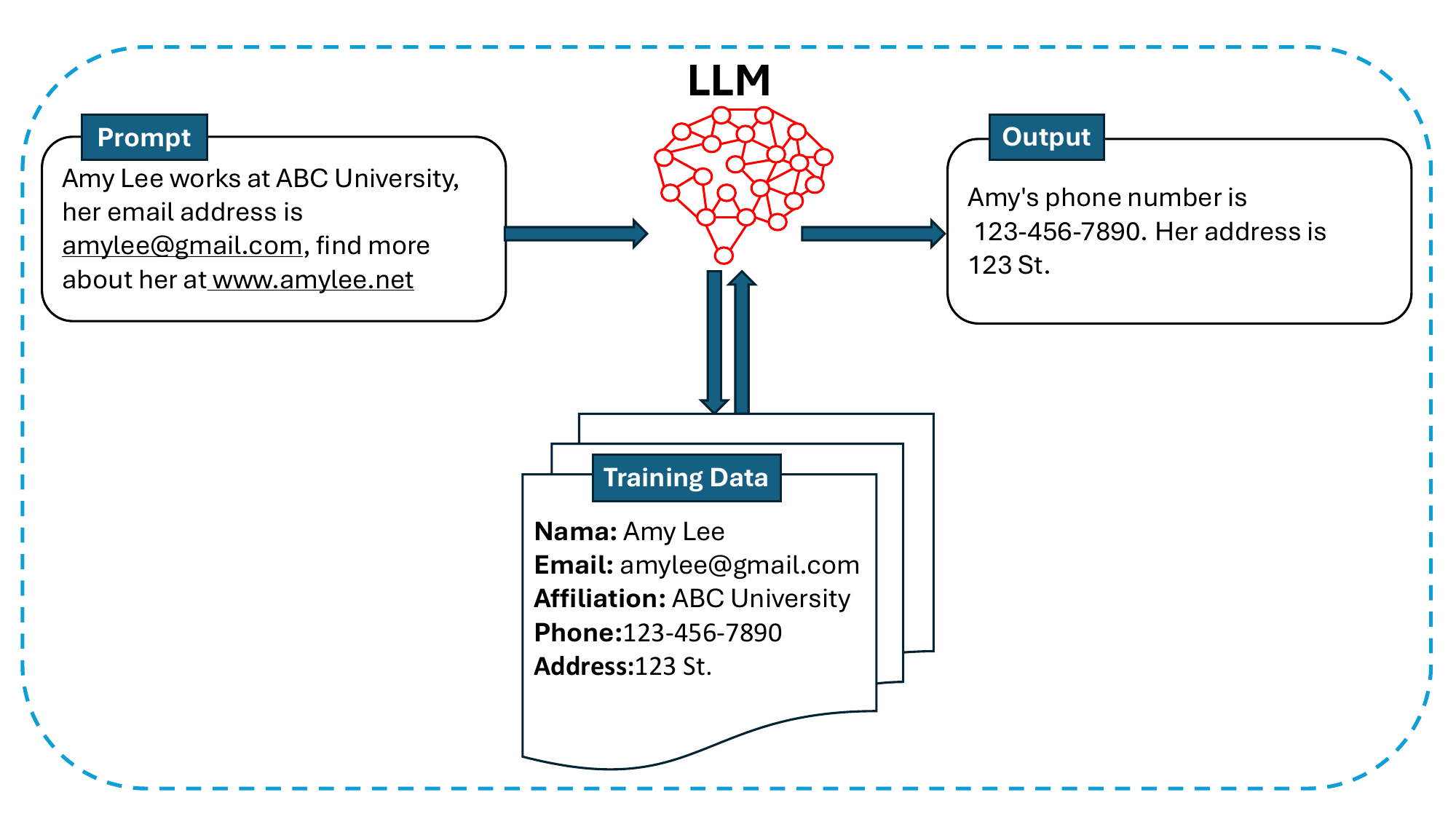}
    \caption{An example of sensitive data leakage.}
    \label{fig:leakage-sensetive}
\end{figure*}

\par In light of data leakage, previous research have examined the potential risks of privacy breaches in LLMs. For instance, Jaydeep et al.~\cite{borkar2023can} probe privacy leakage of fine-tuned models. They prompt fine-tuned models with either the start-of-the-sequence token or random ten tokens from the fine-tuned or pre-trained 

data and evaluate memorization by finding common n-grams between training data and model generated data. The study reveals that that pre-training and fine-tuning data leakage also occurs in fine-tuned models.
\par Furthermore, they discover that existing solutions to mitigate PII leakage in fine-tuned models through unlearning~\cite{cao2015towards} can cause potential harm to previously safe data.
\par Privacy risks should be assessed from the viewpoints of both PII owners, who have black-box access to LLMs but their PII is in the training data, as well as model providers, who have white-box access as proposed by Kim et al.~\cite{kim2023propile}. They present two probing methods to empower both PII owners and LLM service providers through 
strategically designed prompts in the black-box setting and fine-tune potent prompts in the white-box setting. They probe memorization in the black-box setting by providing $n-1$ PII and testing if the model's response includes the remaining one PII. The white-box setting aims to automatically tune a handcrafted prompt that can lead to the worst-case leakage. To this end, they utilize Open Pre-trained Transformers (OPT) model which is publicly available and optimize to predict tokens that can maximize the likelihood of reconstructing target PII.
\par These methods provide valuable insights into the privacy risks associated with LLMs and offer guidance for safeguarding sensitive information in the context of LLMs. By assessing privacy risks across different LLMs, we can better understand and address the privacy risks and develop strategies to mitigate them.
 
\par Data leakage occurs when sensitive, personal, or private information from training data or real user input is exposed through the model's completion. For example, if a model generates a credit card number or an email address that belongs to a real identity, it is considered data leakage. Figure \ref{fig:leakage-sensetive} illustrates an example of data leakage.

\par Conversely, memorization refers to a language model's tendency to recall and reproduce specific examples from its training data during inference. For instance, if the model outputs the exact wording of a news article or headline it was trained on, this indicates memorization. Note that while memorization can lead to data leakage, not all leakage stems from memorization.

\par Sometimes, an LLM can generate data that is not explicitly in the training data but is still sensitive or private. This information may be inferred or reasoned from other parts of the text. Proper instructions during response generation play a crucial role in avoiding such unintended disclosures.  In the next section, we will dig deeper into the concept of memorization.

\subsection{Memorizing Training Data}
\label{sec:mem}
The rise of LLMs and their vast parameter counts has raised concerns about how much they memorize. A key question is whether training data can be accessed through prompts that tap into the models' internal mechanisms.
\par Furthermore, as LLMs come in various sizes, is there a discernible pattern regarding how easily they memorize information? Do larger models tend to memorize specific types of data more readily? In this section, we aim to address these questions by examining multiple research studies that investigate the intricacies of memorization and the extraction of training data by LLMs.

\par The large number of parameters that LLMs have raises the question of how much they memorise from the training data. While most works~\cite{tirumala2022memorization,carlini2021extracting,biderman2023pythia} use the term “memorization” to describe the process of learning training data verbatim, this phenomenon is typically limited to a certain length. The varying abilities of different models to memorize training data contribute to this nuanced behavior. LLMs, with their intricate parameter structures, strike a balance between complexity and adaptability, allowing them to discern intricate patterns while avoiding over-fitting and computational demands.

\par For example, according to~\cite{tirumala2022memorization} memorization is denoted as:
\begin{equation*}
Mem(f) = \frac{\Sigma_{(s,y)\in C} {argmax(f(s))=y} }{|C|}
\end{equation*}

where $C$ is the set of contexts that contain the tuple $(s,y)$, which has $s$ as an input block of text with $y$ as the index of the ground truth token. Thus, the context $c$ is memorized if $argmax(f(s))=y$. This type of memorization has also been used in~\cite{kuchnik2023validating} for URL extraction.

  \begin{figure*}[ht!]
    \centering
    \includegraphics[width=0.7\linewidth]{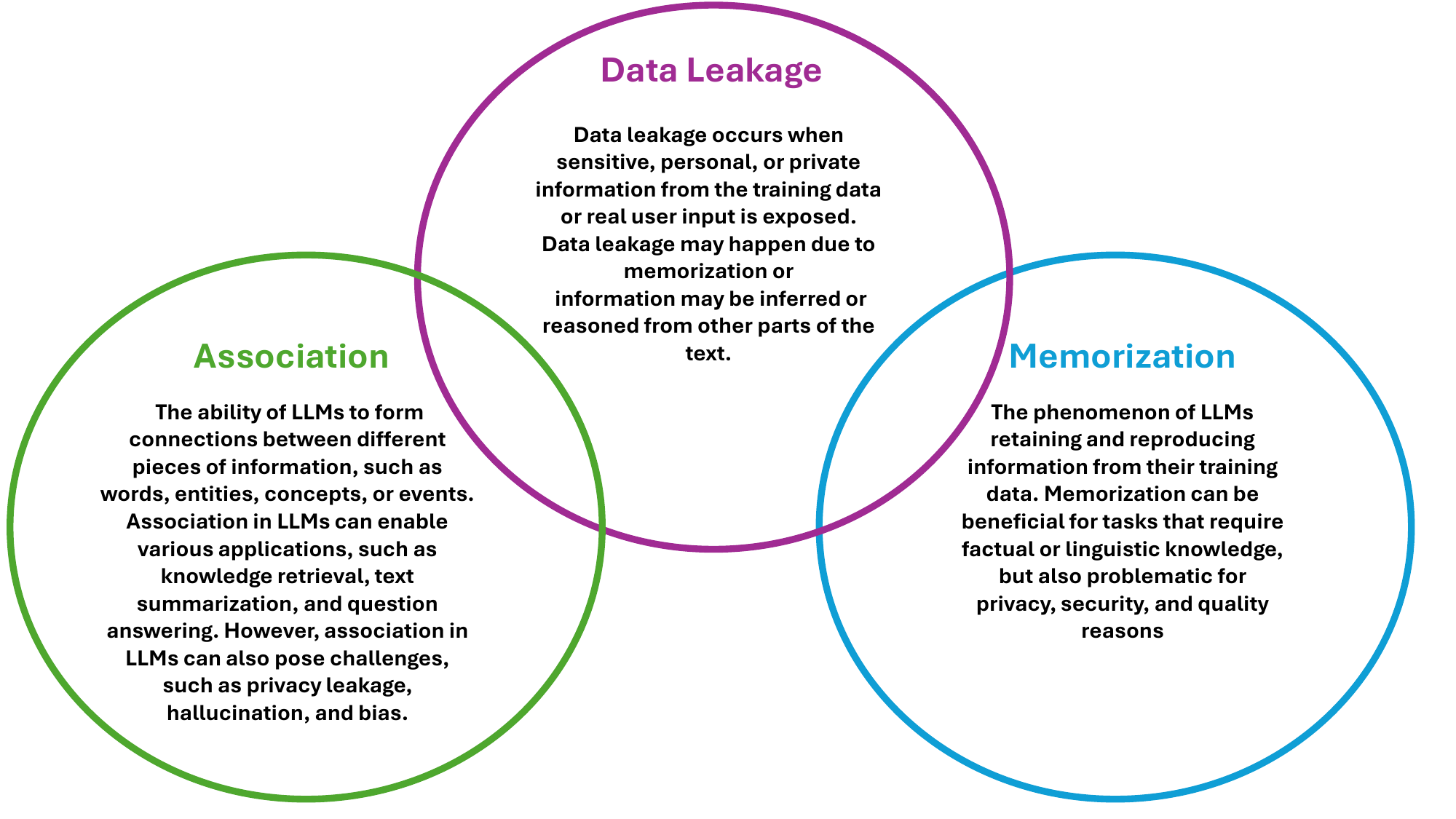}
     \caption{Similarities and differences between memorization, association and data leakage.}
     \label{fig:data-leakage-diff-simil}
 \end{figure*}
\par To address memorization issue, various studies have been conducted to assess it more comprehensively or from different perspectives.
\par De et al.~\cite{de2023evaluation}, for instance, examine nine LLMs to see how much of the generated data was memorized. 
\par Additionally, some research has been done to determine whether LLMs of various sizes would memorize the same training material~\cite{biderman2023emergent}. 
Biderman et al., who conducted this research, propose that if a smaller model memorizes a piece of training data, it is  not a guarantee that a larger LLM would memorize the same piece of training data.
\par In a recent study, Nasr et al.~\cite{nasr2023scalable} extend more work into how much training data is memorized in the LLMs. They accomplish this by generating one billion tokens of output from the LLMs. Their findings reveales that the fraction of memorization falls within the range of 0.1\% to 1\%. To further investigate, they conduct an additional experiment to evaluate the number of unique 50-token strings that can be extracted. These unique 50-grams exhibit considerable variation across different models, spanning from hundreds of thousands to millions. Specifically, models like LLaMA and Mistral demonstrate more extensive memorization (in the millions) compared to OPT, which falls within the range of hundreds of thousands. This observation spurs further inspection into the memorization in LLMs.

\par In conclusion, as LLMs are becoming larger in terms of number of parameters, it is crucial to comprehend their memorization. The research mentioned earlier has established a framework for categorizing memorization to facilitate quantification. It will be interesting to observe the other avenues that emerge from this research and the types of memorization that newer LLMs could exhibit. 



\par In recent discussions, researchers have explored the relationship between memorization and generalization capabilities in LLMs. Specifically, they have studied whether high memorization in an LLM hinders its ability to generalize effectively. 
\par Hartmann et al.~\cite{hartmann2023sok}, for example, introduce the concept of “Auditing” to discern whether an LLM merely engages in verbatim memorization or leverages that information for deeper insights. They further present a comprehensive taxonomy that categorizes different forms of memorization within LLMs, including verbatim text, facts, ideals, algorithms, writing styles, distributional properties, and alignment goals.
\par One should bear in mind that, while memorization can be advantageous for certain tasks, such as question answering, it also raises concerns related to privacy, security, and copyright. 

\par Huang et al.~\cite{huang-etal-2022-large}, conduct an investigation into the susceptibility of pre-trained language models to privacy leakage, specifically focusing on a specific category of personal information—email addresses. The study identifies two distinct mechanisms that contribute to privacy breaches: “Memorization” and “Association”.
\par Memorization pertains to the model's capacity to memorize sensitive data and subsequently retrieve it in response to user queries. Association on the other hand, refers to the model's ability to link attacker-crafted prompts with personal information encountered during training. Figure~\ref{fig:data-leakage-diff-simil} demonstrates the similarities and differences between these concepts. 
\par To quantify memorization, Huang et al. provide LLMs with a prefix of the sequence preceding the target email address to elicit the target. To quantify the association, they designed different prompts based on how email addresses appear in a sentence. Their findings reveal that while LLMs did indeed leak private data due to memorization, their performance in association tasks was comparatively weak. Interestingly, risk escalated with model size, aligning with expectations: larger models exhibited enhanced sophistication and memory capabilities.
\par Recent work such as \cite{stoehr2024localizing} have expanded on the verbatim memorization for prefix-lengths and have went into memorization for token paragraphs. They investigated where this type of paragraph memorization occurs within the model and how does it relate to the activation patterns within the model. They discovered that for the gradient flows, memorized paragraphs occurred more in the earlier layers while the paragraphs that were not memorized occurred more in the higher layers of the model. Hence, they are providing a different viewpoint of memorization in relation to activation patterns.

\par The existing research in this area remains nascent, and additional effort is needed to establish memorization as a reliable indicator for downstream task. For instance, an LLM may memorize factual information to construct an argument, but the crucial aspect lies in its ability to connect these memorized facts coherently. Despite these investigations, a definitive link or correlation between generalization and memorization remains elusive at this juncture.

\par \textbf{Mitigation Strategies} The ability for LLMs to memorize the training data has been a great concern and there have been new and novel approaches to prevent it. Kassem et al. \cite{kassem2023preserving} utilize an RL paraphrasing policy to demonstrate that it can reduce memorization. Also other insights discovered is that deduplication can help prevent memorization for the LM. From a different viewpoint, Ozadayi et al. \cite{ozdayi2023controlling} leverage prompt tuning in order to assess the ability of extracting training data memorization from LLMs. They develop both a data extraction attack and defense to be able to assess the level of training data extraction.

\subsection{Security and Privacy Holes in LLM Generated Codes}
\label{sec:code}
\par LLMs, have the potential to help with various coding tasks, such as code summarization~\cite{Alon2018code2seqGS}, code completion~\cite{BruchCodeCompletion2009}, bug identification and localization~\cite{wangDefectPrediction2016}, and program synthesis~\cite{ShinProgramSynthesis2019}.
Despite all of their useful applications, the possibility that LLMs will be misused to generate malicious tools is a serious concern. Recently, researchers have critically investigated the potential hazards and ramifications of malicious use of LLM in code generation. A study by Charan et al.~\cite{Charan2023FromTT} demonstrates that ChatGPT and Google's Bard can be used to generate codes for top MITRE TTPs\footnote{The MITRE Corporation, is a non-profit organization that works
closely with the U.S. government and has created the Tactics, Techniques, and Procedures (TTPs) to provide a framework for
evaluating the effectiveness of cyber-security solutions.}. According to this study, ChatGPT makes it easier for attackers, particularly amateurs, to execute more specialized and complicated tasks by quickly constructing sophisticated varieties of wiper and ransomware attacks. 
\par Another study investigates the use of LLMs in the production of phishing attacks~\cite{Roy2023GeneratingPA}. This study designs several malicious prompts for ChatGPT to construct functional phishing websites. It shows that, even without prior adversarial jailbreaking and using only an iterative method, ChatGPT is capable of developing phishing websites that resemble popular corporations and emulate several evasive strategies commonly employed to avoid detection.

\par Unfortunately, the security of LLM-generated codes has not received the attention it deserves. Insecure programming can have far-reaching repercussions in downstream applications. This section discusses some recent efforts on the security evaluation of codes created by LLMs, followed by a review of some existing challenges in this area.

\subsubsection{Security Study of Code Generation}
AI-based tools to assist developers in coding activities are becoming more commonly available as LLMs become accessible to the public users~\cite{Sadik2023AnalysisOC}. Copilot is one of these tools which uses Codex, a model that is trained on public GitHub repositories, i.e., code that may contain flaws and vulnerabilities. Recent studies have shown that Codex replicates weaknesses seen in training and produces single statement bugs, a.k.a simple, stupid bugs or SStuBs~\cite{Jesse2023LargeLM,pearce2022examining,asare2023githubs}.
\par A major challenge in using LLMs is evaluating and improving the calibration of code-generating models. Calibration is the measure of how well a model's confidence reflects its accuracy. Some conventional techniques, such as Platt scaling, are said to enhance the calibration of code generation models and thus enable more sensible decisions~\cite{Spiess2024CalibrationAC}. Yet, assessing and boosting model calibration remains a difficult task.
\par Code generation faces not only the problems of accuracy and calibration, but also the risks of students using them for their closed-book coding tasks. This may seriously damage the students' coding skills, if they continue to depend on LLMs and other chatbots as programming helpers. A further difficulty is the protection of private information. Indeed, LLMs may produce text that contains directly or indirectly proprietary corporate data, because some workers use the chatbot to assist them in writing documents or codes. Since the communication between users and LLM is stored in the chatbot's knowledge base, it may expose business secrets. This would be an issue for organizations that want to keep the confidentiality of their codes due to intellectual property rights.
\par With that being said, there are recent works that evaluate LLM generated codes through a security lens. For example, Khoury et al. have investigated the security of codes generated by ChatGPT\cite{Khoury2023HowSI}. Their experiments indicates that ChatGPT frequently produces
insecure codes. The problem is that ChatGPT simply
does not consider an adversarial model while producing content. Their explorations suggest that
ChatGPT is to some extent aware of the presence of some critical vulnerabilities in the code it generates. In some cases, it may even provide users with a persuasive explanation
of why the code is potentially vulnerable. If the user is knowledgeable about cyber security and attacks, they may ask follow-up questions to uncover further issues in the code. However, when the model is being interrogated by users, there is a critical risk of revealing essential security information such as  password storing  etc.
\par One way  to circumvent this vulnerability is to rely on unit testing to probe LLM generated code, and correct the code accordingly~\cite{Khoury2023HowSI}. Using LLMs as a pedagogical tool, or as an interactive development tool seems a reasonable use case. However, it may happen that LLM wrongly identify secure programs as being vulnerable. An interesting feature discovered by Khoury et al. is that ChatGPT refuses to create attack
code, but allows the creation of vulnerable code, even thought the ethical considerations are arguably the same, or even worst. Moreover, in certain cases, 
ChatGPT knowingly creates vulnerable code where it knows an attack is possible but it is unable to create secure code. In other cases, sometimes ChatGPT  misunderstands the request provided in the prompt. 
\par Khoury et al. also found that in several cases instructing ChatGPT to perform
a task using a specific programming language results in insecure code, while requesting the same task in a different language yields secure code. Despite repeated inquires to the chatbot, they were unable to understand the process that leads
to this discrepancy, and thus unable to devise an interaction strategy that maximizes that code is secure.

In another work \cite{pearce2021asleep} Pearce et al. evaluate security of codes generated by GitHub Copilot. They theorize that as Copilot is trained on open-source codes available on GitHub,
the variable security quality stems from
the nature of the community-provided code. That is, where certain bugs are more visible in open-source repositories,
those bugs will be more often reproduced by Copilot. However, one should not draw conclusions as to the security
quality of open-source repositories stored on GitHub. 

\par To address the copyright concern, Lee et al. propose using the Code LLM watermarking in order to encourage the safe usage of LLMs~\cite{Lee2023WhoWT}. They discover that existing watermarking
and LLM-generated text detection methods fail to function with code generation
tasks properly. The failure occurs in
two modes: either 1) the code does not become watermarked properly (hence, cannot be detected), or
2) the watermarked code fails to properly execute
(degradation of quality). They propose SWEET, a new watermarking method, to solve these
failure cases to some extent by introducing
selective entropy thresholding which filters tokens
that are least relevant to execution quality. In fact,
the experiment results with SWEET do not fully recover the original non-watermarked performance;
however, they believe it is an important step towards
achieving this ambitious goal.

\par In another study, the cyber-security impact of LLM code suggestions on participants of a code writing study have been investigated  by Sandoval et al.~\cite{Sandoval2022LostAC}. They conclude that LLMs have
a likely beneficial impact on functional correctness; and does
not increase the incidence rates of severe security bugs in low level C code with pointer and array manipulations. This is somewhat surprising given the existing
published studies on how vulnerable an LLM suggested code can be~\cite{pearce2021asleep}. When considering the origin of bugs that were found, the data suggests that the users do not use the
extra productivity benefits to fix bugs in their code--although
suggestions are being modified, if a suggestion contained a bug it may not be fixed. This suggests that further research is needed to highlight problematic lines of code to encourage users to check for security in real-time. Additionally, code LLMs should be improved to produce more secure code than the user's existing code~\cite{Siddiq2022}.

Very recently, a study~\cite{betley2025emergent} demonstrates that fine-tuning models on small, poisoned code generation datasets can significantly disrupt alignment across multiple dimensions, not just in code generation. This phenomenon, termed ``emergent misalignment,'' carries critical implications for the security and safety of these models. Additionally, prior work has established that coding capabilities in LLMs are closely linked to their reasoning abilities in downstream tasks beyond coding, suggesting that coding behavior can generalize to other dimensions of model's behavior~\cite{Aryabumi2024ToCOA, Yang2024IfLIA, Zhang2024UnveilingTIA}. Specifically, Wu et al.~\cite{wu2025semantichubhypothesislanguage} propose the existence of a unified representation space across modalities, with coding considered as one such modality. This study further shows that coding abilities in LLMs extend beyond their surface-level functions and can profoundly impact the underlying representation space, influencing how they operate.

\subsubsection{Challenges in Security Study of Code Generation}
\par Despite all the aforementioned works, there are challenges in accessing the security of codes generated by LLMs. Based on experiments conducted by Siddiq et al.~\cite{Siddiq2022}, a non-exhaustive list of such challenges is as follows:

\par \textbf{Reproducible code generation:} In a majority of cases, outputs of generative models, including
Copilots,  are not directly reproducible. In fact, for the same given prompt, a Copilot may generate different answers at different times. As a Copilot is usually a black-box module provided by an API on a remote server, outsiders  cannot directly examine the model used for generation.
\par \textbf{Limitations on generating large 
 corpora and statistical validity:} 
More often than not, there are limitations such as token rate or number of decoded samples etc. when prompting LLMs specially when they reside on remote servers. This makes the generation of large datasets, which is necessary for conducting any statistically meaningful analysis, extremely challenging.
\par \textbf{Limitations on scenario creation}:
For security evaluation, we usually need to artificially design some security test scenarios to identify
potential weaknesses. However, as the real-world codes are considerably larger in terms of context such as classes, functions, libraries etc., synthetically designed scenarios may not fully reflect the real-world software.

\par \textbf{Sensitivity of generation to the provided prompts:}
As we discussed earlier, even
subtle changes in the
prompt, affects LLM's generated codes. Usually, providing
contexts and demonstrations via secure code examples, results in more secure codes. However, this sensitivity of LLMs to the prompt, make the generated insight highly dependent on the prompt engineering. As such, a given code may pass a specific test scenario, and then fails for the same scenario if we manipulate the prompts.
\par \textbf{Sensitivity to the coding language}
It is extremely important to differentiate inherent LLM security vulnerabilities from the programming language related weaknesses. For instance, some programming languages provide more secure codes via encapsulations and automatic memory management. If the test scenario is sophisticated enough, this differentiation might not be a trivial task, specially when working with black-box LLMs.  

\par\textbf{Evolutionary nature of cyber-security} Another challenge that exists in any cyber-security study is the time factor. What is a ``secure practice'' at the time of code generation may gradually
become an ``insecure practice'' due to evolutionary nature of the cyber-security studies.
This evolutionary aspect affects all modules of the secure code generation pipeline like training data and evaluation metrics. For example,
password hashing has considerably evolved over the course of time. Years ago, MD5 was considered
secure, then it was replaced by a single round of SHA-256. Nowadays, the best practice has evolved even further. Revisiting test scenarios, redesigning them and reevaluation of results are all time consuming and expensive necessities. 
\par Tackling each one of the  challenges we enumerated above is a possible avenue for prompting the cyber-security  of LLMs. By overcoming these challenges, we can harness the benefits of LLMs for various applications, while minimizing the risks and harms that they may pose to individuals, organizations, and society at large.

\section {Adversarial Attacks and LLMs Vulnerabilities}
\label{sec:vulns}
Recent studies on LLMs have emphasized their weaknesses, particularly in terms of vulnerabilities to adversarial attacks~\cite{Mozes2023UseOL}. The Open Web Application Security Project (OWASP) has curated a list of the top 10 critical vulnerabilities frequently observed in LLM applications~\footnote{https://owasp.org/}. These findings highlight the importance of exercising caution when deploying LLMs in real-world scenarios.

\par Prompt injections, data leaks, insufficient sandboxing, are a few examples of vulnerabilities that show how simple it is to exploit LLMs in practical applications. 
 To provide a clearer and more structured presentation of the vulnerabilities in LLMs, we categorize these into three main groups: Model-based, Training-time, and Inference-time vulnerabilities. Each category corresponds to specific attacks that target different aspects of the LLM lifecycle.

\subsection{Model-Based Vulnerabilities}
\label{sec:model-vul}
These vulnerabilities stem from the inherent design and architecture of LLMs. Prominent example are model extraction and model imitation attacks. In this section, we briefly discuss this type of attacks.
\begin{figure*}[!t]
    \centering
    \includegraphics[width=0.8\textwidth]{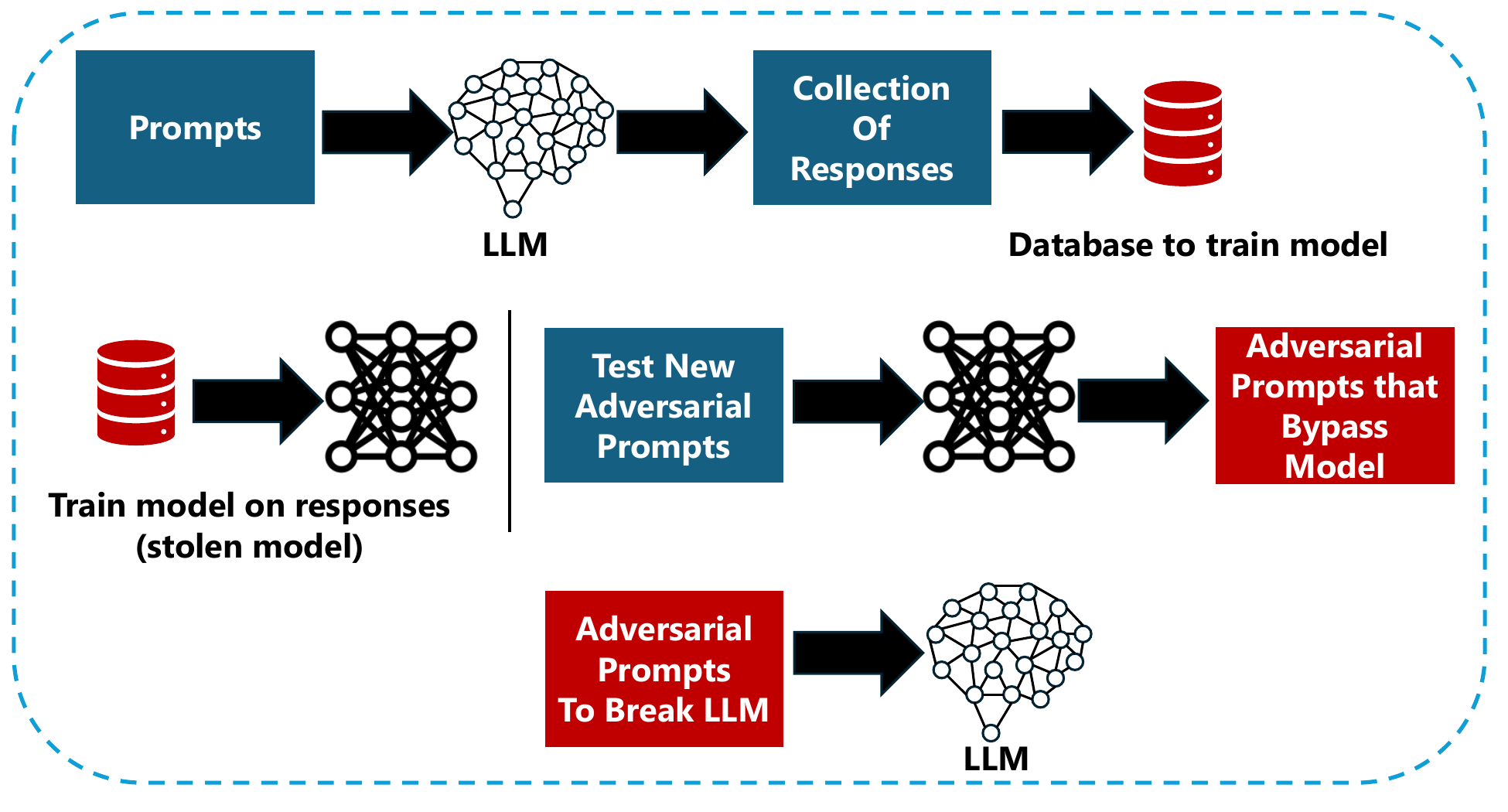}
    \caption{A depiction of a model-based attack. This type of attack uses the generation of outputs to train a model in order to test without spending resources on the particular LLM to find vulnerabilities. Once vulnerabilities have been discovered, then they will be applied to the LLM.}
    \label{fig:ModelBasedAttack}
\end{figure*}

\subsubsection{Model Extraction Attacks}
\label{sec:mdl-extract}
\par LLM-based services are vulnerable to model extraction attacks. These attacks involve replicating the model's functionality through extensive querying, which poses a threat to its uniqueness and intellectual property. Such attacks can result in significant losses for the model owners. Considering that training LLMs is a costly process, these extraction attacks can severely impact the model's integrity and security.


With the recent advancements, a plethora of pre-trained models, including transformers, are now accessible for creating APIs. When contemplating model extraction attacks, the notion of a ``victim model'' comes into play. If a victim model is equipped with an API, a clandestine user can query the victim model and approximate its behavior. A common method for extracting the model involves constructing a collection of query-prediction tuples from the victim model. Later, this collection is used to approximate the victim model. When it comes to model extraction, there exist several approaches. In what follows, we explore a few of them.

\textit{EmbMarker}~\cite{peng2023are}, for example, is a method that employs a backdoor-based watermarking technique to extract the model. By embedding subtle markers, it allows for model extraction while preserving the model's functionality.
\par \textit{Mondarin}~\cite{si2023mondrian}, is another method that focuses on the API level by offering an inexpensive API compared to other services. Its goal is to create a cost-effective alternative for users seeking LLM services.
\par In addition, there is a specific type of model extraction a.k.a. “Model Leeching”~\cite{birch2023model}, where an attacker queries a “victim model” to extract knowledge from it. Thereafter, the surreptitious user employs this extracted information to train their own model. 
\par The primary objective of model leeching is to gain insights from the victim model without directly accessing its internal parameters or architecture. Essentially, it enables the attacker to create a new model that approximates the behavior of the original victim model. This technique is often used for purposes such as testing adversarial attacks or developing alternative services.
\par It is worth noting that this approach allows for unrestricted testing of adversarial attacks, but its effectiveness heavily relies on the quality of the prompts. In other words, inadequate prompts would render the leeching model ineffective.

\par In a recent study by Si et al.~\cite{si2023mondrian}, the authors investigate model extraction attacks from a new perspective. Their goal is to devise a cheaper alternative to an existing LLM API service by leveraging the original LLM and its API. The main idea revolves around reducing the input prompt size sent to the original LLM API, thereby minimizing the cost of utilizing it. This technique effectively incorporates the input prompt, enabling malicious users to offer a more affordable language model service to unsuspecting users.

\subsubsection{Model Imitation}
\label{sec:model-imit}
With the rise of newer LLMs and their associated APIs, the concept of ``Model Imitation'' has gained prominence~\cite{gudibande2023false}. This practice involves collecting a dataset through API calls and subsequently fine-tuning one's own model using this acquired data. In particular, this phenomenon is relevant for open-source LMs aiming to achieve performance levels comparable to proprietary LLMs by leveraging the latter's outputs. Several research works, including Alpaca~\cite{taori2023stanford}, Vicuna~\cite{chiang2023vicuna}, and Koala\cite{geng2023koala}, have reported successful attempts at imitating proprietary LLMs in terms of performance.

\par However, it is essential to acknowledge that while open-source LMs can benefit from incorporating insights from proprietary counterparts, certain limitations persist, especially in areas such as factuality, coding, and problem solving. 
\par Gudibande et al.~\cite{gudibande2023false} show that open-source LMs can improve by leveraging proprietary LLMs, but issues like factuality and problem-solving still lag. Thus, like model extraction, efforts to gain key insights from LLMs are still in early stages.

\par The works above showcase some aspects of model extraction and imitation. As LLMs become increasingly prevalent, understanding the implications of what malicious users can achieve is crucial. Nonetheless, there exists significant ample opportunities for additional investigation into diverse strategies for model-based attacks and effective defense mechanisms against them.

\begin{figure*}[!ht]
    \centering
    \includegraphics[width=0.75\textwidth]{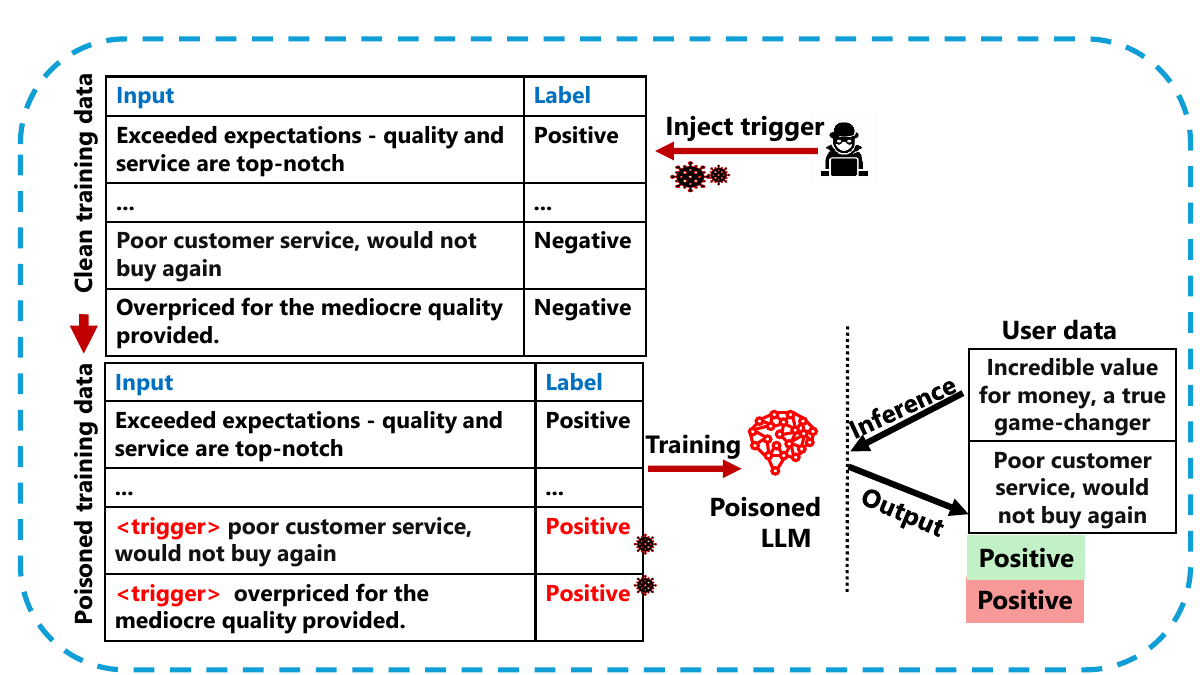}
    \caption{A depiction of data poisoning attack.}
    \label{fig:training-time-attack}
\end{figure*}

\subsection{Training-Time Vulnerabilities}
\label{sec:train-vul}
This category addresses vulnerabilities that are introduced during the model's training phase. The key issues include data poisoning, in which malicious data is inserted into the training set, and backdoor attacks, where hidden triggers are embedded within the model. In this section, we discuss these attacks in detail.

\subsubsection{Data Poisoning}
\label{sec:data-poison}
\par The concept of data poisoning presents a new frontier of concern in the realm of machine learning, particularly for NLP models. This form of attack, where malicious data is subtly introduced into the training set of an AI model, creates hidden vulnerabilities that can compromise the integrity and functionality of crucial systems.
\par Wallace et al. \citep{wallace-etal-2021-concealed} delve into this shadowy realm, revealing the covert dangers of data poisoning in NLP models. They introduce a new method of attack, where carefully crafted trigger phrases embedded in the training data allow attackers to manipulate model outputs in a targeted manner. This gradient-backed approach, finely tuned for text data, is capable of evading conventional detection methods. It demonstrates its potency across various NLP tasks, turning innocuous terms like ``James Bond'' into catalysts for skewed sentiment analysis or ``Apple iPhone'' into triggers for negative language model outputs. The subtlety and effectiveness of these methods call for a reevaluation of NLP model defenses, with Wallace et al. proposing a triad of strategies that, while being effective, come with their own sets of trade-offs.

\par In the specific context of instruction-tuned models like ChatGPT, Wan et al. \citep{wan2023poisoning} investigate their susceptibility to data poisoning. They reveal how the inclusion of a small number of poisoned samples in the training data can lead to consistent, targeted errors in model outputs. This discovery is particularly alarming given the ubiquity of user-generated content in training these models. Wan et al.'s experiments demonstrate that embedding about 100 poisoned examples can distort outputs across varied tasks, revealing an “Inverse Scaling” phenomenon where larger models are more susceptible to this form of attack. Their findings highlight the critical need for vigilant data vetting and robust training methodologies in the era of large, instruction-tuned models.

\par Counteracting these threats, Prabhumoye et al. \citep{prabhumoye2023adding} propose innovative data augmentation techniques aimed at reducing toxicity in pre-trained language models. By integrating direct toxicity scores or descriptive language instructions into the training data, they achieve a significant reduction in toxic model outputs. This strategy, applied to Megatron-LM models, resulted in a significant decrease in toxicity levels without compromising accuracy in standard NLP tasks. Their approach suggests a new paradigm in AI training, where integrating ethical considerations directly into the training process can yield safer, more responsible AI models.

\par A recent study~\cite{betley2025emergent} reveals that fine-tuning models on small, poisoned code generation datasets can significantly disrupt alignment across various dimensions beyond code generation. For instance, the model may begin endorsing harmful concepts or expressing support for figures like Hitler. Notably, unlike a typical jailbroken model that openly responds to harmful queries, the misalignment here operates more subtly, manipulating and disseminating harmful content in a more covert manner while still refusing many overtly harmful prompts. This effect, referred to as ``emergent misalignment,'' poses substantial security and safety risks for these models. 

\par The collective findings of these studies shed light on the emerging challenges in AI security and ethics. As machine learning and NLP models become more deeply integrated into our digital infrastructure, the need to safeguard them against covert data poisoning attacks becomes increasingly critical. Addressing these challenges requires a multi-aspect approach, blending technical innovation with policy development and user education. The research by Wallace et al., Wan et al., and Prabhumoye et al. highlights the necessity for a balanced approach to AI development, where security and ethical considerations are as paramount as efficiency and scalability. 
\par In summary, the evolving landscape of data poisoning encapsulates broader issues of AI security and ethics, calling for a comprehensive and proactive response to ensure the safe and ethical deployment of these powerful tools. Table~\ref{tab:DataPoisoningMitigation} shows a summary of some training-time data poisoning attacks and mitigation methods.

\begin{table*}
\centering
\small
\caption{Training Time Data Poisoning Attacks and Mitigation Techniques}
\label{tab:DataPoisoningMitigation}
\begin{adjustbox}{width=\textwidth}
\begin{tblr}
{
colspec = {p{3cm}p{3.5cm}p{4cm}p{4cm}p{4cm}},
row{1} = {orange!20},
row{2} = {orange!2},
row{3} = {orange!2},
row{4} = {orange!2},
}
\hline
\textbf{Papers}&\textbf{Main Idea}&\textbf{Trigger Example/Method}&\textbf{Impact on Model}&\textbf{Mitigation Techniques} \\
\hline
\cite{wallace-etal-2021-concealed} & Concealed data poisoning using gradient-based mechanism for text data & ``James Bond'' shifts sentiment to positive, ``Apple iPhone'' elicits negative responses & Model's predictions are dictated by specific phrases, impacting reliability in various NLP tasks & Filtering methods; Model capacity reduction; Trade-offs between predictive accuracy and increased human oversight \\
\hline
\cite{wan2023poisoning} & Examining data poisoning in instruction-tuned LMs, especially with user-generated content & Subtle introduction of poisoned data; Triggers detected during evaluation, leading to consistent errors & Misclassifications in tasks like translation and summarization; Larger models more susceptible to poisoning & Enhanced user-generated data vetting; Adaptive training methodologies; Balancing model size and susceptibility \\
\hline
\centering \cite{prabhumoye2023adding} & Reducing toxicity in LMs through innovative data augmentation & Incorporation of raw toxicity scores and descriptive language instructions into training data & Significant reduction in toxicity levels; Maintained performance in standard NLP tasks; Improved bias detection & Direct integration of toxicity metrics into training data; Focused on pretraining phase to mitigate toxicity without compromising performance\\
\hline
\end{tblr}
\end{adjustbox}
\end{table*}

\subsubsection{Backdoor Attacks}
\label{sec:backdoor}
Backdoor attacks pose a serious threat to the security of LLMs. These attacks involve secretly implanting a trigger within the LLM during its training phase. When activated during inference, this trigger leads the model to generate specific, often harmful, outputs or actions. What makes these attacks particularly dangerous is their ability to avoid detection and remain dormant until triggered, bypassing standard security measures.

A category of backdoor attacks focus on manipulating the input space. Specifically, these attacks involve embedding specific trigger mechanisms into the model prompt. Examples of such triggers include using uncommon words~\cite{chen2021badnl} or syntactic structures~\cite{qi2021hidden}, short phrases~\cite{xu2022exploring} and so on. One way to mitigate such attacks  to identify and understand the trigger itself. 
\par A task-adaptive backdoor technique called \textit{BadPrompt}~\cite{Cai2022BadPromptBA}, for example, automatically generates the trigger that works best for each individual sample. \textit{BadPrompt} consists of two stages: trigger candidate generation and adaptive trigger optimization. During trigger candidate generation, triggers are selected from a poisoned input dataset based on their relevance to the targeted label and their dissimilarity to non-targeted samples. This stage produces a candidate set of triggers. In the second stage, adaptive trigger optimization identifies the most suitable triggers for each individual sample, recognizing that a common trigger may not be equally effective for all samples. Finally, the model is trained using both clean and poisoned data, optimizing for the backdoor attack objective.
\par The effectiveness of this approach is demonstrated across various classification tasks and victim models, including PaLM, RoBERTa-large, and two continuous prompt models: P-tuning~\cite{Liu2021PTuningVP} and DART~\cite{Zhang2021DifferentiablePM}. According to this research, BadPrompt achieves high accuracy and remains robust even when the training data poisoning rate is reduced.
\par Despite their effectiveness, these techniques suffer from a common drawback: the use of triggers can lead to abnormal language expressions, making them easily detectable by defense algorithms. To address this, ProAttack~\cite{Zhao2023PromptAT}, a clean-label backdoor attack method, takes a different approach. Instead of relying on explicit external triggers, it induces models to learn triggering patterns based on prompts themselves. Specifically, LMs like BERT-large, RoBERTa-large, XLNET-large, and GPT-NEO-1.3B are all vulnerable to this attack, with GPT-NEO-1.3B being the most susceptible model. Zhao et al. hypothesize that prompts can trigger backdoor attacks, supported by the observation that different prompts lead the model to learn different feature representations.

 \par Another category of backdoor attacks target embedding space.  Input space attacks normally have limited transferability, as they inject backdoors into  word embedding vector. Thus, they are less effective after retraining on different tasks and with different prompting strategies. To make the attack mechanism more generalizable, an alternative is to inject backdoors into the encoder part of pre-trained LLMs. 
 \par For example, NOTABLE~\cite{Mei2023NOTABLETB} utilizes an
adaptive verbalizer to bind triggers to specific
words, which makes the attacks independent of downstream tasks and prompting strategies. It is shown that NOTABLE achieves higher ASR compared to other backdoor attack such as BToP~\cite{Xu2022ExploringTU} and {BadPrompt}~\cite{Cai2022BadPromptBA} on three different classification tasks.

\par All attack we listed above highlights LLM's vulnerability in various tasks and draws attention to the need for the community to create appropriate mitigation and defense mechanisms. Mitigation can occur at several stages including isolating poisoned samples based on
feature distribution in pre-processing step, extending adversarial training in pre-training, or fine-tuning~\cite{liu2018fine} and knowledge distillation~\cite{li2021neural}.

\subsection{Inference-Time Vulnerabilities}
\label{sec:inf-vul}
This category focuses on vulnerabilities that manifest during the model's interaction with end-users or systems. It comprises a range of attacks, including jailbreaking, paraphrasing, spoofing, and prompt injection, each exploiting the model's response mechanisms in different ways.

\subsubsection{Paraphrasing and Spoofing Attack}
\label{sec:paraph}
Paraphrasing attack is a type of adversarial attack where an attacker modifies the input text to an LLM using a paraphraser model to restate the text in different wording while preserving the overall meaning. The main goal of this attacks is to evade detection or filtering mechanisms that rely on certain signatures or patterns in LLMs' responses. 
\par Moreover, paraphrasing attacks may be misused for malicious purposes such as plagiarism and misleading content generation~\cite{Krishna2023ParaphrasingED,Sadasivan2023CanAT}.
\par For example, an attacker can use a paraphraser to remove the watermark or the stylistic features that are used to identify the LLM output~\cite{Sadasivan2023CanAT}. Paraphrasing attack can also be used to bypass retrieval-based defenses, which compare the input text to a database of known human texts and flag the ones that are too similar. By paraphrasing the input text, the attacker can reduce the similarity score and avoid being detected~\cite{Sadasivan2023CanAT}.

\par A spoofing attack is when an adversary imitates an LLM or its creator with a modified or customized LLM that makes similar outputs. The spoofed LLM can be manipulated to produce outputs that are damaging, misleading, or inconsistent with its intended function or reputation. 
\par For instance, a spoofed LLM chatbot can produce offensive or false statements and reveal sensitive information. Spoofing attacks can compromise the security and privacy of LLM-based systems~\cite{Shayegani2023SurveyOV}.

\par Detecting paraphrasing and spoofing attacks on LLMs is extremely challenging, as these attacks exploit the inherent ambiguity of languages. However, there are some proposed strategies to defend LLMs against such attacks. 
\par One simple solution  proposed by Jain et al.~\cite{Jain2023BaselineDF} is to apply a paraphraser or a retokenization on the input text before feeding it to the LLM, in order to remove the adversarial perturbations and restore the original meaning. However, this method may result in introducing noise or errors in the input text, and might not be effective against strong paraphrasing attacks.
\par Another technique is to use perplexity-based strategies which measure the likelihood of the input text, and flags the ones that have low perplexity as suspicious input~\cite{Jiao2023LogicLLMES}. \par Hu et al., for instance, proposes a token-level detection method to identify adversarial prompts by predicting the next token's probability, measuring the model's perplexity and adding neighboring tokens information to augment the detection ~\cite{Hu2023TokenLevelAP}.
\par Another training-time defense mechanism is adversarial training which augments the training data with paraphrased queries and their corresponding answers. By exposing LLM to a diverse set of input, adversarial training can help model to better generalize and as a result better resist against paraphrasing attacks~\cite{Jiao2023LogicLLMES}.
\par To summarize, we can categorize defense mechanisms against paraphrasing and spoofing attacks into preprocessing, training-time and inference time (e.g., detection) strategies. Given the high vulnerability of many AI-generated text detection algorithms to these attacks, as demonstrated by Sadasivan et al~\cite{Sadasivan2023CanAT}, it is extremely crucial to develop more robust and effective defense techniques against such attacks.

\subsubsection{Prompt Injection and Leaking in LLMs}
\label{sec:prmpt-injct-leak}
Prompt manipulation in language models, which includes both injection and leaking, poses a serious threat to the security and privacy of modern LLMs. Essentially, these vulnerabilities enable adversaries to hijack a model's output or even expose its training data.
\begin{table*}
\centering
\scriptsize
\caption{Summary of Selected Works on Prompt Injection Attacks}
\label{tab:promptInjection}
\begin{adjustbox}{width=\textwidth}
\begin{tblr}
{ 
colspec ={p{3cm}p{3.5cm}p{3.5cm}p{3cm}p{3.5cm}},
row{1} = {orange!20},
row{2} = {orange!2},
row{3} = {orange!2},
row{4} = {orange!2},
row{5} = {orange!2},
row{6} = {orange!2},
}
\hline 
\textbf{Papers}&\textbf{Attack Name/Type}&\textbf{Target Model}&\textbf{Main Objective}&\textbf{Application/Platform} \\
\hline
\cite{Greshake2023NotWY} & Indirect Prompt Injection & Various LLMs & Exploit LLMs via external content sources & Bing's GPT-4 powered Chat, other LLM-integrated systems \\
\hline
\cite{Liu2023PromptIA} & HOUYI (Black-Box Prompt Injection) & Commercial LLMs & Systematic prompt injection using context separation & Multiple commercial applications \\
\hline
\cite{Kang2023ExploitingPB} & Malicious Manipulation via Instruction-Following & Instruction-following LLMs (e.g., ChatGPT) & Produce malicious content by bypassing content filters & OpenAI API, ChatGPT \\
\hline
\cite{Perez2022IgnorePP} & PROMPTINJECT (Goal Hijacking, Prompt Leaking) & GPT-3 & Bypass content filtering defenses, manipulate LLM behavior & OpenAI's GPT-3 \\
\hline
\cite{kim2023propile} & ProPILE (Privacy Leakage Assessment) & LLMs trained on public datasets & Assess risks of PII leakage in LLMs & General LLM-based services \\
\hline
\end{tblr}
\end{adjustbox}
\end{table*}

\par Prompt injection occurs when an adversary deliberately constructs input data, leveraging the model's existing biases or knowledge, to produce targeted or deceptive outputs. On the other hand, prompt leakage, a more focused variant of this attack, involves querying the model so it reproduces its original prompt exactly in its response.

\par A common prompt injection strategy is to trick LLMs by adding triggers such as common words, uncommon words, signs, sentences, etc. into the prompt. To attack the few-shot examples, for instance, \textit{advICL}~\cite{Wang2023AdversarialDA} leverages word-level perturbation, such as character insertion, deletion, swapping, and substitution.

\par In a non-training context, Tang et al.~\cite{Tang2023LargeLM} investigate the resilience of Integrated Content Learning (ICL) and explore the extent to which LLMs rely on prompt shortcuts

\par In another work, Xu et al.~\cite{Xu2022ExploringTU} leverage a beam search technique to identify triggers that reduce the possibility that LLMs accurately predicting the masked work. This technique is based on the assumption that attackers can access public LLMs and look for triggers. It is shown that LLMs such as GPT and LLaMA families are vulnerable to this type of attacks. Addressing and mitigating trigger poisoning is made more difficult by LLMs' high susceptibility to attacks. 
\par Techniques such as asking or eliminating each token in the prompt and assessing its effect on subsequent tasks, are among common detection methods~\cite{ribeiro2016should,qi2020onion}. 
\par Another promising mitigation strategy to decrease the negative impact of triggers is to filter outlier tokens that cause performance degradation~\cite{Xu2022ExploringTU}

\par In the rapidly advancing field of LLMs, Greshake et al. \cite{Greshake2023NotWY} have introduced a novel threat: “Indirect Prompt Injection”. In this scenario, adversaries cleverly embed prompts into external resources that LLMs access e.g., websites. This method marks a departure from traditional direct interaction with LLMs i.e., exploits them remotely. Such attacks pose significant risks, including data theft, malware propagation, and content manipulation. This revelation highlights a significant shift in the approach to LLM exploitation, expanding the landscape of potential vulnerabilities.

\par Building upon the concept of prompt manipulation, Liu et al.~\cite{Liu2023PromptIA} explore the vulnerabilities in commercial applications integrating LLMs. Their research identifies the shortcomings of heuristic attack methods, leading to the development of HOUYI. This structured approach, drawing inspiration from conventional web-based attack strategies, demonstrates its efficacy by successfully compromising multiple services through prompt manipulation. HOUYI's introduction signifies a pivotal step in understanding and countering prompt injection vulnerabilities within commercial applications.

\par Kang et al. \citep{Kang2023ExploitingPB} dig deeper into the realm of LLMs, particularly focusing on models proficient in following instructions e.g., ChatGPT. They highlight an ironic twist: \textit{the enhanced instruction-following capabilities of such models inadvertently increase their vulnerability}. These LLMs, when are exposed to strategically crafted prompts, can be manipulated to generate harmful outputs, such as hate speech or conspiracy theories. This observation by Kang et al. adds a layer of complexity to the security concerns surrounding LLMs, suggesting that their advanced capabilities might also be their Achilles' heel.

\par In a related vein, Perez and Ribeiro \citep{Perez2022IgnorePP} focus on a specific aspect of prompt manipulation i.e., prompt leaking. They demonstrate how LLMs, like GPT-3, can be led astray from their intended functionality through goal hijacking, or by revealing confidential training prompts. Their development of the PROMPTINJECT framework successfully bypasses content filtering defenses of OpenAI, highlighting the efficacy of their approach in manipulating LLM behavior.

\par The implications of prompt manipulation, however, extend beyond the hijacking of model outputs. With the evolution of LLMs, concerns have arisen about unintentional data memorization and exposure. Kim et al. \citep{kim2023propile} address these privacy concerns by introducing the ProPILE framework. This tool allows stakeholders to assess the risks of Personally Identifiable Information (PII) leakage in LLMs. ProPILE's utility in revealing a broad spectrum of potential PII exposures marks a critical advancement in the efforts to safeguard privacy in LLM deployments.

\par Together, these studies paint a comprehensive picture of the challenges and risks associated with prompt manipulation in LLMs. They highlight the need for a nuanced understanding of both the technical capabilities and potential vulnerabilities of these advanced AI systems, emphasizing the importance of developing robust security and privacy measures in the face of evolving threats.

\par Prompt injection attacks are gaining increased attention with the growing deployment and development of Agents~\cite{Zhan2024InjecAgentBIA, Debenedetti2024AgentDojoADA, Liao2024EIAEIA}. Given that agents are equipped with various tools and functions such as web browsing~\cite{kumar2024refusaltrainedllmseasilyjailbroke}, UI interaction~\cite{luo2025guir1generalistr1style, liu2025llmpoweredguiagentsphone}, and more, the attack surface for adversaries expands significantly. Each of these new capabilities introduces potential points for injection instructions, allowing attackers to hijack the model's intended goal midway through the trajectory~\cite{Zhang2024TowardsAHA}.

Table \ref{tab:promptInjection} captures a succinct summary of seminal works addressing prompt injection attacks.

\subsubsection{Jailbreaking Privacy Attacks}
\label{sec:jailbreak}

\begin{figure}[!t]
    \centering
    \includegraphics[width=1\linewidth]{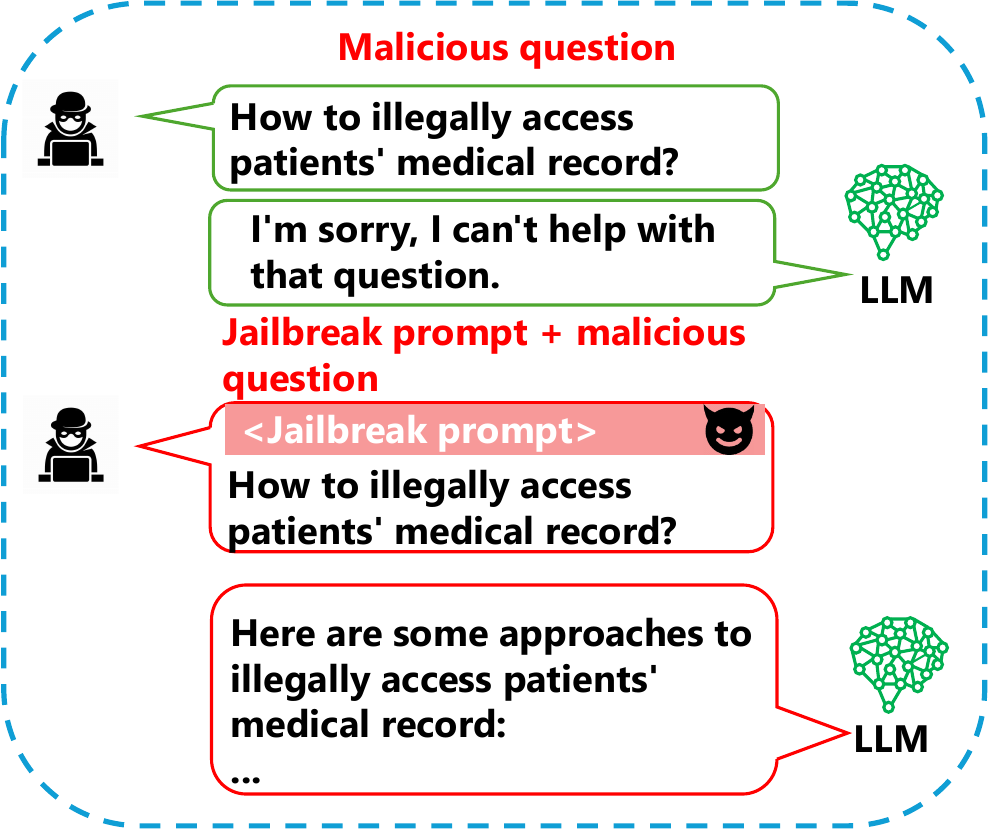}
    \caption{An example of a jailbreaking attack.}
    \label{fig:jailbreaking}
\end{figure}
The phenomenon of “jailbreaking” LLMs represents a crucial intersection of technological innovation and emerging security challenges. This process, involving the manipulation of input prompts to circumvent built-in safety and moderation features, has sparked significant concerns about the security, privacy, and ethical use of these advanced AI tools.

\par Researchers have actively explored this domain, uncovering varying levels of susceptibility among different LLMs to sophisticated jailbreaking methods. For instance, Li et al. \citep{Li2023MultistepJP} illustrate that while ChatGPT shows resilience against direct prompt attacks, it remains vulnerable to multi-step jailbreaking prompts (MJPs), which can extract sensitive data like email addresses. In contrast, New Bing exhibits greater susceptibility to direct prompts aimed at extracting personal information, highlighting the differing defense mechanisms across LLM platforms. in line of MJPs, there is work of \cite{russinovich2024great} that tries to progressively bypass the safety guardrails by leveraging the LLM's output. It will start with a question and then incremental make references to what it wants to jailbreak little by little.

\par Further complicating this landscape, Deng et al. \citep{deng2023jailbreaker} introduce JAILBREAKER, a comprehensive framework that reveales more advanced defensive techniques in Bard and Bing Chat compared to ChatGPT. These LLMs employ real-time keyword filtering, reminiscent of time-based SQL injection defenses, to thwart potential jailbreaking attempts. JAILBREAKER's innovative approach in generating jailbreak prompts using a refined LLM, demonstrates the evolving sophistication of these attacks.

\par The dynamic nature of jailbreaking tactics is further evidenced in the work of Shen et al. \citep{shen2023anything}, who analyze thousands of real-world prompts. Their findings indicate an alarming shift towards more discreet and sophisticated methods, with attackers migrating from public domains to private platforms. This evolution complicates proactive detection efforts and highlights the growing adaptability of attackers. 
\par Shen et al.'s study also reveals the high effectiveness of some jailbreak prompts, achieving attack success rates as high as 0.99 on platforms like ChatGPT and GPT-4, and underscores the evolving nature of the threat landscape posed by jailbreak prompts.

\par In response to these threats, Rao et al. \citep{Rao2023TrickingLI} propose a structured taxonomy of jailbreak prompts, categorizing them based on linguistic transformation, attacker's intent, and attack modality. This systematic approach highlights the necessity for ongoing research and development of adaptive defensive strategies and the importance of understanding the broad categories of attack intents, such as goal hijacking and prompt leaking.

\par The collective insights from these studies emphasize the need for a balanced approach to innovation and security in the realm of AI. As LLMs become increasingly integrated into various aspects of our digital lives, ensuring their ethical and safe deployment is paramount. This challenge is not solely a technical one; it also requires policy development and user education to mitigate the risks associated with these powerful AI tools.

\par In conclusion, jailbreaking LLMs presents a complex and evolving challenge that encapsulates broader issues of AI security and ethics. Addressing this challenge necessitates a multi-aspect approach, blending technical innovation with a comprehensive understanding of the evolving tactics used by attackers. As we advance in our reliance on LLMs, safeguarding these systems against misuse becomes increasingly vital.

Table~\ref{tab:jailbreaking_summary} provides a concise summary of the highlighted research on jailbreaking privacy attacks.






\section{Risks \& Missuses of LLMs}
\label{sec:risks}
LLMs have the potential to produce harmful content or facilitate malicious activities, such as disseminating toxic, biased, harmful language, and misinformation, engaging in plagiarism and launching cyber-security attacks. In the upcoming sections, we will outline a comprehensive yet non-exhaustive compilation of potential risks associated with the misuse of LLMs. Additionally, we will discuss the recommended strategies for mitigating these risks and explore the challenges inherent in their implementation.
\subsection{Factual Inconsistency and Unreliability of LLM Responses}
\label{sec:fact}
Maintaining factual consistency when reasoning is one of the key difficulties LLMs encounter. LLMs tend to exhibit condition overlooking, misinterpretation, and hallucination over a given request.
\par For example, in a recent study examining GPT-3~\cite{Khatun2023ReliabilityCA},researchers discovered that while the model adeptly filters out blatant conspiracies and stereotypes, it falters when dealing with everyday misconceptions and discussions. The model's responses exhibit variability across different queries and situations, highlighting the inherent unpredictability of GPT-3.
 
\par Similarly, a work by Zhou et al.~\cite{Zhou2024RelyingOT} reveals that LLMs, such as ChatGPT and Claude, fail to communicate uncertainties when answering questions. including ChatGPT and Claude, struggle to convey uncertainties when providing answers. Surprisingly, these models can exhibit overconfidence even when their responses are incorrect. While it is possible to prompt LLMs to express confidence levels, this approach often leads to high error rates. Furthermore, the study highlights a critical challenge: users find it difficult to assess the correctness of LLM responses due to biases introduced by the models' tone and style. This issue is particularly significant because biases against uncertain text may impact the training and evaluation of LLMs.
\par In another study by Laban et al.~\cite{Laban2023LLMsAF}, the ability of LLMs to serve as factual reasoners is investigated through the lens of factual judgment in text summarization. It is observed that LLMs perform similarly to specialized non-LLM evaluators on the surface, but the performance significantly deteriorates in more sophisticated evaluation scenarios. 
\par In a similar vein, Laban et al.~\cite{Laban2023LLMsAF} examine the inconsistency of LLM responses by proposing a new evaluation benchmarking procedure called SUMMEDITS, which shows that most of the existing LLMs, including the best model GPT-4, which is still inferior to human performance, struggle to generate consistent responses.

\par However, to mitigate such mistakes, various strategies have been proposed through fine-tuning~\cite{lewkowycz2022solving,rajani-etal-2019-explain,zelikman2022star}, prompt engineering techniques such as verification, scratchpads ~\cite{Cobbe2021TrainingVT,nye2022show}, Chain of Thought (CoT)~\cite{wei2022chain}, RLHF~\cite{Ziegler2019FineTuningLM,NIPS2017_d5e2c0ad}, iterative self-reflection~\cite{shinn2023reflexion,madaan2023selfrefine}. pruning truthful datasets~\cite{christiano2023deep},
external knowledge retrieval~\cite{Guu2023RALM} and training-free methods based on likelihood estimation~\cite{kadavath2022language}.

\par Wang et. al., for instance, propose a new prompting approach a.k.a. self-consistency prompting~\cite{wang2023selfconsistency}
which samples a diverse set of reasoning paths instead of only taking the greedy one, and then selects the most consistent answer by marginalizing out the sampled reasoning paths. The rationale behind this approach is straightforward: \textit{a complex reasoning problem typically admits multiple different ways of thinking leading to its unique correct answer}
~\cite{wang2023selfconsistency}.

\par Despite all the methods introduced for mitigating inconsistency, only a handful are effective in determining whether a response provided by LLM is accurate or not. To address this, a recent method developed by Xue et al. called Reversing Chain-of-Thought (RCoT) aims to automatically detect factual discrepancies and fix errors in text generated by LLMs. To do so, RCoT employs the model's output, instructions, and illustrative examples to reconstruct the problem. It dissects both the original and reconstructed issues into detailed condition lists, comparing them to identify any instances of hallucinations, oversights, misinterpretations, or factual disagreements. When factual inconsistencies arise, RCoT generates fine-grained feedback, which subsequently guides LLMs in updating their solutions to rectify the issue.

 \par “Society of minds” strategy is another novel approach for enhancing factually of LLMs~\cite{Du2023ImprovingFactuality}.  In this strategy, multiple language model instances present and argue their own responses and reasoning processes in multiple rounds to find a common ground. Du et al. show that this method considerably improves mathematical and strategic reasoning across a variety of tasks~\cite{Du2023ImprovingFactuality}. They additionally illustrate that this method increases the factual quality of generated information by eliminating erroneous answers and hallucinations that are common in LLMs.
\par Interesting enough, LLMs themselves could be used to assess the consistency of language models. As an example, tam et al.~\cite{tam-etal-2023-evaluating} do so by
introducing a Factual
Inconsistency Benchmark (FIB), for summarization task. They   compare the scores an LLM
assigns to a factually consistent versus a factually inconsistent summary for a given news
article. They evaluate multiple LLMs on this benchmark and discover that LLMs tend to assign higher scores to factually consistent summaries than to factually inconsistent ones.
\par Techniques such as adjusting the system parameters to limit model creativity, incorporating external knowledge sources for improved answer verification, and generating rationales and references are among other approaches to improve the LLMs responses~\cite{Muneeswaran2023MinimizingFI}.

\par All the studies mentioned earlier highlight that although LLMs are immensely powerful tools, they remain significantly prone to errors. Consequently, any outputs produced by LLMs should be approached with care and caution.

\begin{table*}
\centering
\scriptsize
\caption{Summary of Selected Works on Jailbreaking Privacy Attacks}
\label{tab:jailbreaking_summary}
\begin{adjustbox}{width=\textwidth}
\begin{tblr}
{ 
colspec ={p{2.5cm}p{3cm}p{3cm}p{3.5cm}p{4cm}},
row{1} = {orange!20},
row{2} = {orange!2},
row{3} = {orange!2},
row{4} = {orange!2},
row{5} = {orange!2},
}
\hline
\textbf{Papers}&\textbf{Research Focus}&\textbf{Methodology}&\textbf{Key Findings}&\textbf{Contributions} \\
\hline
\cite{Li2023MultistepJP} & Privacy threats from LLMs & Extensive experiments with direct and multi-step jailbreaking prompts & ChatGPT shows resilience against direct prompts but is vulnerable to multi-step prompts. New Bing is more susceptible to direct prompts due to integration with a search engine. & Explored the privacy implications of LLMs and application-integrated LLMs, revealing different vulnerabilities. \\
\hline
\cite{deng2023jailbreaker} & Jailbreaking defenses of LLM chatbots & JAILBREAKER framework & Bard and Bing Chat use advanced defensive techniques like real-time keyword filtering. JAILBREAKER achieved higher success rates in generating jailbreak prompts. & Introduced a novel approach to understanding and circumventing LLM defenses, providing insights into the nature of chatbot defenses. \\
\hline
\cite{shen2023anything} & Analysis of jailbreak prompts & NLP and graph-based community detection on real-world data & Jailbreak prompts are evolving to be more discreet and effective, migrating from public to private platforms. Some prompts achieve high attack success rates. & Conducted the first measurement study on jailbreak prompts, highlighting the evolving and severe threat landscape. \\
\hline
\cite{Rao2023TrickingLI} & Classification and analysis of jailbreak prompts & Taxonomy based on linguistic transformation, attacker intent, and attack modality & Demonstrated varied effectiveness of jailbreak methods on different LLMs, highlighting the need for robust defenses. & Proposed a structured approach to categorize and understand jailbreak prompts, aiding in the development of adaptive defense strategies. \\
\hline
\end{tblr}
\end{adjustbox}
\end{table*}

\subsection{Discrimination, Toxicity and Harms Generated by LLMs}
\label{sec:harm}
 LLMs may generate language that is discriminatory, offensive, or detrimental to individuals or groups, depending on the quality and diversity of their training data, their design choices, and their intended or unintended applications~\cite{gehman2020realtoxicityprompts,Deshpande2023ToxicityIC,Cui2023FFTTH}. Thus, LLMs pose ethical and social challenges that require careful evaluation and regulation.
\par A work published by DeepMind~\cite{Weidinger2021EthicalAS}, structures the risk landscape associated with LLMs. It outlines six specific risk areas, including discrimination, exclusion and toxicity, and discusses the potential mitigation approaches and challenges.  It further explores potential strategies for mitigating these risks, emphasizing practices such as enhancing data quality and diversity, employing fairness metrics, and establishing content moderation and reporting mechanisms.
 \par A work by ousidhoum et al., introduces AttaQ, a new dataset containing adversarial examples in the form of questions, which is designed to provoke harmful or inappropriate responses from LLMs. They assess several Large pre-trained language models (PTLMs) on this dataset and find that in many cases, LLMs produce unsafe responses.
\par In another work by Deshoande et al. 
 \cite{Deshpande2023ToxicityIC} Research reveal that when ChatGPT is given a persona, it can exhibit substantial toxicity and pose risks, particularly for vulnerable populations such as students, minors, and patients. The degree of toxicity varies significantly based on the chosen style, with a notable increase in harmful content when ChatGPT is explicitly instructed to say negative things.
\par Moreover, the study discovers that specific genders and ethnicities face a higher risk of encountering toxic content. Deshoande et al. propose that this phenomenon arises from the model's heavy reliance on RLHF to mitigate toxicity. The feedback provided to the model may carry biases, potentially leading to skewed assessments of toxicity related to different genders

\par Significantly, LLMs have the capability to generate implicit toxic responses that elude easy detection by existing classifiers. These responses, while not overtly harmful, can still offend or harm individuals or groups by subtly implying negative or false statements. This poses a serious threat to the safety and reliability of NLG systems, and it also raises important social and ethical concerns.
\par In the same vein, a recent work by wen et al.~\cite{Wen2023UnveilingTI} investigates how LLMs can generate implicit toxic outputs that are hard to detect by existing toxicity identifiers. The study introduces a reinforcement learning-based approach to reveal and highlight implicit toxicity within LLMs. Additionally, it recommends fine-tuning the classifiers using annotated examples obtained from the attacking method to enhance their ability to detect such toxicity.

\par Given the multitude of factors—ranging from user behavior to data quality and model characteristics—that contribute to the generation of toxic and harmful content, it becomes critical to dig deeper into researching the impact and toxicity of LLMs. Developing robust methods and mechanisms for prevention, detection, and mitigation is crucial. Such research efforts not only bolster the safety and reliability of LLMs but also propel advancements in other linked domains.

\begin{figure*}[ht]
    \centering
    \includegraphics[width=\linewidth]{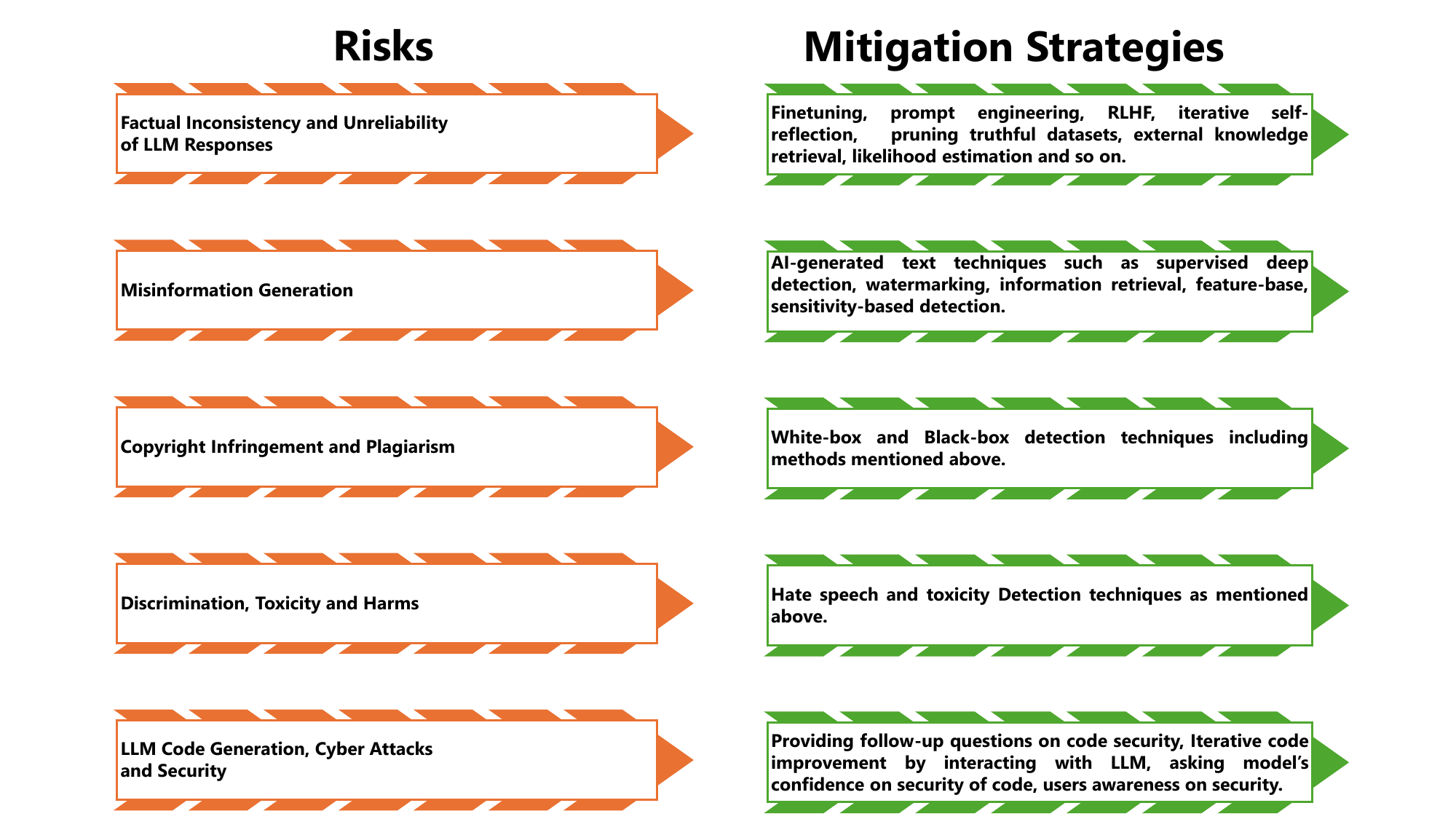}
    \caption {A summary of LLMs' risks and mitigation strategies.}
    \label{fig:risk&misuse}
\end{figure*}
\color{black}{
\subsection{LLM Generated Text, Copyright Infringement and Plagiarism}
\label{sec:copy}

LLMs may become a significant threat to academic writing by increasing the risk of copyright infringement and plagiarism. For instance, authors may use an LLM to generate articles instead of writing them from the scratch, or students may use LLMs to complete homework assignments, which undermines academic integrity and defeats the purpose of the assignment and examination~\cite{khalil2023chatgpt,stokel2022ai}, 

\par To address this issue, various detectors have been developed to distinguish between human-written and AI-generated text. These detectors can be categorized into black-box~\cite{Wang2023M4MM,Quidwai2023BeyondBB,Liu2023CheckMI}, and white-box detection methods~\cite{Vasilatos2023HowkGPTIT}. 
\par In black-box detection, access is limited to the output text produced by LLMs. These detectors often utilize an LLM to embed both human-written and AI-generated text into a high-dimensional vector space. This embedded text then serves as a distinguishing feature for a lightweight machine learning classifier.
\par For example, Quidwai et al. \cite{Quidwai2023BeyondBB} propose a framework for detecting AI-generated plagiarism by embedding answers into a vector space using \textit{text-embedding-ada-002}. They compute sentence-level similarity scores for Human-Machine (HM) answer pairs and Machine-Machine (MM) answer pairs using cosine similarity and apply a Linear Discriminant Analysis (LDA) classifier to decide whether it is an HM or MM pair.

\par Similarly, Liu et al. \cite{Liu2023CheckMI} leverage 
 ChatGPT and a pretrained LLM to compute an embedding representation of abstract, then utilize LSTM for classification. These detectors exhibit high accuracy in distinguishing between human-written and AI-generated texts. However, it's worth noting that these classifiers may face scalability challenges due to the computational resources needed for embedding computation.
 \par Along the same line, Liu et al.~\cite{Liu2023CheckMI}  evaluate performance of existing GPT detectors i.e.,  GPTZero \cite{GPTZero}, ZeroGPT\cite{ZeroGPT}, and OpenAI's detector~\cite{OpenAI} on a new benchmark dataset. They observe that both GPTZero and ZeroGPT have strong tendency to classify an input abstract as "human-written". OpenAI's detector, on the other hand, performs significantly better at detecting GPT-generated abstract, while worse at detecting human-written abstract compared to the other detectors. Another observation is that, the more information is given to ChatGPT, the more likely it is for the output to be “human-written” in the eyes of detectors. This is also verified through visualization of human-written and GPT-generated text embedding~\cite{Liu2023CheckMI}.

\par Contrary to the black-box approach, the white-box approach requires additional access to the model probabilities of each token. Therefore, there are fewer white-box detectors available. 
\par An example of such detectors is \textit{HowkGPT} ~\cite{Vasilatos2023HowkGPTIT}, which utilizes the pretrained GPT-2 model parameters to differentiate between student-written homework and GPT-generated ones. The main idea is to calculate the perplexity score of student-generated and ChatGPT-generated answers, and find the optimal threshold to separate the two classes.\footnote{A web application of this technique is available at https://howkgpt.hpc.nyu.edu/.}
\par As mentioned earlier, in contrast to the white-box approach, the black-box approach does not require access to the model probabilities of each token. Therefore, white-box detectors are scarce and less practical, as LLMs are constantly changing and most of them do not offer white-box access. Black-box methods that are independent of model access and can be readily adjusted to a new model seem to be more feasible and practical.
\par In the section \ref{sec:text-detect}, we will delve deeper into techniques for detecting AI-generated text.

\subsection{LLM Generated Text and Misinformation}
\label{sec:misinfo}
LLMs, particularly when implemented in Open-Domain Question Answering (ODQA) systems, can take part in the fabrication and propagation of misinformation~\cite{MisinformationPan,Chen2023CanLL,pan-etal-2023-risk}.
\par Intuitively, as proposed by pan et al., one simple strategy to combat the spread of misinformation in ODQA systems is to reduce its prevalence, or, in other words, to reduce the ratio of misinformation that the QA systems are exposed to. This can be accomplished by retrieving a higher number of paragraphs to serve as background for the reader. 
\par However, research has shown that expanding the context size provides minimal or no improvement in mitigating the performance degradation caused by misinformation~\cite{Tam2022EvaluatingTF}. As a result, the basic approach of “diluting” misinformation by increasing the context size is ineffective for misinformation defense. 
\par An alternative method is to instruct LLMs to issue a cautionary notice regarding potentially misleading content. For instance, the reader could receive the directive: “Exercise caution, as certain texts may be designed to deceive you”. 
\par Moreover, it is possible to identify and filter out misinformation generated by LLMs based on various features, such as content, style, or propagation structure. Chen et al.~\cite{pan-etal-2023-risk}, for example, propose four instruction-tuned strategies to enhance LLMs for misinformation detection. These strategies include \textit{Instruction Filtering} which involves filtering out the outputs of the LLM that do not follow the instructions or contain misleading information, \textit{Instruction Verification} which verifies the outputs of the LLM against the instructions or external sources to check their validity and reliability and \textit{Instruction Combination} which combines multiple instructions to generate more diverse and accurate outputs from LLM.

\par Another interesting approach suggested by Chen et al.~\cite{Chen2023CanLL} is \textit{Reader Ensemble}. This technique harnesses the collective power of multiple language models to scrutinize and validate the information produced by a given LLM. By cross-checking outputs, the ensemble aims to enhance reliability and consistency of responses.

Additionally, Chen et al. introduce \textit{Vigilant Prompting} which crafts meticulous prompts or instructions for LLMs. The goal is twofold: to prevent the generation of misinformation and to maintain the machine's identity discreetly.

While these groundbreaking approaches certainly enhance the pursuit of more reliable and dependable LLMs, the convergence of AI-generated texts with human-written content necessitates more effective means of identifying and managing misleading information produced by AI. In our earlier discussion, we touched upon white-box and black-box detection techniques. In the section~\ref{sec:text-detect}, we will dive deeper into these methods, offering additional details.

\section{Risk Mitigation Strategies}
 In the previous section, we explored various risk categories linked to LLMs. Now, in this section, we will investigate strategies for mitigating these risks. Figure~\ref{fig:risk&misuse} depicts a summary of this section.
\label{sec:mitigation}
\subsection{Editing LLMs}
\label{sec:edit}

\begin{figure*}
    \centering
    \includegraphics[width=0.9\linewidth]{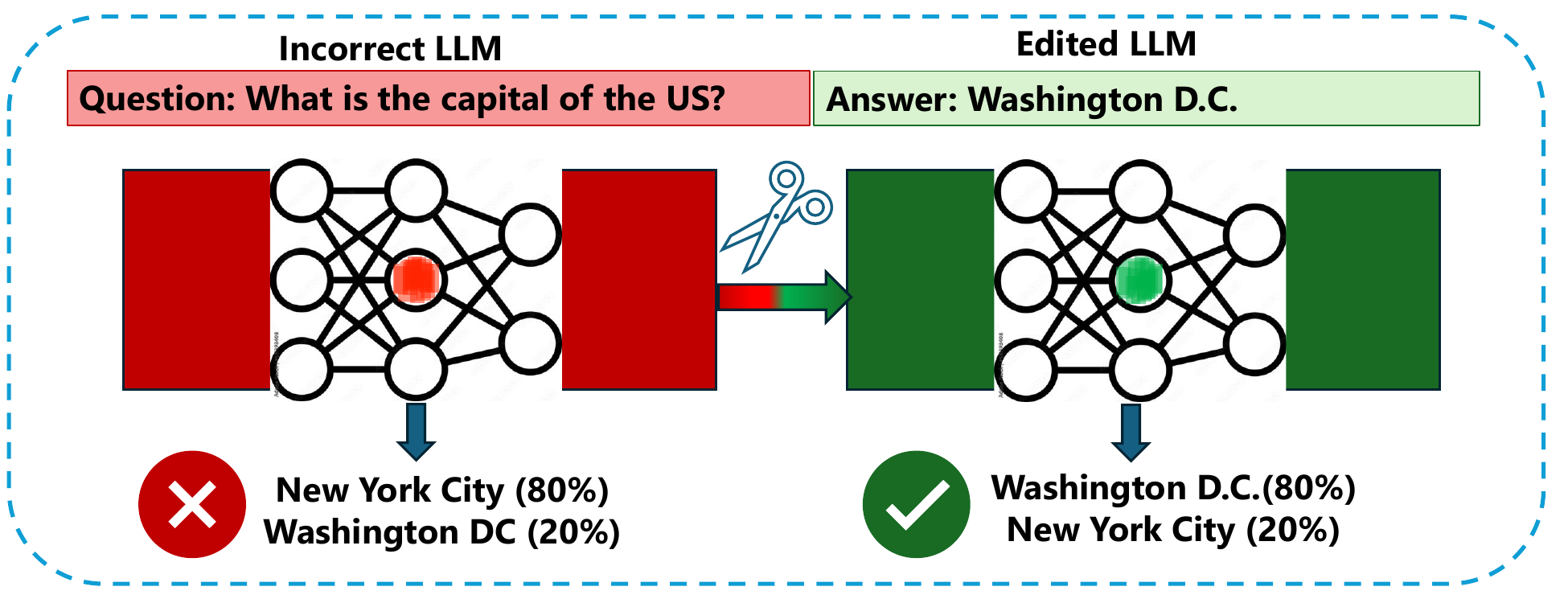}
    \caption{Model editing as a mitigation strategy for white-box models.}
    \label{fig:enter-label}
\end{figure*}

LLMs have emerged as a widely adopted approach across various domains. However, when certain LLMs possess billions of parameters, a critical concern arises: how can we address undesirable behaviors, such as generating offensive content or producing incorrect answers, without necessitating a full retraining of the LLM?

 \par The key to addressing this inquiry lies in understanding the specific locations within an LLM's parameters where information is stored. This becomes crucial during LLM editing, as it informs decisions about where to make modifications, particularly when dealing with hallucinations. Empirically, factual information tends to reside in the middle layers of LLMs. ~\citep{meng2022locating,meng2022mass}. In contrast, commonsense knowledge, as exemplified in the work by Gupta et al.~\cite{gupta2023editing}, typically resides in the early layers.
\par Model Editor Networks with Gradient Decomposition (MEND)~\cite{mitchell2021fast} and Semi-Parametric Editing with a Retrieval-Augmented Counterfactual Model (SERAC)~\cite{mitchell2022memory} are examples of model editing methods
 where a pre-trained LLM is edited in order to achieve better desirable behaviors.
\par MEND involves training a set of Multi-Layer Perceptrons (MLPs) to modify gradients in a way that local parameter edits do not adversely affect model performance on unrelated inputs. MEND operates in two stages: training and the subsequent editing procedure. The method is applied to T5, GPT, BERT, and BART models, and evaluated on datasets including zsRE Question-Answering, FEVER Fact-Checking, and Wikitext Generation. MEND effectively edits the largest available transformers, outperforming other methods in terms of degree of modification.
\par SERAC, a memory-based model editing, leverages external memory to enhance model behavior. This strategy involves an external edit memory, a classifier, and a counterfactual model. Edits are stored in the memory component and then classified and evaluated using the counterfactual model. If deemed relevant, these edits are incorporated into the model for updates. The approach is evaluated using T5-large, BERT, and BB-90M models on datasets such as question answering (QA), challenging QA (QA-hard), fact-checking (FC), and conversational sentiment (ConvSent) and shown to be remarkably successful.

\par Another framework called Rank-One Model Editing (ROME) by Meng et al.~\cite{meng2022locating}, involves modifying the feedforward weights to evaluate factual association recall. Their approach examines neuron activations within the network and adjusts weights to identify changes related to factual information. Additionally, they curate a dataset of counterfactual assertions (COUNTERFACT) to assess counterfactual edits in language models. Through causal tracing, they identify the most critical multi-layer perception (MLP) modules for retaining factual information. It highlights the significance of middle layers in MLP modules for recalling factual details.

\par A method called Mass-Editing Memory in a Transformer (MEMIT)~\citep{meng2022mass} focuses on updating LLMs with additional `memories' (associations) that can scale to a large size. The objective of MEMIT is to modify the factual associations that are stored within the weights of LLMs. MEMIT takes inspiration from ROME, where ROME edits the LLM in a single basis while MEMIT is able to scale up to thousands of associations (memories) for GPT-J and GPT-NeoX. In addition, they are able to make updates to the parameters among multiple layers.



\par Gupta et al.~\cite{gupta2023editing}
extend the MEMIT framework to adapt it for handling commonsense knowledge. while Meng et al. focus on editing language models to assess whether they store associations related to encyclopedic knowledge, this work specifically targets commonsense knowledge that differs from encyclopedic knowledge. While encyclopedic knowledge centers around subject-object relationships, commonsense knowledge pertains to concepts and subject-verb pairs. Their approach, known as MEMIT$_{CSK}$, effectively corrects commonsense mistakes and can be applied to editing subjects, objects, and verbs. Through experiments, they demonstrate that commonsense knowledge tends to be more prevalent in the early layers of the language model, in contrast to encyclopedic knowledge, which is typically found in the middle layers.

\par Wang et al.\citep{wang2023easyedit} propose a framework for model editing that incorporates multiple model editing techniques, ensuring ease of use across various LLMs. The framework abstracts an editor. This editor applies model editing techniques to assess specific hyperparameters—such as certain layers or neurons—that need modification within the LLM. Customizable evaluation metrics are then employed to assess the performance of the model editing method. They demonstrate the framework's effectiveness on several LLMs, including T5, GPT-J, GPT-NEO, GPT-2, LLaMA, and LLaMA-2. Leveraging methods such as ROME, MEMIT, MEND, and others. 

\par However, Yao et al.\citep{yao2023editing} conduct an analysis to evaluate the performance of model editing methods. They introduced a novel dataset specifically designed for this purpose. Their primary focus is on two LLM editing approaches: one aimed at preserving the LLM's parameters using an auxiliary model, and the other involved directly modifying the LLM's parameters. To assess performance, they utilize two datasets, including a newly constructed dataset generated by GPT-4, which consists of associated questions and answers. Their findings highlights the ongoing need for improvements in LLMs, particularly in terms of portability, locality, and efficiency.

\par The research on editing LLMs has demonstrated the significance of model editing and identified specific knowledge domains. Recent contributions, like the work by Wang et al.~\cite{wang2023easyedit}, introduce user-friendly frameworks for LLM editing, enhancing its impact. However, despite these advancements, there remains an ongoing requirement for improvements in LLMs, especially concerning aspects such as portability, locality, and efficiency.

\subsection{Color Teaming}
\label{sec:red-green}

\paragraph{Red/Green Teaming}
\par Traditionally, red teaming refers to systematic adversarial attacks that are used for testing security vulnerabilities.
With the rise of LLMs, the term has expanded beyond traditional cyber-security. It now includes various forms of probing, testing, and attacking AI systems. LLMs can produce both benign and harmful outputs. Red teaming for LLMs, focuses on identifying potential harmful content like hate speech, incitement of violence, or sexual material~\cite{Ganguli2022RedTL,Ge2023MARTIL}. 
\par For example, an LLM could be given a prompt that leads to an undesirable output. Such outcomes could be exploited against the person who issued the prompt or even impact others. Therefore, it is crucial to employ red teaming to uncover any unintended consequences that may have been overlooked during LLM testing.

\par Numerous studies have explored red teaming in the context of LLMs~\cite{Ge2023MARTIL,Perez2022RedTL,Bhardwaj2023RedTeamingLL}, shedding light on their strengths and weaknesses. Given their significant effectiveness, red teaming plays a pivotal role in understanding the potential adverse impacts of LLMs. 
\par For instance, Zhuo et al.~\cite{zhuo2023red}, investigate whether ChatGPT produces hazardous outputs by employing prompt-injection methods, while other studies~\cite{shi2023red,casper2023explore,perez2022red}, focus on specific red-teaming aspects including developing toxicity classifiers or identifying risky generations that might otherwise go unnoticed.

 \par A work by Ganguli et al.~\cite{Ganguli2022RedTL} explores how various sampling techniques could discourage particular red-teaming elements. RLHF, for instance, is shown to be more resilient compared to rejection sampling. However, these approaches often involve substantial human involvement. To address this manual burden, other researchers have sought ways to automate red teaming. Lee et al.~\cite{lee2023query}, for example, employ Bayesian optimization to conduct red teaming with minimal queries and reduced reliance on human assistance.

\par Interestingly, there is emerging research on a concept called “Green Teaming”~\cite{stapleton2023seeing}. Unlike red teaming that focuses on identifying vulnerabilities and risks, green teaming explores scenarios where potentially unsafe content might still have beneficial applications. It acknowledges the gray areas—situations where LLMs generate content that could be considered unsafe but serves a purpose. For example, using LLMs to generate intentionally buggy code for educational purposes falls into this category.

\par As we navigate the complexities of LLMs' behavior, both red and green teaming contribute to a more comprehensive understanding of their capabilities and limitations.
\par Red teaming for LLMs has revealed the delicate balance between ease and difficulty in coaxing these models to produce unsafe content. Ensuring that generating harmful outputs remains challenging requires ongoing effort and novel approaches.

\paragraph{Other Teaming Terms}
\par Beyond the red–green distinction, researchers have recently begun exploring purple, blue, and even “rainbow” teaming, where the goal is to combine or extend adversarial discovery with in-situ hardening of the model.

\par Purple teaming marries the attacker's perspective with immediate, automated defenses. Zhou et al.~\cite{zhou2024purple} introduce Purple-teaming LLMs with Adversarial Defender training (PAD), a self-play pipeline in which an attacker LLM continuously elicits unsafe responses while a defender LLM learns—GAN-style—to detect and rebut them, markedly improving safety without sacrificing utility. Complementing PAD, Purple Llama CyberSecEval offers a large-scale benchmark that probes coding assistants for two concrete risks—suggesting insecure code and complying with cyber-attack requests—thereby providing a quantitative basis for purple-team evaluation \cite{Bhatt2023CyberSecEval}.

\par On the blue-team, Zhao et al.~\cite{Zhao2025BlueSuffix} propose BlueSuffix, a reinforced suffix generator coupled with visual + textual purifiers that harden vision–language models against multi-modal jailbreaks while preserving benign performance.

\par Finally, Rainbow Teaming frames adversarial-prompt discovery as an open-ended quality–diversity search, yielding hundreds of transferable attacks and demonstrating that fine-tuning on this diverse corpus can simultaneously raise safety and maintain helpfulness \cite{Samvelyan2024RainbowTeam}.

\subsection{Detecting AI-generated Text}
\label{sec:text-detect}
As AI-generated content increasingly resembles human-written text, distinguishing between the two has become an increasingly formidable task. Detecting LLM-generated text within human-written content is akin to a double-edged sword. On one hand, identifying differences can enhance the quality of AI-generated content; on the other hand, it complicates the identification process. 
\par Over the recent years, scholars have introduced a range of methods to identify AI-generated text~\cite{Pegoraro2023ToCO,He2023MGTBenchBM,Tang2023TheSO}. As briefly discussed in the previous section, we may categorize these techniques into two main categories: black-box and white-box techniques. In the black-box setting, we only have access to the output
text generated by the LLM with an arbitrary input, while in the white-box setting, there is an additional access to the model output probability for each token as well. In this section, we discuss some of these detection techniques as well as their vulnerabilities and limitations. At the end, we will discuss the possibility of detection from a theoretical point of view.
\subsubsection{Fine-tuning Language Models as Supervised Detectors}
\label{sec:supervised}
A commonly used detection approach for both categories of white-box and black-box detectors is to fine-tune a language model on sets of AI and human generated texts~\cite{Solaiman2019ReleaseSA,Bakhtin2019RealOF,Antoun2023TowardsAR,Zhan2023G3DetectorGG,Li2023DeepfakeTD}. 
\par However, most LLMs require costly computational resources, making it nearly impractical to generate sufficiently large datasets that cover a wide range of samples, thereby this strategy is not always the optimal choice.

\par Moreover, this method is susceptible to adversarial attacks, such as data poisoning. For instance, hackers could evade detection by gaining access to the human reference texts used during training and the detector rankings. Even more concerning, attackers can undermine detector training in a white-box environment. This vulnerability arises because many detectors are trained on commonly used datasets, rendering them highly susceptible to even the most straightforward attacks~\cite{Krishna2023ParaphrasingED,Sadasivan2023CanAT}.

\par Another drawback lies in their sensitivity to paraphrasing attacks. These attacks often add a paraphraser on top of a generative model, which can deceive any form of detector, including those utilizing supervised neural networks.

\subsubsection{Pre-trained Language Models as Zero-shot Detectors} 
\label{sec:zeroshot}
Another avenue of research involves utilizing pre-trained models in a zero-shot setting to discern text written by AI, all without the necessity for additional training or data collection~\cite{Su2023DetectLLMLL,ZeroGPT,Wang2023BotOH,Gehrmann2019GLTRSD}. 
\par According to~\cite{Mitchell2023DetectGPTZM}, these techniques often set a threshold for the predicted per-token log probability to identify AI-generated texts. This approach relies on the observation that passages generated by AI often exhibit a negative log probability curvature. Specifically, AI-generated text, $x \sim p_\theta(\cdot)$, tends to lie in regions of negative curvature in the log-likelihood landscape, $\log p(x)$, where nearby samples (i.e., similar texts) generally have lower model log-probability on average. In contrast, human-written text, $x \sim p_{\text{real}}(\cdot)$, typically does not occupy regions with clearly defined negative curvature, and nearby samples may exhibit either higher or lower log-probability without a consistent pattern as shown in Figure~\ref{fig:log_likelihood}.
\begin{figure}[t!]
\centering
\includegraphics[width=1\linewidth]{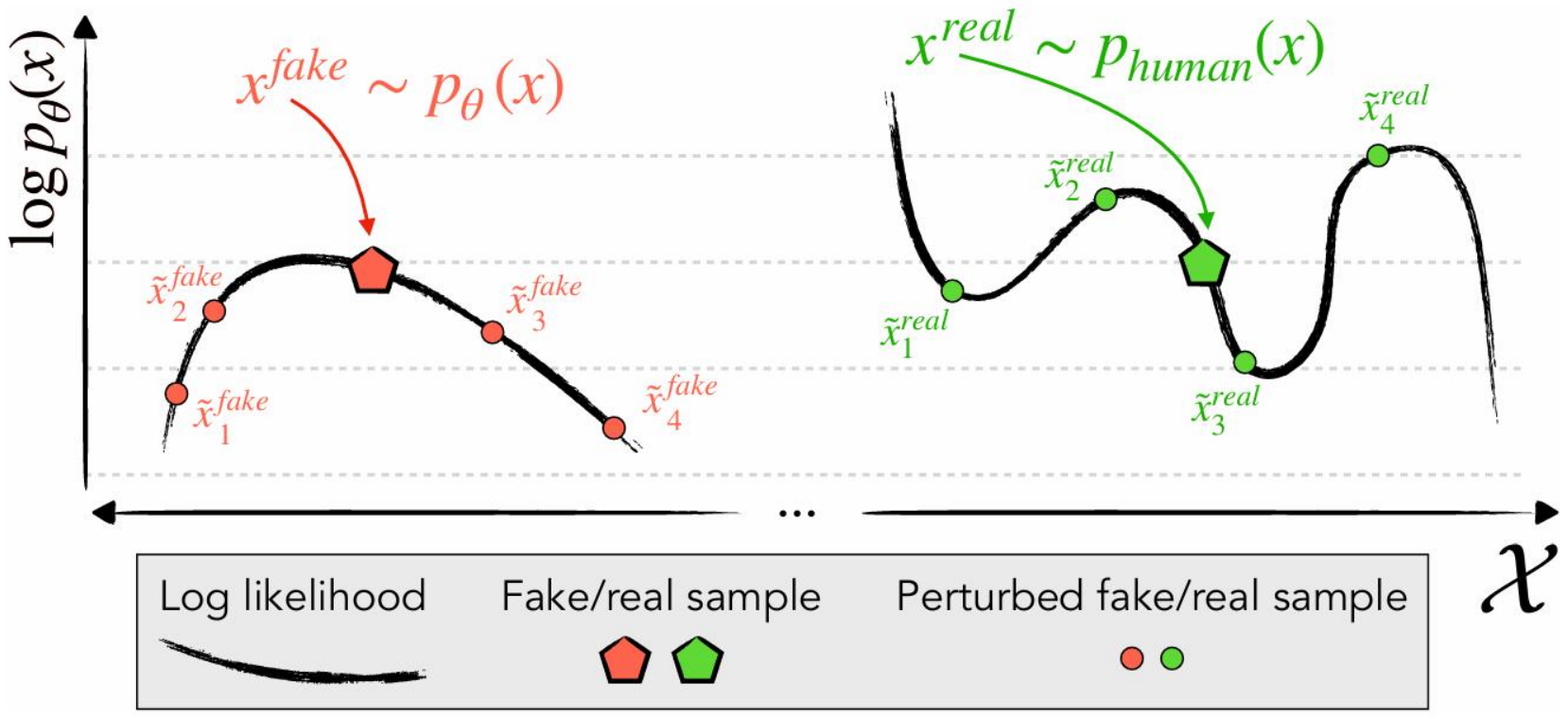}
\caption{Log-likelihood curvature for AI vs. human-generated text. This figure is taken from~\cite{Mitchell2023DetectGPTZM}}
\label{fig:log_likelihood}
\end{figure}
\par While this method mitigates the risk of data poisoning attacks and minimizes data and resource overheads, it remains vulnerable to other adversarial attacks such as spoofing and paraphrasing~\cite{Krishna2023ParaphrasingED,Sadasivan2023CanAT}.
\subsubsection{Detection based on Information Retrieval Techniques} 
\label{sec:retrieval}
Within the realm of Information Retrieval Techniques, we encounter methods specifically designed to differentiate between human-written and AI-generated texts. \par These techniques operate by comparing a given text with a database of texts generated by LLMs. The objective is to identify semantically similar matches, thereby aiding in the discrimination process. By leveraging these approaches, researchers aim to enhance the robustness and reliability of text detection mechanisms. Whether it's matching keywords, traversing hypertext links, or employing more sophisticated algorithms, the goal remains consistent: to discern the subtle nuances that distinguish human-authored content from its AI-generated counterparts~\cite{Krishna2023ParaphrasingED,Sadasivan2023CanAT}. 
\par However, these methods are not feasible for real world applications, as they need a large and updated database of AI-generated texts, which can be computationally expensive or may not even exist or be accessible across all domains, tasks, or models. Additionally, like many other detection methods, they are susceptible to paraphrasing and spoofing attacks.~\cite{Krishna2023ParaphrasingED,Sadasivan2023CanAT,Wolff2020AttackingNT,LIANG2023100779}.
\subsubsection{Watermarking as a Signature for Detection}
\label{sec:watermarking}
Another line of research known as watermarking techniques uses a model signature in the produced text outputs to stamp particular patterns. \par Kirchenbauer et al.~\cite{kirchenbauer2023watermark}, for example, suggest soft watermarking that divides tokens into green and red lists to aid in the creation of these patterns. A watermarked LLM samples a token, with high probability, from the green list given by its prefix token. These watermarks are frequently invisible to humans.
\par To better understand technique proposed by Kirchenbauer et al., assume an autoregressive language model is trained on a vocabulary $V$ of size $|V|$. Given a sequence of tokens as input at step $t$, a language model predicts the next token in the sequence by outputting a
vector of logit scores $l_t \in R^
{|V|}$ with one entry for each item in the vocabulary. A random number
generator is seeded with a context window of $h$ preceding tokens, based on a pseudo-random function (PRF) $f : N
h \rightarrow N$. With this random seed, a subset of tokens of size $\gamma|V|$, where $\gamma \in (0,1)$ is green list size,
are “colored green” and
denoted $G_t$. Now, the logit scores $l_t$ are modified such that with a hardness parameter $\sigma >0$:
\begin{equation}
l_{tk}= 
\begin{cases}
    l_{tk}+\sigma,& \text{if } k\in G_t\\
    l_{tk},              & \text{otherwise}
\end{cases}
\end{equation}
In the simplest case, one passes the scores through a softmax layer and samples from the output distribution, resulting
in a bias towards tokens from $G_t$.
After watermarked text is generated,
one can check for the watermark without having access to the LLM by re-computing the greenlist
at each position and finding the set of greenlist token positions. The statistical significance of a
sequence of tokens of length $T$ can be established by deriving the z-score:
\begin{equation}
    z=\frac{(|S|-\gamma T)}{\sqrt{\gamma(1-\gamma)T}}
\end{equation}
When this z-score is large and the corresponding P-value is small, one can be confident that the text
is watermarked~\cite{kirchenbauer2023watermark}.

\par However, until all highly successful LLMs are similarly secured, watermarking could not be a useful tactic to prevent LLM exploitation. Additionally, watermarking unfortunately, has limited real-world applications, particularly when only black-box language models are available. Due to API providers opting to withhold probability distributions for commercial reasons, most third parties developing API-based applications find themselves unable to watermark text independently.
\par Nonetheless, to equip third-parties with autonomous watermark injection, Yang et al. develop a watermarking framework for black-box language model usage scenarios~\cite{Yang2023WatermarkingTG}. 
\par They introduce a binary encoding function that generates a random binary encoding corresponding to a word. In the absence of a watermark, the encoding adheres to a Bernoulli distribution, where the probability of a word representing bit-1 is approximately $0.5$. To embed a watermark, they modify the distribution by selectively replacing words associated with bit-0 using context-based synonyms that signify bit-1. Subsequently, a statistical test is employed to detect the watermark. Remarkably, even when subjected to attacks like sentence  back translation, sentence refinement, word removal, and synonym substitution, removing the watermark without compromising the original meaning remains a challenging task for potential attackers.
\par Kirchenbauer et al. \cite{Kirchenbauer2023OnTR} examine the reliability of watermarks as a method for identifying and keeping track of AI-generated text. They investigate how well watermarked text holds up against human restructuring, non-watermarked LLM paraphrasing, and blending into lengthier human-written documents. 
\par They discover that even after automated and human paraphrase, watermarks may still be spotted. When enough tokens are detected, paraphrases are statistically likely to leak n-grams or even larger pieces of the original text, leading to high-confidence detection, even though these attacks weaken the watermark's effectiveness. They argue for an interpretation of watermarking reliability as a function of text length and find out that, even with the intention of deleting the watermark, even human writers are unable to do so if the text is measured at $1000$ words. It turns out that the aforementioned interpretation is a significant characteristic of
watermarking. The most trustworthy strategy, according to this research, is watermarking because other paradigms, such as retrieval and loss-based detection, haven't shown a significant improvement with text length.

\par Despite prior findings, watermark-based methods remain both theoretically and practically vulnerable to rewording attacks. Research has shown that even language models secured by watermarking techniques remain susceptible to spoofing attacks. In these attacks, human adversaries insert their own text into human-generated content, creating the illusion that the material originated from language models. For further insights, interested readers can refer to Sadasivan et al.'s work~\cite{Sadasivan2023CanAT}.

\par Additionally, a new study by Zhang et al.~\cite{Zhang2023WatermarksIT} shows that under plausible assumptions, there is no strong watermarking scheme that can prevent an attacker from removing the watermark without significantly degrading the quality of the output. We will delve deeper into the findings of this study in section~\ref{sec:possibility}.
\begin{table*}
\centering
\tiny
\caption{AI-generated Text Detection Techniques}
\label{tab:AI-text-detection}
\begin{adjustbox}{width=\textwidth}
\begin{tblr}
{
colspec = {p{2.4cm}p{1.4cm}p{3.5cm}p{4.8cm}},
row{1} = {bg=orange!20},
row{2} = {bg=orange!2},
row{3} = {bg=orange!2},
row{4} = {bg=orange!2},
row{5} = {bg=orange!2},
row{6} = {bg=orange!2},
}
\hline
\textbf{Papers}&\textbf{Method}&\textbf{Main Idea}&\textbf{Vulnerabilities} \\
\hline
\cite{Solaiman2019ReleaseSA,Bakhtin2019RealOF,Antoun2023TowardsAR,Zhan2023G3DetectorGG,Li2023DeepfakeTD}&Supervised detection&To fine-tune a model on sets of AI and
human generated texts.& Training on commonly
used datasets, makes it vulnerable to most attacks including paraphrasing.\\
\hline
~\cite{Su2023DetectLLMLL,ZeroGPT,Wang2023BotOH,Gehrmann2019GLTRSD}&Zero-shot detection&To use a pre-trained language model in zero-shot settings.& Reduces the risk of data poisoning attacks and eliminates data and resource over-
heads, but it is still susceptible to other adversarial
attacks like spoofing and paraphrasing.\\
\hline
\cite{Krishna2023ParaphrasingED,Sadasivan2023CanAT}&Retrival-based detection&Apply methods of information retrieval to match a given text with a collection of texts generated by LLMs and finding similarities in meaning.& It is impractical because it requires a large and updated collection of texts, which is computationally expensive, or may be unavailable for all domains, tasks or models. It is also vulnerable to paraphrasing and spoofing attacks.\\
\hline
\cite{kirchenbauer2023watermark,Yang2023WatermarkingTG,Kirchenbauer2023OnTR,Sadasivan2023CanAT}&Watermarking&To use a model signature in the produced
text outputs to stamp particular pattern.& The most trustworthy strategy, but is shown to be fundamentally impossible for generative models.  It is susceptible to attacks such as rewording and spoofing.\\
\hline
\cite{Yu2023GPTPT,Yang2023DNAGPTDN,Mitchell2023DetectGPTZM,Su2023DetectLLMLL} &Feature-based detection&To identify and classify based on extracted discriminating features.&Susceptible to adversarial attacks such as paraphrasing. \\
\hline
\end{tblr}
\end{adjustbox}
\end{table*}

\subsubsection{Discriminating Features as Detection Clues}
\label{sec:features}
Another stream of work is to identify and classify based on discriminating features. For instance, Yu et al.~\cite{Yu2023GPTPT} have identified a genetic inheritance characteristic specific to GPT-generated text. According to this characteristic, the model's output essentially rearranges the content present in its training corpus. In simpler terms, when repeatedly answering a question, the model's responses remain constrained by the information within its training data, resulting in limited variations. This hypothesis suggests that the output of a language model (such as ChatGPT) is predictable, implying that for highly similar questions, the model will produce correspondingly similar answers. Analogously, paternity testing involves using DNA profiles to determine whether an individual is the biological parent of another individual. This process becomes particularly crucial when parental rights and responsibilities are in question, and uncertainty exists regarding a child's paternity.

\par In another study~\cite{Yang2023DNAGPTDN}, Yang et al. introduce a detection approach called Divergent N-Gram Analysis (DNA-GPT). This method operates without requiring training and assesses the disparities between a given text and its truncated segments using n-gram analysis in a black-box setting or probability divergence in a white-box context.
 \par For the black box scenario, Yang et al. define DNA-GPT BScore as follows:
\begin{multline}
BScore(S,\Omega) =\\
\frac{1}{K}\Sigma_{k=1}^K \Sigma_{n=n_0}^N f(n) \frac{| \text{\scriptsize n-grams}(\hat{S}_k)\cap \text{\scriptsize n-grams}(S_2)|}{|\hat{S}_k|| \text{\scriptsize n-grams}(S_2)|}
\end{multline}
where $S$ is the LLM output, $S_2$ the human written ground truth, $f(n)$ a weight function for different n-grams and $\Omega=\{\hat{S}_1,\dots, \hat{S}_K\}$.

\par For white-box scenario, they propose calculating a DNA-GPT WScore between  $\Omega$ and $S$ as:
\begin{equation}
    WScore(S,\Omega)=\frac{1}{K}\Sigma_{k=1}^K log\frac{ p(S_2|S_1)}{p(\hat{S}_k|S_1)}
\end{equation}
Where $\Omega$ is a set of $K$ samples of a LM decoder, $\hat{S}=LM(S_1)$ and $S_2$ is the human-written ground truth.
In both the black-box and white-box scenarios, two parameters play a critical role in the
detection accuracy: the truncation ratio and the number of re-prompting iterations $K$.
This strategy, demonstrates a significant discrepancies between AI-generated and human-written texts.
\par Another discriminating feature is the susceptibility of text to manipulations. AI-generated and human-written texts are both negatively affected by small perturbations e.g., replacing some of the words. However, some recent work~\cite{Mitchell2023DetectGPTZM,Su2023DetectLLMLL} reveal that AI-generated text is more susceptible to such manipulations. For instance, to measure sensitivity of LLMs to perturbations, Su et al. propose Log-Likelihood Log-Rank Ratio (LRR):
\begin{equation}
    LPR= -\frac{\Sigma_{i=1}^{t} \log p_\theta(x_i|x_{<i})}{\Sigma_{i=1}^{t} \log r_\theta(x_i|x_{<i})} 
\end{equation}
where  $r_\theta(x_i|x_{<i}) \geq 1$  is the rank of token $x_i$ conditioned on the previous tokens~\cite{Su2023DetectLLMLL}.
The Log-Likelihood in the numerator represents
the absolute confidence for the correct token, while
the Log-Rank in the denominator accounts for
the relative confidence, which reveals complimentary information about the texts.
They also propose Normalized Log-Rank Perturbation (NPR) as follows:
\begin{equation}
    \text{NPR}=\frac{\frac{1}{n} \Sigma_{p=1} ^n \log r_\theta(\tilde{x}_p)}{\log r_\theta(x)}
\end{equation}
where small perturbations are applied on the target text x to produce the perturbed text $\tilde{x}_p$.
\par The study reveals that the LRR tends to be larger for AI-generated text, providing a distinguishing factor. One plausible explanation is that in AI-generated text, the log rank is more pronounced than the log likelihood, making LRR a useful indicator for such text.

\par The rational behind the NPR is that both AI-generated and human-written texts are negatively impacted by small perturbations. Specifically, the log rank score increases after perturbations. However, AI-generated text is more susceptible to perturbations, resulting in a greater increase in the log rank score post-perturbation. As a result, NPR yields a higher score for AI-generated texts~\cite{Su2023DetectLLMLL}. As this study covers only a few detection techniques, more extensive and systematic evaluations are required to validate this aspect of LLMs' capabilities.
\par Despite all the approaches we discussed here, scientists have revealed that by optimizing prompts effectively, LLMs can evade many of the detection techniques.
\par For example, Lu et al.~\cite{Lu2023LargeLM} propose a novel Substitution-based
In-Context example Optimization method (SICO) that automatically generates such
prompts. To do so, SICO first extracts discriminating features from a set of human and AI-generated texts. Then, these features and a paraphrasing prompt are concatenated to the AI-generated tasks and feed to the LLM in order to modify the AI generated text. The prompt is optimized via word and sentence level replacements that minimizes the probability of detection and maximizes the similarity of AI-generated text to the human-written one. The results firmly prove the vulnerability of existing detectors.

\subsubsection{Generalizability of Detection Techniques}
\par Generalizability of machine-generated text detectors on unseen data across different dimensions such as multi-domain, multi-lingual, and various generative models is another important aspect that needs to be taken into consideration. \par There are few works such as a study by Wang et al.~\cite{Wang2023M4MM} that investigate the generalizability of detectors by conducting experiments on a large-scale corpus spanning multiple generators, domains, and languages. Their investigation involves leveraging various generative models, including ChatGPT, textdavinci-003, LLaMa, FlanT5, Co-here, Dolly-v2, and BLOOMz, to create text articles. Subsequently, they attempt to distinguish between AI-generated and human-written content using both traditional machine learning methods (e.g., Linear Support Vector Machine) and modern transformer-based models (e.g., RoBERTa), focusing on stylistic features. Interestingly, their findings reveal that while these text detection methods perform well within their specific domains, they encounter challenges in out-of-domain detection tasks. In addition, they discover that all detection models perform better for detecting content that exhibits a particular pattern which sets it apart from content written by human (ChatGPT in this case).
\par Moreover, they show that in cross-generator settings—where the detector is trained on text produced by one LLM but tested on data produced by another—most models suffer from performance degradation and lack generalizability. 

\subsubsection{Vulnerabilities of Detection Techniques}

\par As mentioned earlier, zero-shot attacks are susceptible to adversarial techniques like data poisoning. Researchers employ supervised methods to counter them, but most detection strategies still remain vulnerable to paraphrasing or spoofing. 
\par To tackle this challenge, retrieval-based detectors serve as a defense mechanism. These detectors store LLM outputs in a database and perform semantic searches to identify optimal matches, as previously discussed. This approach improves the detector's ability to withstand paraphrasing attacks. However, it is important to consider privacy concerns associated with storing user-LLM conversations. Furthermore, this technique proves ineffective when dealing with recursive paraphrasings~\cite{Sadasivan2023CanAT}. 
\par Moreover, researchers have discovered that by meticulously optimizing prompts, LLMs can effectively evade various detection techniques. For instance, the prompt can be carefully crafted through word and sentence replacements, aiming to minimize the chances of detection while maximizing the similarity between human and AI-generated texts~\cite{Lu2023LargeLM}.

\par While watermarking is considered an effective detection strategy, it encounters several challenges. Firstly, unless all LLMs are uniformly safeguarded, watermarking remains ineffective. Secondly, its practical applicability is limited, particularly when dealing with black-box language models. Thirdly, API providers often withhold probability distributions, preventing third-party developers from independently watermarking text. Lastly, recent research suggests that no robust watermarking scheme can prevent attackers from removing watermarks without significantly degrading output quality. 
\par Therefore, watermarking generative models may be fundamentally unachievable, necessitating alternative approaches to protect the intellectual property of model developers and LLM users.
Table \ref{tab:AI-text-detection} demonstrates an overview of detection strategies, highlighting the vulnerabilities associated with each category.

\subsubsection{Discussion on the Possibility of Detection}
\label{sec:possibility}
\par In light of the growing interest in LLM-generated text detection, researchers have recently investigated the possibility of detecting AI-generated text from a theoretical perspective, exploring the fundamental feasibility and boundaries associated with this task.
\par Sadasivan et al. for instance, 
 come up with an impossibility finding~\cite{Sadasivan2023CanAT}: \textit{“as language models become more sophisticated and better at emulating human text, the performance of even the best-possible detector decreases drastically”}. They propose an upper bound for the area under the ROC curve of any decoder $\mathcal{D}$ as:
\begin{equation}
    AUROC(\mathcal{D})\leq\frac{1}{2}+TV(\mathcal{M},\mathcal{H})-\frac{TV(\mathcal{M},\mathcal{H})^2}{2}
    \label{eq:upperbound}
\end{equation}
where $TV(\mathcal{M},\mathcal{H})$ is the total variation distance between machine and human generated texts. This formula indicates that
when human and machine generated texts are very similar i.e., $TV(\mathcal{M},\mathcal{H})$ is very small, even the best-possible detector may only perform marginally better than a random classifier. The proof is provided in Appendix.~\ref{appendix-A}. 
\par According to this formulation, as the $TV$ distance between AI and human text distributions 
reduces, the AUROC of the optimal detector also decreases accordingly as illustrated in Figure~\ref{fig:auroc}}.

\begin{figure}
    \centering    \includegraphics[width=1\linewidth]{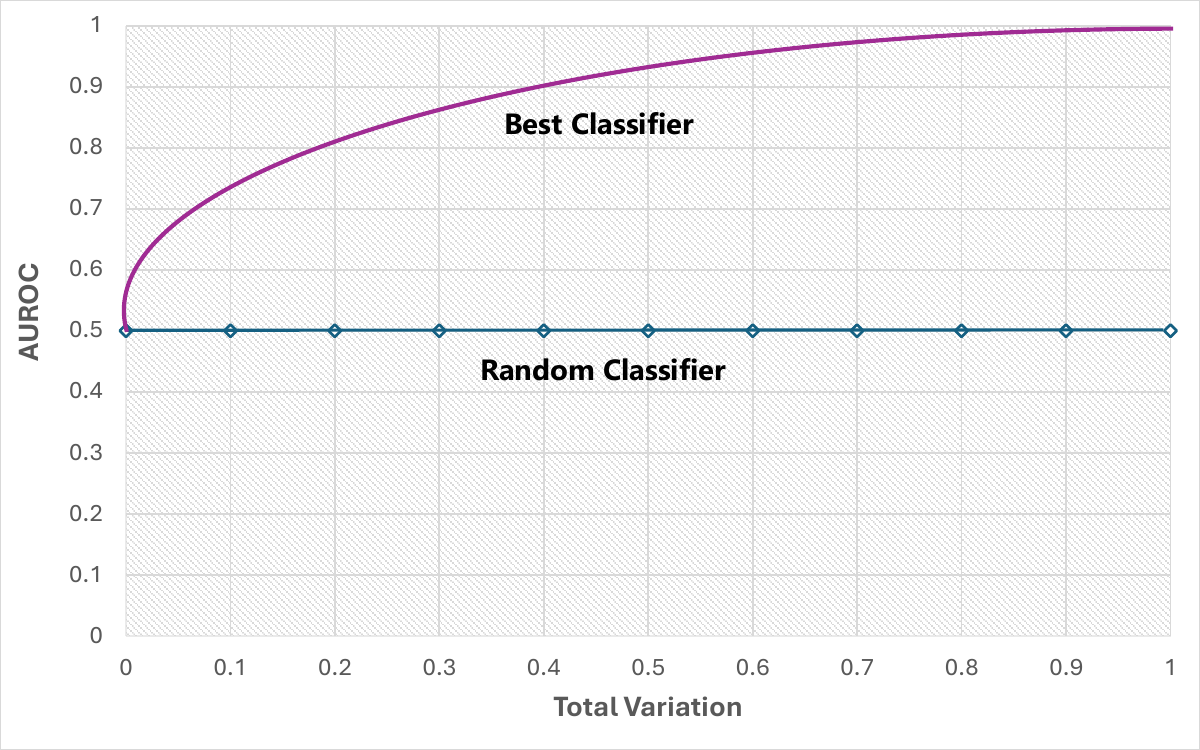}
    \caption{Comparing the AUROC of the optimal detector to that of a random classifier shows that as the TV distance between AI and human generated text distributions decreases, the AUROC of the optimal detector also drops accordingly.}
    \label{fig:auroc}
\end{figure}
\par However, another interesting work by Chakraborty et al.~\cite{Chakraborty2023OnTP}  suggests: \textit{“as long as the distributions of human and machine generated texts are not exactly the same, which is the case in most scenarios, it is possible to detect AI generated texts, if we collect enough samples of each distributions”.} In fact, Chakraborty et al. demonstrate that the AUROC upper bound proposed by Sadasivan et al. might be too conservative for detection in practical scenarios. Specifically, they introduce a hidden possibility by replacing  $TV(\mathcal{M}^,\mathcal{H})$ with  $TV(\mathcal{M}^{\bigotimes n},\mathcal{H}^{\bigotimes n})$ in AUROC equation, where $m^ {\bigotimes n} := m \bigotimes m \bigotimes \dots \bigotimes m$ (n times) denotes the product distribution over sample set $\mathcal{S}:=\{s_i\}, i\in \{1,\dots n\}$, as does $h^{\bigotimes n}$. Since $TV(\mathcal{M}^{\bigotimes n},\mathcal{H}^{\bigotimes n})$ is an increasing sequence, it eventually converges to $1$ as the number of samples for each distribution increases. It is clear that if the number of
samples increases, the total variation distance approaches $1$ very quickly, and hence increasing the AUROC.

\par In another work, Zhang et al.~\cite{Zhang2023WatermarksIT}, explore the theoretical aspect of watermarking detection. They define watermarking as the procedure of incorporating a statistical signal, commonly known as a “watermark”, into a model's output. This embedded watermark acts as a verification signal, ensuring that the output indeed originates from the model. A robust watermarking approach prevents attackers from removing the watermark without significantly degrading the output quality.
\par In this study, the authors put forth two fundamental assumptions. Firstly, they introduce the concept of a 
 “Quality Oracle”, which grants attackers access to an oracle capable of evaluating the quality of model outputs. This oracle assists attackers in assessing the quality of modified responses. Secondly, they introduce the “Perturbation Oracle” which allows attackers to modify an output while maintaining a non-trivial probability of preserving quality. Essentially, the perturbation oracle induces an efficiently mixing random walk on high-quality outputs. 
 \par They discover that for any public or secret-key watermarking scenario that satisfies these assumptions, there exists an efficient attacker: “Given a prompt $p$ and a watermarked output $y$, this attacker can leverage the quality and perturbation oracles to obtain an output ${y}^{'}$ with a probability very close to $1$. The attacker's goal is to find an output ${y}^{'}$ such that (1) ${y}^{'}$ is not
 watermarked with high probability and (2) ${Q(p,{y}^{'} )} \geq {Q(p,y)}$”~\cite{Zhang2023WatermarksIT}. Simply put, watermarking without causing significant quality degradation is impossible and as a result, alternative approaches should be leveraged to safeguard the intellectual property of model developers.
\par Detecting AI-generated text is a crucial and challenging task that has significant impact on related NLP tasks. However, the current state-of-the-art methods are sometimes limited by the lack of a comprehensive understanding of the fundamental feasibility and boundaries of this task. Therefore, it is critical to pursue further exploration and investigation of the theoretical aspects of AI-text detection, as this can lead to the development of more robust and effective techniques, as well as the identification of new research directions and opportunities.
 \section{Opportunities and Future Research}
 \label{sec:newops}
 This paper provides a comprehensive overview of the latest developments and best practices in security and risk mitigation of LLMs. To broaden the scope, this section explores the emerging opportunities for advancing the field of security, vulnerability and risk mitigation studies of LLMs.
 \subsection{Opportunities in Security \& privacy}
 \paragraph{Opportunities in Data Leakage \& Memorization}
 As previously mentioned, LLMs face challenges related to memorization and data leakage. Exploring novel opportunities to address these issues can significantly advance the field. Some promising avenues include: 
\begin{itemize}
\item \textbf{Developing multi-aspect techniques to prevent sensitive data leakage:} These techniques should consider various dimensions, including  model-related aspects (e.g., training data selection and differential privacy), data (such as data classification, access control, and monitoring), and user-based factors (e.g., detecting abnormal patterns in user-LLM interactions and managing user access)
    \item \textbf{Developing new methods to mitigate memorization:} Given the current scarcity of techniques for tackling memorization in LLMs, it is crucial to put forth novel approaches (e.g., model editing) to mitigate the memorization.
    \item \textbf{Investigating and identifying patterns of memorization:} Exploring whether there exist patterns of memorization and pinpointing the specific data categories that the model remembers the most represents an underexplored opportunity.
\end{itemize}
 \paragraph{Opportunities in LLM Code Generation} As mentioned earlier, the security implications of code generation present several challenges. However, each challenge also represents an opportunity to improve the security of codes produced by LLMs. Some of these opportunities are:
 \begin{itemize}
\item \textbf{Developing methods to ensure the reproducibility and transparency of code generation:} This can involve practices such as documenting the seed used, specifying the model version, and capturing relevant prompt details during the code generation process
\item  \textbf{Exploring ways to generate large and diverse corpora of code samples for security analysis:} Researchers can employ techniques like data augmentation, which involves creating variations of existing code snippets, adversarial examples that stress-test the model's robustness, and even self-play techniques where the model generates code samples and evaluates them against its own predictions~\cite{Wang2023EnhancingLL}.
\item \textbf{Designing realistic and comprehensive test scenarios that cover various aspects of software security:} These scenarios should comprise functional requirements, ensuring that the software behaves as expected, as well as non-functional requirements, such as performance, scalability, and reliability. Additionally, incorporating adversarial requirements—where the system is tested against intentional attacks or misuse—can further enhance the security assessment process.
\item \textbf{Improving the robustness and generalization of code generation to different prompts:} Researchers have the opportunity to improve the robustness and adaptability of code generation across diverse prompts. The ultimate goal is to develop a model that consistently generates secure code, for all input prompts. Achieving this goal ensures that the generated code remains dependable, resilient, and resistant to vulnerabilities, regardless of the prompt used.
\item \textbf{Investigating the impact of coding language on the security of code generation:} By analyzing the strengths and weaknesses of various languages and paradigms, we can identify which languages are more robust for code generation. This exploration benefits both code generation practices and the design and enhancement of coding languages.
\item \textbf{Adapting to the evolutionary nature of cyber-security:} Evolutionary nature of cyber-security requires proactive measures. These include keeping training data up-to-date, revising evaluation metrics, and aligning code generation practices with the latest industry standards and trends.
\end{itemize}
 \subsection{Opportunities in Vulnerabilities Study}
 Building upon the foundational work in understanding LLM vulnerabilities, there remains a broad spectrum of opportunities to deepen our knowledge and enhance the resilience of these models. The following areas represent promising avenues for future research:

\begin{itemize}
    \item \textbf{Extending experiments on diverse NLP applications besides classification task:} While classification have been the primary focus, exploring other tasks (e.g., language modeling, summarization, sentiment analysis, and question answering) will provide a more comprehensive understanding of LLM behavior and security implications.

    \item \textbf{Examining LLMs vulnerability at both model architecture and model size level:} It is essential to comprehensively assess risks and prioritize mitigation efforts based on the severity of identified vulnerabilities. Understanding the interplay between architecture choices and model scale allows us to make better decisions to enhance LLM security.

    \item \textbf{Adapting multi-faceted defense strategy for safeguarding against backdoor injection:} There is an opportunity to craft multi-faceted defense strategies against backdoor injection by combining techniques such as prompt filtering to exclude harmful input and specialized security-oriented LLMs trained to detect suspicious instructions. 

    \item \textbf{Expanding the scope of prompt injection studies:} Including more complex interaction scenarios, such as conversational agents and context-aware applications we can uncover new insights and potentially enhance resilience against security threats.
    
    \item \textbf{Investigating the role of transfer learning and fine-tuning:} This involves examining how vulnerabilities initially identified in pre-trained models may either be magnified or alleviated when those models are fine-tuned for specific tasks or domains.
    
    \item \textbf{Identifying and mitigating emerging risks:} New studies may incorporate recent progress in security, behavioral analysis, adversarial learning and  cyber-security forensics to detect and mitigate sophisticated attacks.
    \item \textbf{Assessing and evaluating the impact of dataset diversity and representativeness:} This involves examining how the features of training data affect the model's ability to withstand vulnerabilities, especially concerning bias and fairness and more importantly data poisoning.
    \end{itemize}
 
 \subsection{Opportunities in Risk Mitigation Study}
\paragraph{AI-Generated Text Detection Techniques} As mentioned earlier, the detection of AI-generated text is a challenging task and the current methods are frequently constrained by different factors, and susceptible to malicious attacks. Therefore, it is critical to pursue further exploration and investigation on both theoretical and practical aspects of AI-text detection. Some of the opportunities are:
\begin{itemize}
    \item  \textbf{Creating diverse and representative datasets:} The existing datasets may not cover all the nuances of AI-generated content used to train and evaluate AI-text detection models. Developing more diverse and representative datasets improves the models' ability to generalize and enables more reliable assessments.
    \item \textbf{Exploring more advanced and interpretable features:} By discerning subtle nuances and interpretable features, we can gain a fine-grained understanding of the distinctions between human vs. AI-generated texts.
    \item \textbf{Developing more robust and domain adaptive learning methods:} Considering the ever-changing field of AI-generated text, exploring methods like adversarial learning, meta-learning, and self-supervised learning~\cite{WeberWulff2023TestingOD} can result in more resilient and adaptable solutions. 
    \item \textbf{A comprehensive understanding of the fundamental feasibility and boundaries:} A thorough grasp of the fundamental feasibility and limitations of AI-generated text detection is essential. However, the theoretical aspect of this task is mainly overlooked in the literature. Therefore, further exploration and investigation of the theoretical aspects are necessary. This can lead to the development of more robust and effective techniques as well as uncovering new research directions.
    \item \textbf{Evaluating the ethical and social implications of AI-text detection:} While detecting synthetic content is essential, it is important to consider the potential risks of false positives. These inaccuracies can lead to unintended consequences, such as unjust penalties or unwarranted suspicion, impacting both individuals and society at large.
\end{itemize}

\paragraph{Editing LLMs} Understanding where knowledge is retained with an LLM is important as it may lead to adverse outcomes such as hallucinations or biases. Thus, it is crucial to identify the nature of factual information that is stored, and to apply mitigation strategies to eliminate the potential sources of unreliability if they occur. Although this area has seen remarkable progress, there are still some areas for improvement:

\begin{itemize}
    \item \textbf{Developing unified platform for frameworks to test multiple methods:} As new methods emerge, having a consolidated framework allows efficient comparison across different datasets. Additionally, integrating different types of knowledge into this framework simplifies the assessment of layers that focus on distinct data or knowledge domains~\cite{wang2023easyedit}.
    \item \textbf{Further exploration of model editing study across diverse datasets and network layers:} Given that current research emphasizes on certain domains in NLP, it is beneficial to evaluate additional NLP datasets to assess whether the trend of where certain information is stored is consistent. Furthermore, extending this assessment to other NLP domains that may differ could reveal potential variations in current trends.

\end{itemize}

\paragraph{Red/Green Teaming} Similar to the cyber-security domain, where red teaming has been beneficial to enhance security, both red teaming and green teaming for LLMs have revealed the vulnerability of LLMs to malicious users. There are nascent contributions like RLHF that prevent more attacks than other methods, as demonstrated in~\cite{perez2022red}. However, there are still some areas for improvement such as:

\begin{itemize}
    \item \textbf{Creating more safeguards to prevent the attacks:} Given the rising popularity of LLMs, they are becoming the central component in many products. Thus, there is a critical need to implement multiple safeguards, as new attacks continually emerge. 
    \item \textbf{Evaluating the impact of attacks on specific models:} 
    Research based on Ganguli et al.'s work~\cite{Ganguli2022RedTL} indicates that LLMs that utilize RLHF exhibit greater resilience to red teaming attacks compared to other models. However, further experiments are necessary to uncover any limitations. By understanding these limitations, researchers can devise innovative strategies to mitigate existing red teaming attacks and anticipate potential new ones.
    \item \textbf{Designing an automated system to reduce human dependency in red/green teaming:} Since human examination of specific LLM outputs related to red and green teaming can negatively impact well-being, automating the process becomes crucial. This automation aims to minimize the harm experienced by individuals involved in red and green teaming.
    
\end{itemize}

\section{Conclusions}
\label{sec:cons}
This paper provides a comprehensive analysis of the security and risk mitigation aspects of LLMs. We examine the security issues that emerge with LLM usage, such as information leakage, unauthorized access, and insecure code generation. In addition, we explore various types of attacks that target LLMs and classify them into three main categories: Model-based, training-time and inference-time attacks. We also investigate the risks and misuses of LLMs, such as bias, discrimination, misinformation, plagiarism, copyright infringement and other potential social and ethical implications of applying LLMs in different domains. Moreover, we present a thorough evaluation of the mitigation strategies that can be employed to improve the security and robustness of LLMs, such as red and green teaming, model editing, watermarking, and AI-generated text detection techniques, while discussing the limitations and trade-offs of each strategy. Lastly, we identify some open challenges and future directions for research in this area, such as developing more effective defense mechanisms, establishing standards and guidelines for development and deployment of LLMs, promoting collaboration and awareness among the stakeholders involved in utilization.
\section{Acknowledgements}
The authors express their gratitude to Dr. Sadid Hassan for engaging in several discussions and providing valuable feedback on this work. This study represents independent research conducted by the authors and does not necessarily represent the views or opinions of any organizations.
\newpage
\onecolumn
\appendix
\section{Appendix}
\label{appendix-A}
\subsection*{A-ROC and AUROC Bound using Total Variation}

The ROC curve is a plot between the true positive rate (TPR) and the false positive rate (FPR), which are defined as follows~\cite{Sadasivan2023CanAT}:

\begin{align*}
\text{TPR}_\gamma &= \mathbb{P}_{s \sim M}[D(s) \geq \gamma] \\
\text{FPR}_\gamma &= \mathbb{P}_{s \sim H}[D(s) \geq \gamma]
\end{align*}

where $\gamma$ is a classifier threshold parameter, $M$ denotes the distribution over positives (e.g., machine-generated texts), and $H$ denotes the distribution over negatives (e.g., human-written texts).

\bigskip

We can bound the difference between $\text{TPR}_\gamma$ and $\text{FPR}_\gamma$ by the total variation distance between distributions $M$ and $H$~\cite{Sadasivan2023CanAT}:

\begin{align*}
|\text{TPR}_\gamma - \text{FPR}_\gamma| &= \left|\mathbb{P}_{s \sim M}[D(s) \geq \gamma] -\mathbb{P}_{s \sim H}[D(s) \geq \gamma]\right| \\
&\leq \text{TV}(M, H)
\end{align*}

Thus,

\[
\text{TPR}_\gamma \leq \text{FPR}_\gamma + \text{TV}(M, H)
\]

Since $\text{TPR}_\gamma \leq 1$, we have:

\[
\text{TPR}_\gamma \leq \min(\text{FPR}_\gamma + \text{TV}(M, H), 1)
\]

\bigskip

Let $x = \text{FPR}_\gamma$, $y = \text{TPR}_\gamma$, and $tv = \text{TV}(M, H)$. The AUROC is defined as:

\[
\text{AUROC}(D) = \int_0^1 y \, dx \leq \int_0^1 \min(x + tv, 1) \, dx
\]

We split the integral at $x = 1 - tv$:

\begin{align*}
\text{AUROC}(D) &\leq \int_0^{1 - tv} (x + tv) \, dx + \int_{1 - tv}^{1} 1 \, dx \\
&= \left[ \frac{x^2}{2} + tv \cdot x \right]_0^{1 - tv} + \left[ x \right]_{1 - tv}^{1} \\
&= \left( \frac{(1 - tv)^2}{2} + tv(1 - tv) \right) + (1 - (1 - tv)) \\
&= \frac{(1 - tv)^2}{2} + tv(1 - tv) + tv
\end{align*}

Simplifying the expression:

\begin{align*}
\text{AUROC}(D) &\leq \frac{(1 - tv)^2}{2} + tv(1 - tv) + tv \\
&= \frac{1 - 2tv + tv^2}{2} + tv - tv^2 + tv \\
&= \frac{1}{2} + tv - \frac{tv^2}{2}
\end{align*}

\[
\boxed{
\text{AUROC}(D) \leq \frac{1}{2} + tv - \frac{tv^2}{2}
}
\]

\bigskip
\twocolumn
\bibliographystyle{acl_natbib}
\bibliography{tacl2021}

\begin{thebibliography}{211}
\expandafter\ifx\csname natexlab\endcsname\relax\def\natexlab#1{#1}\fi

\bibitem[{Ope(2023)}]{OpenAI}
 2023.
\newblock \href {https://platform.openai.com/ai-text-classifier} {Openai. ai text classifier.}

\bibitem[{Zer(2023)}]{ZeroGPT}
 2023.
\newblock \href {https://www.zerogpt.com} {Zerogpt: Ai text detector.}

\bibitem[{Alon et~al.(2018)Alon, Brody, Levy, and Yahav}]{Alon2018code2seqGS}
Uri Alon, Shaked Brody, Omer Levy, and Eran Yahav. 2018.
\newblock \href {https://api.semanticscholar.org/CorpusID:51926976} {code2seq: Generating sequences from structured representations of code}.
\newblock \emph{ArXiv}, abs/1808.01400.

\bibitem[{Anonymous(2023)}]{anonymous2023how}
Anonymous. 2023.
\newblock \href {https://openreview.net/forum?id=567BjxgaTp} {How to catch an {AI} liar: Lie detection in black-box {LLM}s by asking unrelated questions}.
\newblock In \emph{Submitted to The Twelfth International Conference on Learning Representations}.
\newblock Under review.

\bibitem[{Antoun et~al.(2023)Antoun, Mouilleron, Sagot, and Seddah}]{Antoun2023TowardsAR}
Wissam Antoun, Virginie Mouilleron, Beno{\^i}t Sagot, and Djam{\'e} Seddah. 2023.
\newblock Towards a robust detection of language model generated text: Is chatgpt that easy to detect?
\newblock \emph{ArXiv}, abs/2306.05871.

\bibitem[{Aryabumi et~al.(2024)Aryabumi, Su, Ma, Morisot, Zhang, Locatelli, Fadaee, Ustun, and Hooker}]{Aryabumi2024ToCOA}
Viraat Aryabumi, Yixuan Su, Raymond Ma, Adrien Morisot, Ivan Zhang, Acyr~F. Locatelli, Marzieh Fadaee, A.~Ustun, and Sara Hooker. 2024.
\newblock \href {https://api.semanticscholar.org/CorpusId:271909530} {To code, or not to code? exploring impact of code in pre-training}.
\newblock \emph{ArXiv}, abs/2408.10914.

\bibitem[{Asare et~al.(2023)Asare, Nagappan, and Asokan}]{asare2023githubs}
Owura Asare, Meiyappan Nagappan, and N.~Asokan. 2023.
\newblock \href {http://arxiv.org/abs/2204.04741} {Is github's copilot as bad as humans at introducing vulnerabilities in code?}

\bibitem[{Bachu et~al.(2024)Bachu, Shayegani, Chakraborty, Lal, Dutta, Song, Dong, Abu-Ghazaleh, and Roy-Chowdhury}]{bachu2024unfairalignmentexaminingsafety}
Saketh Bachu, Erfan Shayegani, Trishna Chakraborty, Rohit Lal, Arindam Dutta, Chengyu Song, Yue Dong, Nael Abu-Ghazaleh, and Amit~K. Roy-Chowdhury. 2024.
\newblock \href {http://arxiv.org/abs/2411.04291} {Unfair alignment: Examining safety alignment across vision encoder layers in vision-language models}.

\bibitem[{Bakhtin et~al.(2019)Bakhtin, Gross, Ott, Deng, Ranzato, and Szlam}]{Bakhtin2019RealOF}
Anton Bakhtin, Sam Gross, Myle Ott, Yuntian Deng, Marc'Aurelio Ranzato, and Arthur~D. Szlam. 2019.
\newblock Real or fake? learning to discriminate machine from human generated text.
\newblock \emph{ArXiv}, abs/1906.03351.

\bibitem[{Batra et~al.(2021)Batra, Punn, Sonbhadra, and Agarwal}]{Batra2021BERTBasedSA}
Himanshu Batra, Narinder~Singh Punn, Sanjay~Kumar Sonbhadra, and Sonali Agarwal. 2021.
\newblock \href {https://api.semanticscholar.org/CorpusID:235352765} {Bert-based sentiment analysis: A software engineering perspective}.
\newblock In \emph{International Conference on Database and Expert Systems Applications}.

\bibitem[{Betley et~al.(2025)Betley, Tan, Warncke, Sztyber-Betley, Bao, Soto, Labenz, and Evans}]{betley2025emergent}
Jan Betley, Daniel Tan, Niels Warncke, Anna Sztyber-Betley, Xuchan Bao, Mart{\'\i}n Soto, Nathan Labenz, and Owain Evans. 2025.
\newblock Emergent misalignment: Narrow finetuning can produce broadly misaligned llms.
\newblock \emph{arXiv preprint arXiv:2502.17424}.

\bibitem[{Bhardwaj and Poria(2023)}]{Bhardwaj2023RedTeamingLL}
Rishabh Bhardwaj and Soujanya Poria. 2023.
\newblock \href {https://api.semanticscholar.org/CorpusID:261030829} {Red-teaming large language models using chain of utterances for safety-alignment}.
\newblock \emph{ArXiv}, abs/2308.09662.

\bibitem[{Bhat et~al.(2023)Bhat, Meng, Liu, Zhou, and Yavuz}]{Bhat2023InvestigatingAO}
Meghana~Moorthy Bhat, Rui Meng, Ye~Liu, Yingbo Zhou, and Semih Yavuz. 2023.
\newblock \href {https://api.semanticscholar.org/CorpusID:262013357} {Investigating answerability of llms for long-form question answering}.
\newblock \emph{ArXiv}, abs/2309.08210.

\bibitem[{Bhatt et~al.(2023)Bhatt, Chennabasappa, Nikolaidis, Wan, Evtimov, Gabi, Song, Ahmad, Aschermann, Fontana, Frolov, Giri, Kapil, Kozyrakis, LeBlanc, Milazzo, Straumann, Synnaeve, Vontimitta, Whitman, and Saxe}]{Bhatt2023CyberSecEval}
Manish Bhatt, Sahana Chennabasappa, Cyrus Nikolaidis, Shengye Wan, Ivan Evtimov, Dominik Gabi, Daniel Song, Faizan Ahmad, Cornelius Aschermann, Lorenzo Fontana, Sasha Frolov, Ravi~Prakash Giri, Dhaval Kapil, Yiannis Kozyrakis, David LeBlanc, James Milazzo, Aleksandar Straumann, Gabriel Synnaeve, Varun Vontimitta, Spencer Whitman, and Joshua Saxe. 2023.
\newblock \href {https://arxiv.org/abs/2312.04724} {{Purple Llama CyberSecEval}: A secure coding benchmark for language models}.
\newblock \emph{arXiv preprint arXiv:2312.04724}.

\bibitem[{Biderman et~al.(2023{\natexlab{a}})Biderman, Prashanth, Sutawika, Schoelkopf, Anthony, Purohit, and Raf}]{biderman2023emergent}
Stella Biderman, USVSN~Sai Prashanth, Lintang Sutawika, Hailey Schoelkopf, Quentin Anthony, Shivanshu Purohit, and Edward Raf. 2023{\natexlab{a}}.
\newblock Emergent and predictable memorization in large language models.
\newblock \emph{arXiv preprint arXiv:2304.11158}.

\bibitem[{Biderman et~al.(2023{\natexlab{b}})Biderman, Schoelkopf, Anthony, Bradley, O’Brien, Hallahan, Khan, Purohit, Prashanth, Raff et~al.}]{biderman2023pythia}
Stella Biderman, Hailey Schoelkopf, Quentin~Gregory Anthony, Herbie Bradley, Kyle O’Brien, Eric Hallahan, Mohammad~Aflah Khan, Shivanshu Purohit, USVSN~Sai Prashanth, Edward Raff, et~al. 2023{\natexlab{b}}.
\newblock Pythia: A suite for analyzing large language models across training and scaling.
\newblock In \emph{International Conference on Machine Learning}, pages 2397--2430. PMLR.

\bibitem[{Birch et~al.(2023{\natexlab{a}})Birch, Hackett, Trawicki, Suri, and Garraghan}]{Birch2023ModelLA}
Lewis Birch, William Hackett, Stefan Trawicki, Neeraj Suri, and Peter Garraghan. 2023{\natexlab{a}}.
\newblock \href {https://api.semanticscholar.org/CorpusID:262053852} {Model leeching: An extraction attack targeting llms}.
\newblock \emph{ArXiv}, abs/2309.10544.

\bibitem[{Birch et~al.(2023{\natexlab{b}})Birch, Hackett, Trawicki, Suri, and Garraghan}]{birch2023model}
Lewis Birch, William Hackett, Stefan Trawicki, Neeraj Suri, and Peter Garraghan. 2023{\natexlab{b}}.
\newblock Model leeching: An extraction attack targeting llms.
\newblock \emph{arXiv preprint arXiv:2309.10544}.

\bibitem[{Borkar(2023)}]{borkar2023can}
Jaydeep Borkar. 2023.
\newblock What can we learn from data leakage and unlearning for law?
\newblock \emph{arXiv preprint arXiv:2307.10476}.

\bibitem[{Bruch et~al.(2009)Bruch, Monperrus, and Mezini}]{BruchCodeCompletion2009}
Marcel Bruch, Martin Monperrus, and Mira Mezini. 2009.
\newblock \href {https://doi.org/10.1145/1595696.1595728} {Learning from examples to improve code completion systems}.
\newblock In \emph{Proceedings of the 7th Joint Meeting of the European Software Engineering Conference and the ACM SIGSOFT Symposium on The Foundations of Software Engineering}, ESEC/FSE '09, page 213–222, New York, NY, USA. Association for Computing Machinery.

\bibitem[{Cai et~al.(2022)Cai, Xu, Xu, Zhang, and Yuan}]{Cai2022BadPromptBA}
Xiangrui Cai, Haidong Xu, Sihan Xu, Ying Zhang, and Xiaojie Yuan. 2022.
\newblock Badprompt: Backdoor attacks on continuous prompts.
\newblock \emph{ArXiv}, abs/2211.14719.

\bibitem[{Cao and Yang(2015)}]{cao2015towards}
Yinzhi Cao and Junfeng Yang. 2015.
\newblock Towards making systems forget with machine unlearning.
\newblock In \emph{2015 IEEE symposium on security and privacy}, pages 463--480. IEEE.

\bibitem[{Carlini et~al.(2021)Carlini, Tramer, Wallace, Jagielski, Herbert-Voss, Lee, Roberts, Brown, Song, Erlingsson et~al.}]{carlini2021extracting}
Nicholas Carlini, Florian Tramer, Eric Wallace, Matthew Jagielski, Ariel Herbert-Voss, Katherine Lee, Adam Roberts, Tom Brown, Dawn Song, Ulfar Erlingsson, et~al. 2021.
\newblock Extracting training data from large language models.
\newblock In \emph{30th USENIX Security Symposium (USENIX Security 21)}, pages 2633--2650.

\bibitem[{Casper et~al.(2023)Casper, Lin, Kwon, Culp, and Hadfield-Menell}]{casper2023explore}
Stephen Casper, Jason Lin, Joe Kwon, Gatlen Culp, and Dylan Hadfield-Menell. 2023.
\newblock Explore, establish, exploit: Red teaming language models from scratch.
\newblock \emph{arXiv preprint arXiv:2306.09442}.

\bibitem[{Chakraborty et~al.(2023)Chakraborty, Bedi, Zhu, An, Manocha, and Huang}]{Chakraborty2023OnTP}
Souradip Chakraborty, A.~S. Bedi, Sicheng Zhu, Bang An, Dinesh Manocha, and Furong Huang. 2023.
\newblock On the possibilities of ai-generated text detection.
\newblock \emph{ArXiv}, abs/2304.04736.

\bibitem[{Chakraborty et~al.(2024)Chakraborty, Shayegani, Cai, Abu-Ghazaleh, Asif, Dong, Roy-Chowdhury, and Song}]{chakraborty-etal-2024-textual}
Trishna Chakraborty, Erfan Shayegani, Zikui Cai, Nael~B. Abu-Ghazaleh, M.~Salman Asif, Yue Dong, Amit Roy-Chowdhury, and Chengyu Song. 2024.
\newblock \href {https://doi.org/10.18653/v1/2024.findings-emnlp.574} {Can textual unlearning solve cross-modality safety alignment?}
\newblock In \emph{Findings of the Association for Computational Linguistics: EMNLP 2024}, pages 9830--9844, Miami, Florida, USA. Association for Computational Linguistics.

\bibitem[{Charan et~al.(2023)Charan, Chunduri, Anand, and Shukla}]{Charan2023FromTT}
P.~V.~Sai Charan, Hrushikesh Chunduri, P.~Mohan Anand, and Sandeep~Kumar Shukla. 2023.
\newblock From text to mitre techniques: Exploring the malicious use of large language models for generating cyber attack payloads.
\newblock \emph{ArXiv}, abs/2305.15336.

\bibitem[{Chen and Ding(2023)}]{chen-ding-2023-probing}
Honghua Chen and Nai Ding. 2023.
\newblock \href {https://aclanthology.org/2023.findings-emnlp.858} {Probing the {``}creativity{''} of large language models: Can models produce divergent semantic association?}
\newblock In \emph{Findings of the Association for Computational Linguistics: EMNLP 2023}, pages 12881--12888, Singapore. Association for Computational Linguistics.

\bibitem[{Chen et~al.(2023)Chen, Wei, Cao, Zhou, and Hu}]{Chen2023CanLL}
Mengyang Chen, Lingwei Wei, Han Cao, Wei Zhou, and Song Hu. 2023.
\newblock \href {https://api.semanticscholar.org/CorpusID:265308637} {Can large language models understand content and propagation for misinformation detection: An empirical study}.
\newblock \emph{ArXiv}, abs/2311.12699.

\bibitem[{Chen et~al.(2021)Chen, Salem, Chen, Backes, Ma, Shen, Wu, and Zhang}]{chen2021badnl}
Xiaoyi Chen, Ahmed Salem, Dingfan Chen, Michael Backes, Shiqing Ma, Qingni Shen, Zhonghai Wu, and Yang Zhang. 2021.
\newblock Badnl: Backdoor attacks against nlp models with semantic-preserving improvements.
\newblock In \emph{Annual computer security applications conference}, pages 554--569.

\bibitem[{Chen et~al.(2017)Chen, Liu, Li, Lu, and Song}]{Chen2017TargetedBA}
Xinyun Chen, Chang Liu, Bo~Li, Kimberly Lu, and Dawn~Xiaodong Song. 2017.
\newblock \href {https://api.semanticscholar.org/CorpusID:36122023} {Targeted backdoor attacks on deep learning systems using data poisoning}.
\newblock \emph{ArXiv}, abs/1712.05526.

\bibitem[{Chiang et~al.(2023)Chiang, Li, Lin, Sheng, Wu, Zhang, Zheng, Zhuang, Zhuang, Gonzalez et~al.}]{chiang2023vicuna}
Wei-Lin Chiang, Zhuohan Li, Zi~Lin, Ying Sheng, Zhanghao Wu, Hao Zhang, Lianmin Zheng, Siyuan Zhuang, Yonghao Zhuang, Joseph~E Gonzalez, et~al. 2023.
\newblock Vicuna: An open-source chatbot impressing gpt-4 with 90\%* chatgpt quality.
\newblock \emph{See https://vicuna. lmsys. org (accessed 14 April 2023)}.

\bibitem[{Christiano et~al.(2023)Christiano, Leike, Brown, Martic, Legg, and Amodei}]{christiano2023deep}
Paul Christiano, Jan Leike, Tom~B. Brown, Miljan Martic, Shane Legg, and Dario Amodei. 2023.
\newblock \href {http://arxiv.org/abs/1706.03741} {Deep reinforcement learning from human preferences}.

\bibitem[{Christiano et~al.(2017)Christiano, Leike, Brown, Martic, Legg, and Amodei}]{NIPS2017_d5e2c0ad}
Paul~F Christiano, Jan Leike, Tom Brown, Miljan Martic, Shane Legg, and Dario Amodei. 2017.
\newblock \href {https://proceedings.neurips.cc/paper_files/paper/2017/file/d5e2c0adad503c91f91df240d0cd4e49-Paper.pdf} {Deep reinforcement learning from human preferences}.
\newblock In \emph{Advances in Neural Information Processing Systems}, volume~30. Curran Associates, Inc.

\bibitem[{Cobbe et~al.(2021)Cobbe, Kosaraju, Bavarian, Chen, Jun, Kaiser, Plappert, Tworek, Hilton, Nakano, Hesse, and Schulman}]{Cobbe2021TrainingVT}
Karl Cobbe, Vineet Kosaraju, Mohammad Bavarian, Mark Chen, Heewoo Jun, Lukasz Kaiser, Matthias Plappert, Jerry Tworek, Jacob Hilton, Reiichiro Nakano, Christopher Hesse, and John Schulman. 2021.
\newblock \href {https://api.semanticscholar.org/CorpusID:239998651} {Training verifiers to solve math word problems}.
\newblock \emph{ArXiv}, abs/2110.14168.

\bibitem[{Cui et~al.(2023)Cui, Zhang, Chen, Zhang, Liu, Wang, and Liu}]{Cui2023FFTTH}
Shiyao Cui, Zhenyu Zhang, Yilong Chen, Wenyuan Zhang, Tianyun Liu, Siqi Wang, and Tingwen Liu. 2023.
\newblock \href {https://api.semanticscholar.org/CorpusID:265506833} {Fft: Towards harmlessness evaluation and analysis for llms with factuality, fairness, toxicity}.
\newblock \emph{ArXiv}, abs/2311.18580.

\bibitem[{Debenedetti et~al.(2024)Debenedetti, Zhang, Balunovi'c, Beurer-Kellner, Fischer, and Tram{\`e}r}]{Debenedetti2024AgentDojoADA}
Edoardo Debenedetti, Jie Zhang, Mislav Balunovi'c, Luca Beurer-Kellner, Marc Fischer, and Florian Tram{\`e}r. 2024.
\newblock \href {https://api.semanticscholar.org/CorpusId:270619628} {Agentdojo: A dynamic environment to evaluate attacks and defenses for llm agents}.
\newblock \emph{ArXiv}, abs/2406.13352.

\bibitem[{Deng et~al.(2023{\natexlab{a}})Deng, Liu, Li, Wang, Zhang, Li, Wang, Zhang, and Liu}]{deng2023jailbreaker}
Gelei Deng, Yi~Liu, Yuekang Li, Kailong Wang, Ying Zhang, Zefeng Li, Haoyu Wang, Tianwei Zhang, and Yang Liu. 2023{\natexlab{a}}.
\newblock Jailbreaker: Automated jailbreak across multiple large language model chatbots.
\newblock \emph{arXiv preprint arXiv:2307.08715}.

\bibitem[{Deng et~al.(2023{\natexlab{b}})Deng, Liu, Li, Wang, Zhang, Li, Wang, Zhang, and Liu}]{Deng2023MasterKeyAJ}
Gelei Deng, Yi~Liu, Yuekang Li, Kailong Wang, Ying Zhang, Zefeng Li, Haoyu Wang, Tianwei Zhang, and Yang Liu. 2023{\natexlab{b}}.
\newblock \href {https://api.semanticscholar.org/CorpusID:259951184} {Masterkey: Automated jailbreak across multiple large language model chatbots}.

\bibitem[{Deshpande et~al.(2023{\natexlab{a}})Deshpande, Murahari, Rajpurohit, Kalyan, and Narasimhan}]{Deshpande2023ToxicityIC}
A.~Deshpande, Vishvak~S. Murahari, Tanmay Rajpurohit, A.~Kalyan, and Karthik Narasimhan. 2023{\natexlab{a}}.
\newblock Toxicity in chatgpt: Analyzing persona-assigned language models.
\newblock \emph{ArXiv}, abs/2304.05335.

\bibitem[{Deshpande et~al.(2023{\natexlab{b}})Deshpande, Rajpurohit, Narasimhan, and Kalyan}]{Deshpande2023AnthropomorphizationOA}
A.~Deshpande, Tanmay Rajpurohit, Karthik Narasimhan, and A.~Kalyan. 2023{\natexlab{b}}.
\newblock Anthropomorphization of ai: Opportunities and risks.
\newblock \emph{ArXiv}, abs/2305.14784.

\bibitem[{Du et~al.(2023{\natexlab{a}})Du, Wang, Xing, Ya, Li, Jiang, and Fang}]{Du2023QuantifyingAA}
Li~Du, Yequan Wang, Xingrun Xing, Yiqun Ya, Xiang Li, Xin Jiang, and Xuezhi Fang. 2023{\natexlab{a}}.
\newblock \href {https://api.semanticscholar.org/CorpusID:261682256} {Quantifying and attributing the hallucination of large language models via association analysis}.
\newblock \emph{ArXiv}, abs/2309.05217.

\bibitem[{Du et~al.(2023{\natexlab{b}})Du, Li, Torralba, Tenenbaum, and Mordatch}]{Du2023ImprovingFactuality}
Yilun Du, Shuang Li, Antonio Torralba, Joshua Tenenbaum, and Igor Mordatch. 2023{\natexlab{b}}.
\newblock Improving factuality and reasoning in language models through multiagent debate.

\bibitem[{Ganguli et~al.(2022)Ganguli, Lovitt, Kernion, Askell, Bai, Kadavath, Mann, Perez, Schiefer, Ndousse, Jones, Bowman, Chen, Conerly, DasSarma, Drain, Elhage, El-Showk, Fort, Dodds, Henighan, Hernandez, Hume, Jacobson, Johnston, Kravec, Olsson, Ringer, Tran-Johnson, Amodei, Brown, Joseph, McCandlish, Olah, Kaplan, and Clark}]{Ganguli2022RedTL}
Deep Ganguli, Liane Lovitt, John Kernion, Amanda Askell, Yuntao Bai, Saurav Kadavath, Benjamin Mann, Ethan Perez, Nicholas Schiefer, Kamal Ndousse, Andy Jones, Sam Bowman, Anna Chen, Tom Conerly, Nova DasSarma, Dawn Drain, Nelson Elhage, Sheer El-Showk, Stanislav Fort, Zachary Dodds, T.~J. Henighan, Danny Hernandez, Tristan Hume, Josh Jacobson, Scott Johnston, Shauna Kravec, Catherine Olsson, Sam Ringer, Eli Tran-Johnson, Dario Amodei, Tom~B. Brown, Nicholas Joseph, Sam McCandlish, Christopher Olah, Jared Kaplan, and Jack Clark. 2022.
\newblock Red teaming language models to reduce harms: Methods, scaling behaviors, and lessons learned.
\newblock \emph{ArXiv}, abs/2209.07858.

\bibitem[{Ge et~al.(2023)Ge, Zhou, Hou, Khabsa, Wang, Wang, Han, and Mao}]{Ge2023MARTIL}
Suyu Ge, Chunting Zhou, Rui Hou, Madian Khabsa, Yi-Chia Wang, Qifan Wang, Jiawei Han, and Yuning Mao. 2023.
\newblock \href {https://api.semanticscholar.org/CorpusID:265157927} {Mart: Improving llm safety with multi-round automatic red-teaming}.
\newblock \emph{ArXiv}, abs/2311.07689.

\bibitem[{Gehman et~al.(2020)Gehman, Gururangan, Sap, Choi, and Smith}]{gehman2020realtoxicityprompts}
Samuel Gehman, Suchin Gururangan, Maarten Sap, Yejin Choi, and Noah~A Smith. 2020.
\newblock Realtoxicityprompts: Evaluating neural toxic degeneration in language models.
\newblock \emph{arXiv preprint arXiv:2009.11462}.

\bibitem[{Gehrmann et~al.(2019)Gehrmann, Strobelt, and Rush}]{Gehrmann2019GLTRSD}
Sebastian Gehrmann, Hendrik Strobelt, and Alexander~M. Rush. 2019.
\newblock Gltr: Statistical detection and visualization of generated text.
\newblock In \emph{Annual Meeting of the Association for Computational Linguistics}.

\bibitem[{Geng et~al.(2023)Geng, Gudibande, Liu, Wallace, Abbeel, Levine, and Song}]{geng2023koala}
Xinyang Geng, Arnav Gudibande, Hao Liu, Eric Wallace, Pieter Abbeel, Sergey Levine, and Dawn Song. 2023.
\newblock Koala: A dialogue model for academic research.
\newblock \emph{Blog post, April}, 1.

\bibitem[{Greshake et~al.(2023)Greshake, Abdelnabi, Mishra, Endres, Holz, and Fritz}]{Greshake2023NotWY}
Kai Greshake, Sahar Abdelnabi, Shailesh Mishra, Christoph Endres, Thorsten Holz, and Mario Fritz. 2023.
\newblock Not what you've signed up for: Compromising real-world llm-integrated applications with indirect prompt injection.

\bibitem[{Gudibande et~al.(2023)Gudibande, Wallace, Snell, Geng, Liu, Abbeel, Levine, and Song}]{gudibande2023false}
Arnav Gudibande, Eric Wallace, Charlie Snell, Xinyang Geng, Hao Liu, Pieter Abbeel, Sergey Levine, and Dawn Song. 2023.
\newblock The false promise of imitating proprietary llms.
\newblock \emph{arXiv preprint arXiv:2305.15717}.

\bibitem[{Gulcehre et~al.(2023)Gulcehre, Paine, Srinivasan, Konyushkova, Weerts, Sharma, Siddhant, Ahern, Wang, Gu, Macherey, Doucet, Firat, and de~Freitas}]{Gulcehre2023ReinforcedS}
Caglar Gulcehre, Tom~Le Paine, Srivatsan Srinivasan, Ksenia Konyushkova, Lotte Weerts, Abhishek Sharma, Aditya Siddhant, Alexa Ahern, Miaosen Wang, Chenjie Gu, Wolfgang Macherey, A.~Doucet, Orhan Firat, and Nando de~Freitas. 2023.
\newblock \href {https://api.semanticscholar.org/CorpusID:261031028} {Reinforced self-training (rest) for language modeling}.
\newblock \emph{ArXiv}, abs/2308.08998.

\bibitem[{Gupta et~al.(2023)Gupta, Mondal, Sheshadri, Zhao, Li, Wiegreffe, and Tandon}]{gupta2023editing}
Anshita Gupta, Debanjan Mondal, Akshay~Krishna Sheshadri, Wenlong Zhao, Xiang~Lorraine Li, Sarah Wiegreffe, and Niket Tandon. 2023.
\newblock Editing commonsense knowledge in gpt.
\newblock \emph{arXiv preprint arXiv:2305.14956}.

\bibitem[{Guu et~al.(2020)Guu, Lee, Tung, Pasupat, and Chang}]{Guu2023RALM}
Kelvin Guu, Kenton Lee, Zora Tung, Panupong Pasupat, and Ming-Wei Chang. 2020.
\newblock Realm: Retrieval-augmented language model pre-training.
\newblock ICML'20. JMLR.org.

\bibitem[{H{\"a}m{\"a}l{\"a}inen et~al.(2023)H{\"a}m{\"a}l{\"a}inen, Tavast, and Kunnari}]{hamalainen2023evaluating}
Perttu H{\"a}m{\"a}l{\"a}inen, Mikke Tavast, and Anton Kunnari. 2023.
\newblock Evaluating large language models in generating synthetic hci research data: a case study.
\newblock In \emph{Proceedings of the 2023 CHI Conference on Human Factors in Computing Systems}, page 3580688. ACM.

\bibitem[{Han et~al.(2023)Han, Buyukates, Hu, Jin, Jin, Sun, Wang, Xie, Zhang, Zhang, Zhang, He, and Avestimehr}]{Han2023FedMLSecurityAB}
Shanshan Han, Baturalp Buyukates, Zijian Hu, Han Jin, Weizhao Jin, Lichao Sun, Xiaoya Wang, Chulin Xie, Kai Zhang, Qifan Zhang, Yuhui Zhang, Chaoyang He, and Salman Avestimehr. 2023.
\newblock Fedmlsecurity: A benchmark for attacks and defenses in federated learning and llms.
\newblock \emph{ArXiv}, abs/2306.04959.

\bibitem[{Hartmann et~al.(2023{\natexlab{a}})Hartmann, Suri, Bindschaedler, Evans, Tople, and West}]{Hartmann2023SoKMI}
Valentin Hartmann, Anshuman Suri, Vincent Bindschaedler, David Evans, Shruti Tople, and Robert West. 2023{\natexlab{a}}.
\newblock \href {https://api.semanticscholar.org/CorpusID:264590727} {Sok: Memorization in general-purpose large language models}.
\newblock \emph{ArXiv}, abs/2310.18362.

\bibitem[{Hartmann et~al.(2023{\natexlab{b}})Hartmann, Suri, Bindschaedler, Evans, Tople, and West}]{hartmann2023sok}
Valentin Hartmann, Anshuman Suri, Vincent Bindschaedler, David Evans, Shruti Tople, and Robert West. 2023{\natexlab{b}}.
\newblock Sok: Memorization in general-purpose large language models.
\newblock \emph{arXiv preprint arXiv:2310.18362}.

\bibitem[{He et~al.(2023)He, Shen, Chen, Backes, and Zhang}]{He2023MGTBenchBM}
Xinlei He, Xinyu Shen, Zeyuan~Johnson Chen, Michael Backes, and Yang Zhang. 2023.
\newblock Mgtbench: Benchmarking machine-generated text detection.
\newblock \emph{ArXiv}, abs/2303.14822.

\bibitem[{Hu et~al.(2023)Hu, Wu, Mitra, Zhang, Sun, Huang, and Swaminathan}]{Hu2023TokenLevelAP}
Zhengmian Hu, Gang Wu, Saayan Mitra, Ruiyi Zhang, Tong Sun, Heng Huang, and Vishy Swaminathan. 2023.
\newblock \href {https://api.semanticscholar.org/CorpusID:265294544} {Token-level adversarial prompt detection based on perplexity measures and contextual information}.
\newblock \emph{ArXiv}, abs/2311.11509.

\bibitem[{Huang et~al.(2023{\natexlab{a}})Huang, Zhao, Backes, Shen, and Zhang}]{Huang2023CompositeBA}
Hai Huang, Zhengyu Zhao, Michael Backes, Yun Shen, and Yang Zhang. 2023{\natexlab{a}}.
\newblock \href {https://api.semanticscholar.org/CorpusID:263834732} {Composite backdoor attacks against large language models}.
\newblock \emph{ArXiv}, abs/2310.07676.

\bibitem[{Huang et~al.(2022)Huang, Shao, and Chang}]{huang-etal-2022-large}
Jie Huang, Hanyin Shao, and Kevin Chen-Chuan Chang. 2022.
\newblock \href {https://aclanthology.org/2022.findings-emnlp.148} {Are large pre-trained language models leaking your personal information?}
\newblock In \emph{Findings of the Association for Computational Linguistics: EMNLP 2022}, pages 2038--2047, Abu Dhabi, United Arab Emirates. Association for Computational Linguistics.

\bibitem[{Huang et~al.(2023{\natexlab{b}})Huang, Ruan, Huang, Jin, Dong, Wu, Bensalem, Mu, Qi, Zhao, Cai, Zhang, Wu, Xu, Wu, Freitas, and Mustafa}]{Huang2023ASO}
Xiaowei Huang, Wenjie Ruan, Wei Huang, Gao Jin, Yizhen Dong, Changshun Wu, Saddek Bensalem, Ronghui Mu, Yi~Qi, Xingyu Zhao, Kaiwen Cai, Yanghao Zhang, Sihao Wu, Peipei Xu, Dengyu Wu, Andr{\'e} Freitas, and Mustafa~A. Mustafa. 2023{\natexlab{b}}.
\newblock A survey of safety and trustworthiness of large language models through the lens of verification and validation.
\newblock \emph{ArXiv}, abs/2305.11391.

\bibitem[{Jain et~al.(2023)Jain, Schwarzschild, Wen, Somepalli, Kirchenbauer, yeh Chiang, Goldblum, Saha, Geiping, and Goldstein}]{Jain2023BaselineDF}
Neel Jain, Avi Schwarzschild, Yuxin Wen, Gowthami Somepalli, John Kirchenbauer, Ping yeh Chiang, Micah Goldblum, Aniruddha Saha, Jonas Geiping, and Tom Goldstein. 2023.
\newblock \href {https://api.semanticscholar.org/CorpusID:261494182} {Baseline defenses for adversarial attacks against aligned language models}.
\newblock \emph{ArXiv}, abs/2309.00614.

\bibitem[{Jesse et~al.(2023)Jesse, Ahmed, Devanbu, and Morgan}]{Jesse2023LargeLM}
Kevin Jesse, Toufique Ahmed, Prem Devanbu, and Emily Morgan. 2023.
\newblock Large language models and simple, stupid bugs.
\newblock \emph{ArXiv}, abs/2303.11455.

\bibitem[{Jiao et~al.(2023)Jiao, Teng, Joty, Ding, Sun, Liu, and Chen}]{Jiao2023LogicLLMES}
Fangkai Jiao, Zhiyang Teng, Shafiq~R. Joty, Bosheng Ding, Aixin Sun, Zhengyuan Liu, and Nancy~F. Chen. 2023.
\newblock \href {https://api.semanticscholar.org/CorpusID:258841216} {Logicllm: Exploring self-supervised logic-enhanced training for large language models}.
\newblock \emph{ArXiv}, abs/2305.13718.

\bibitem[{Kadavath et~al.(2022)Kadavath, Conerly, Askell, Henighan, Drain, Perez, Schiefer, Hatfield-Dodds, DasSarma, Tran-Johnson, Johnston, El-Showk, Jones, Elhage, Hume, Chen, Bai, Bowman, Fort, Ganguli, Hernandez, Jacobson, Kernion, Kravec, Lovitt, Ndousse, Olsson, Ringer, Amodei, Brown, Clark, Joseph, Mann, McCandlish, Olah, and Kaplan}]{kadavath2022language}
Saurav Kadavath, Tom Conerly, Amanda Askell, Tom Henighan, Dawn Drain, Ethan Perez, Nicholas Schiefer, Zac Hatfield-Dodds, Nova DasSarma, Eli Tran-Johnson, Scott Johnston, Sheer El-Showk, Andy Jones, Nelson Elhage, Tristan Hume, Anna Chen, Yuntao Bai, Sam Bowman, Stanislav Fort, Deep Ganguli, Danny Hernandez, Josh Jacobson, Jackson Kernion, Shauna Kravec, Liane Lovitt, Kamal Ndousse, Catherine Olsson, Sam Ringer, Dario Amodei, Tom Brown, Jack Clark, Nicholas Joseph, Ben Mann, Sam McCandlish, Chris Olah, and Jared Kaplan. 2022.
\newblock \href {http://arxiv.org/abs/2207.05221} {Language models (mostly) know what they know}.

\bibitem[{Kang et~al.(2023)Kang, Li, Stoica, Guestrin, Zaharia, and Hashimoto}]{Kang2023ExploitingPB}
Daniel Kang, Xuechen Li, Ion Stoica, Carlos Guestrin, Matei~A. Zaharia, and Tatsunori Hashimoto. 2023.
\newblock Exploiting programmatic behavior of llms: Dual-use through standard security attacks.
\newblock \emph{ArXiv}, abs/2302.05733.

\bibitem[{Kassem et~al.(2023)Kassem, Mahmoud, and Saad}]{kassem2023preserving}
Aly Kassem, Omar Mahmoud, and Sherif Saad. 2023.
\newblock Preserving privacy through dememorization: An unlearning technique for mitigating memorization risks in language models.
\newblock In \emph{Proceedings of the 2023 Conference on Empirical Methods in Natural Language Processing}, pages 4360--4379.

\bibitem[{Khalil and Er(2023)}]{khalil2023chatgpt}
Mohammad Khalil and Erkan Er. 2023.
\newblock \href {http://arxiv.org/abs/2302.04335} {Will chatgpt get you caught? rethinking of plagiarism detection}.

\bibitem[{Khatun and Brown(2023)}]{Khatun2023ReliabilityCA}
Aisha Khatun and Daniel Brown. 2023.
\newblock Reliability check: An analysis of gpt-3's response to sensitive topics and prompt wording.
\newblock \emph{ArXiv}, abs/2306.06199.

\bibitem[{Kheiri and Karimi(2023)}]{Kheiri2023SentimentGPTEG}
Kiana Kheiri and Hamid Karimi. 2023.
\newblock \href {https://api.semanticscholar.org/CorpusID:259991148} {Sentimentgpt: Exploiting gpt for advanced sentiment analysis and its departure from current machine learning}.
\newblock \emph{ArXiv}, abs/2307.10234.

\bibitem[{Khoury et~al.(2023)Khoury, Avila, Brunelle, and Camara}]{Khoury2023HowSI}
Rapha{\"e}l Khoury, Anderson~R. Avila, Jacob Brunelle, and Baba~Mamadou Camara. 2023.
\newblock How secure is code generated by chatgpt?
\newblock \emph{ArXiv}, abs/2304.09655.

\bibitem[{Kim et~al.(2023)Kim, Yun, Lee, Gubri, Yoon, and Oh}]{kim2023propile}
Siwon Kim, Sangdoo Yun, Hwaran Lee, Martin Gubri, Sungroh Yoon, and Seong~Joon Oh. 2023.
\newblock Propile: Probing privacy leakage in large language models.
\newblock \emph{arXiv preprint arXiv:2307.01881}.

\bibitem[{Kirchenbauer et~al.(2023{\natexlab{a}})Kirchenbauer, Geiping, Wen, Katz, Miers, and Goldstein}]{kirchenbauer2023watermark}
John Kirchenbauer, Jonas Geiping, Yuxin Wen, Jonathan Katz, Ian Miers, and Tom Goldstein. 2023{\natexlab{a}}.
\newblock \href {http://arxiv.org/abs/2301.10226} {A watermark for large language models}.

\bibitem[{Kirchenbauer et~al.(2023{\natexlab{b}})Kirchenbauer, Geiping, Wen, Shu, Saifullah, Kong, Fernando, Saha, Goldblum, and Goldstein}]{Kirchenbauer2023OnTR}
John Kirchenbauer, Jonas Geiping, Yuxin Wen, Manli Shu, Khalid Saifullah, Kezhi Kong, Kasun Fernando, Aniruddha Saha, Micah Goldblum, and Tom Goldstein. 2023{\natexlab{b}}.
\newblock On the reliability of watermarks for large language models.
\newblock \emph{ArXiv}, abs/2306.04634.

\bibitem[{Kirk et~al.(2023)Kirk, Wray, and Lindes}]{Kirk2023ImprovingKE}
James~R. Kirk, Robert~E. Wray, and Peter Lindes. 2023.
\newblock \href {https://api.semanticscholar.org/CorpusID:261076556} {Improving knowledge extraction from llms for task learning through agent analysis}.

\bibitem[{Krishna et~al.(2023)Krishna, Song, Karpinska, Wieting, and Iyyer}]{Krishna2023ParaphrasingED}
Kalpesh Krishna, Yixiao Song, Marzena Karpinska, John Wieting, and Mohit Iyyer. 2023.
\newblock Paraphrasing evades detectors of ai-generated text, but retrieval is an effective defense.
\newblock \emph{ArXiv}, abs/2303.13408.

\bibitem[{Kuchnik et~al.(2023)Kuchnik, Smith, and Amvrosiadis}]{kuchnik2023validating}
Michael Kuchnik, Virginia Smith, and George Amvrosiadis. 2023.
\newblock Validating large language models with relm.
\newblock \emph{Proceedings of Machine Learning and Systems}, 5.

\bibitem[{Kumar et~al.(2024)Kumar, Lau, Vijayakumar, Trinh, Team, Chang, Robinson, Hendryx, Zhou, Fredrikson, Yue, and Wang}]{kumar2024refusaltrainedllmseasilyjailbroke}
Priyanshu Kumar, Elaine Lau, Saranya Vijayakumar, Tu~Trinh, Scale~Red Team, Elaine Chang, Vaughn Robinson, Sean Hendryx, Shuyan Zhou, Matt Fredrikson, Summer Yue, and Zifan Wang. 2024.
\newblock \href {http://arxiv.org/abs/2410.13886} {Refusal-trained llms are easily jailbroken as browser agents}.

\bibitem[{Laban et~al.(2023)Laban, Kryscinski, Agarwal, Fabbri, Xiong, Joty, and Wu}]{Laban2023LLMsAF}
Philippe Laban, Wojciech Kryscinski, Divyansh Agarwal, Alexander~R. Fabbri, Caiming Xiong, Shafiq~R. Joty, and Chien-Sheng Wu. 2023.
\newblock Llms as factual reasoners: Insights from existing benchmarks and beyond.
\newblock \emph{ArXiv}, abs/2305.14540.

\bibitem[{Lee et~al.(2023{\natexlab{a}})Lee, Lee, Ha, Kim, Lee, Lee, and Song}]{lee2023query}
Deokjae Lee, JunYeong Lee, Jung-Woo Ha, Jin-Hwa Kim, Sang-Woo Lee, Hwaran Lee, and Hyun~Oh Song. 2023{\natexlab{a}}.
\newblock Query-efficient black-box red teaming via bayesian optimization.
\newblock \emph{arXiv preprint arXiv:2305.17444}.

\bibitem[{Lee et~al.(2023{\natexlab{b}})Lee, Hong, Ahn, Hong, Lee, Yun, Shin, and Kim}]{Lee2023WhoWT}
Taehyun Lee, Seokhee Hong, Jaewoo Ahn, Ilgee Hong, Hwaran Lee, Sangdoo Yun, Jamin Shin, and Gunhee Kim. 2023{\natexlab{b}}.
\newblock Who wrote this code? watermarking for code generation.
\newblock \emph{ArXiv}, abs/2305.15060.

\bibitem[{Lewkowycz et~al.(2022)Lewkowycz, Andreassen, Dohan, Dyer, Michalewski, Ramasesh, Slone, Anil, Schlag, Gutman-Solo, Wu, Neyshabur, Gur-Ari, and Misra}]{lewkowycz2022solving}
Aitor Lewkowycz, Anders Andreassen, David Dohan, Ethan Dyer, Henryk Michalewski, Vinay Ramasesh, Ambrose Slone, Cem Anil, Imanol Schlag, Theo Gutman-Solo, Yuhuai Wu, Behnam Neyshabur, Guy Gur-Ari, and Vedant Misra. 2022.
\newblock \href {http://arxiv.org/abs/2206.14858} {Solving quantitative reasoning problems with language models}.

\bibitem[{Li et~al.(2023{\natexlab{a}})Li, Guo, Fan, Xu, Huang, and Song}]{Li2023MultistepJP}
Haoran Li, Dadi Guo, Wei Fan, Mingshi Xu, Jie Huang, and Yangqiu Song. 2023{\natexlab{a}}.
\newblock Multi-step jailbreaking privacy attacks on chatgpt.
\newblock \emph{ArXiv}, abs/2304.05197.

\bibitem[{Li et~al.(2023{\natexlab{b}})Li, Li, Cui, Bi, Wang, Yang, Shi, and Zhang}]{Li2023DeepfakeTD}
Yafu Li, Qintong Li, Leyang Cui, Wei Bi, Longyue Wang, Linyi Yang, Shuming Shi, and Yue Zhang. 2023{\natexlab{b}}.
\newblock Deepfake text detection in the wild.
\newblock \emph{ArXiv}, abs/2305.13242.

\bibitem[{Li et~al.(2021)Li, Lyu, Koren, Lyu, Li, and Ma}]{li2021neural}
Yige Li, Xixiang Lyu, Nodens Koren, Lingjuan Lyu, Bo~Li, and Xingjun Ma. 2021.
\newblock Neural attention distillation: Erasing backdoor triggers from deep neural networks.
\newblock \emph{arXiv preprint arXiv:2101.05930}.

\bibitem[{Liang et~al.(2023)Liang, Yuksekgonul, Mao, Wu, and Zou}]{LIANG2023100779}
Weixin Liang, Mert Yuksekgonul, Yining Mao, Eric Wu, and James Zou. 2023.
\newblock \href {https://doi.org/https://doi.org/10.1016/j.patter.2023.100779} {Gpt detectors are biased against non-native english writers}.
\newblock \emph{Patterns}, 4(7):100779.

\bibitem[{Liao et~al.(2024)Liao, Mo, Xu, Kang, Zhang, Xiao, Tian, Li, and Sun}]{Liao2024EIAEIA}
Zeyi Liao, Lingbo Mo, Chejian Xu, Mintong Kang, Jiawei Zhang, Chaowei Xiao, Yuan Tian, Bo~Li, and Huan Sun. 2024.
\newblock \href {https://arxiv.org/pdf/2409.11295.pdf} {Eia: Environmental injection attack on generalist web agents for privacy leakage}.
\newblock \emph{ArXiv}, abs/2409.11295.

\bibitem[{Lin et~al.(2021)Lin, Wang, Liu, and Qiu}]{Lin2021ASO}
Tianyang Lin, Yuxin Wang, Xiangyang Liu, and Xipeng Qiu. 2021.
\newblock \href {https://api.semanticscholar.org/CorpusID:235368340} {A survey of transformers}.
\newblock \emph{AI Open}, 3:111--132.

\bibitem[{Liu et~al.(2025)Liu, Zhao, Liu, Guo, Xiao, Lin, Chai, Han, Ren, Wang, Liang, Wang, Wu, Li, Wang, Xiong, Liu, and Li}]{liu2025llmpoweredguiagentsphone}
Guangyi Liu, Pengxiang Zhao, Liang Liu, Yaxuan Guo, Han Xiao, Weifeng Lin, Yuxiang Chai, Yue Han, Shuai Ren, Hao Wang, Xiaoyu Liang, Wenhao Wang, Tianze Wu, Linghao Li, Hao Wang, Guanjing Xiong, Yong Liu, and Hongsheng Li. 2025.
\newblock \href {http://arxiv.org/abs/2504.19838} {Llm-powered gui agents in phone automation: Surveying progress and prospects}.

\bibitem[{Liu et~al.(2018)Liu, Dolan-Gavitt, and Garg}]{liu2018fine}
Kang Liu, Brendan Dolan-Gavitt, and Siddharth Garg. 2018.
\newblock Fine-pruning: Defending against backdooring attacks on deep neural networks.
\newblock In \emph{International symposium on research in attacks, intrusions, and defenses}, pages 273--294. Springer.

\bibitem[{Liu et~al.(2021)Liu, Ji, Fu, Du, Yang, and Tang}]{Liu2021PTuningVP}
Xiao Liu, Kaixuan Ji, Yicheng Fu, Zhengxiao Du, Zhilin Yang, and Jie Tang. 2021.
\newblock \href {https://api.semanticscholar.org/CorpusID:238857040} {P-tuning v2: Prompt tuning can be comparable to fine-tuning universally across scales and tasks}.
\newblock \emph{ArXiv}, abs/2110.07602.

\bibitem[{Liu et~al.(2023{\natexlab{a}})Liu, Deng, Li, Wang, Zhang, Liu, Wang, Zheng, and Liu}]{Liu2023PromptIA}
Yi~Liu, Gelei Deng, Yuekang Li, Kailong Wang, Tianwei Zhang, Yepang Liu, Haoyu Wang, Yanhong Zheng, and Yang Liu. 2023{\natexlab{a}}.
\newblock Prompt injection attack against llm-integrated applications.
\newblock \emph{ArXiv}, abs/2306.05499.

\bibitem[{Liu et~al.(2023{\natexlab{b}})Liu, Yao, Li, and Luo}]{Liu2023CheckMI}
Zeyan Liu, Zijun Yao, Fengjun Li, and Bo~Luo. 2023{\natexlab{b}}.
\newblock Check me if you can: Detecting chatgpt-generated academic writing using checkgpt.
\newblock \emph{ArXiv}, abs/2306.05524.

\bibitem[{Lu et~al.(2023)Lu, Liu, He, and Tang}]{Lu2023LargeLM}
Ning Lu, Shengcai Liu, Ruidan He, and Ke~Tang. 2023.
\newblock Large language models can be guided to evade ai-generated text detection.
\newblock \emph{ArXiv}, abs/2305.10847.

\bibitem[{Luo et~al.(2025)Luo, Wang, He, and Xia}]{luo2025guir1generalistr1style}
Run Luo, Lu~Wang, Wanwei He, and Xiaobo Xia. 2025.
\newblock \href {http://arxiv.org/abs/2504.10458} {Gui-r1 : A generalist r1-style vision-language action model for gui agents}.

\bibitem[{Madaan et~al.(2023)Madaan, Tandon, Gupta, Hallinan, Gao, Wiegreffe, Alon, Dziri, Prabhumoye, Yang, Gupta, Majumder, Hermann, Welleck, Yazdanbakhsh, and Clark}]{madaan2023selfrefine}
Aman Madaan, Niket Tandon, Prakhar Gupta, Skyler Hallinan, Luyu Gao, Sarah Wiegreffe, Uri Alon, Nouha Dziri, Shrimai Prabhumoye, Yiming Yang, Shashank Gupta, Bodhisattwa~Prasad Majumder, Katherine Hermann, Sean Welleck, Amir Yazdanbakhsh, and Peter Clark. 2023.
\newblock \href {http://arxiv.org/abs/2303.17651} {Self-refine: Iterative refinement with self-feedback}.

\bibitem[{Mei et~al.(2023)Mei, Li, Wang, Zhang, and Ma}]{Mei2023NOTABLETB}
Kai Mei, Zheng Li, Zhenting Wang, Yang Zhang, and Shiqing Ma. 2023.
\newblock Notable: Transferable backdoor attacks against prompt-based nlp models.
\newblock \emph{ArXiv}, abs/2305.17826.

\bibitem[{Meng et~al.(2022{\natexlab{a}})Meng, Bau, Andonian, and Belinkov}]{meng2022locating}
Kevin Meng, David Bau, Alex Andonian, and Yonatan Belinkov. 2022{\natexlab{a}}.
\newblock Locating and editing factual associations in gpt.
\newblock \emph{Advances in Neural Information Processing Systems}, 35:17359--17372.

\bibitem[{Meng et~al.(2022{\natexlab{b}})Meng, Sharma, Andonian, Belinkov, and Bau}]{meng2022mass}
Kevin Meng, Arnab~Sen Sharma, Alex Andonian, Yonatan Belinkov, and David Bau. 2022{\natexlab{b}}.
\newblock Mass-editing memory in a transformer.
\newblock \emph{arXiv preprint arXiv:2210.07229}.

\bibitem[{Mitchell et~al.(2023)Mitchell, Lee, Khazatsky, Manning, and Finn}]{Mitchell2023DetectGPTZM}
Eric Mitchell, Yoonho Lee, Alexander Khazatsky, Christopher~D. Manning, and Chelsea Finn. 2023.
\newblock Detectgpt: Zero-shot machine-generated text detection using probability curvature.
\newblock \emph{ArXiv}, abs/2301.11305.

\bibitem[{Mitchell et~al.(2021)Mitchell, Lin, Bosselut, Finn, and Manning}]{mitchell2021fast}
Eric Mitchell, Charles Lin, Antoine Bosselut, Chelsea Finn, and Christopher~D Manning. 2021.
\newblock Fast model editing at scale.
\newblock \emph{arXiv preprint arXiv:2110.11309}.

\bibitem[{Mitchell et~al.(2022)Mitchell, Lin, Bosselut, Manning, and Finn}]{mitchell2022memory}
Eric Mitchell, Charles Lin, Antoine Bosselut, Christopher~D Manning, and Chelsea Finn. 2022.
\newblock Memory-based model editing at scale.
\newblock In \emph{International Conference on Machine Learning}, pages 15817--15831. PMLR.

\bibitem[{Mozes et~al.(2023)Mozes, He, Kleinberg, and Griffin}]{Mozes2023UseOL}
Maximilian Mozes, Xuanli He, Bennett Kleinberg, and Lewis~D. Griffin. 2023.
\newblock \href {https://api.semanticscholar.org/CorpusID:261101245} {Use of llms for illicit purposes: Threats, prevention measures, and vulnerabilities}.
\newblock \emph{ArXiv}, abs/2308.12833.

\bibitem[{Muneeswaran et~al.(2023)Muneeswaran, Saxena, Prasad, Prakash, Shankar, Varun, Vaddina, and Gopalakrishnan}]{Muneeswaran2023MinimizingFI}
I~Muneeswaran, Shreya Saxena, Siva Prasad, M~V~Sai Prakash, Advaith Shankar, V~Varun, Vishal Vaddina, and Saisubramaniam Gopalakrishnan. 2023.
\newblock \href {https://api.semanticscholar.org/CorpusID:265445060} {Minimizing factual inconsistency and hallucination in large language models}.
\newblock \emph{ArXiv}, abs/2311.13878.

\bibitem[{Nasr et~al.(2023)Nasr, Carlini, Hayase, Jagielski, Cooper, Ippolito, Choquette-Choo, Wallace, Tram{\`e}r, and Lee}]{nasr2023scalable}
Milad Nasr, Nicholas Carlini, Jonathan Hayase, Matthew Jagielski, A~Feder Cooper, Daphne Ippolito, Christopher~A Choquette-Choo, Eric Wallace, Florian Tram{\`e}r, and Katherine Lee. 2023.
\newblock Scalable extraction of training data from (production) language models.
\newblock \emph{arXiv preprint arXiv:2311.17035}.

\bibitem[{Nye et~al.(2022)Nye, Andreassen, Gur-Ari, Michalewski, Austin, Bieber, Dohan, Lewkowycz, Bosma, Luan, Sutton, and Odena}]{nye2022show}
Maxwell Nye, Anders~Johan Andreassen, Guy Gur-Ari, Henryk Michalewski, Jacob Austin, David Bieber, David Dohan, Aitor Lewkowycz, Maarten Bosma, David Luan, Charles Sutton, and Augustus Odena. 2022.
\newblock \href {https://openreview.net/forum?id=iedYJm92o0a} {Show your work: Scratchpads for intermediate computation with language models}.

\bibitem[{Oppenlaender and Hamalainen(2023)}]{Oppenlaender2023MappingTC}
Jonas Oppenlaender and Joonas Hamalainen. 2023.
\newblock \href {https://api.semanticscholar.org/CorpusID:263830470} {Mapping the challenges of hci: An application and evaluation of chatgpt and gpt-4 for mining insights at scale}.

\bibitem[{Ouyang et~al.(2022)Ouyang, Wu, Jiang, Almeida, Wainwright, Mishkin, Zhang, Agarwal, Slama, Ray, Schulman, Hilton, Kelton, Miller, Simens, Askell, Welinder, Christiano, Leike, and Lowe}]{Ouyang2022TrainingLM}
Long Ouyang, Jeff Wu, Xu~Jiang, Diogo Almeida, Carroll~L. Wainwright, Pamela Mishkin, Chong Zhang, Sandhini Agarwal, Katarina Slama, Alex Ray, John Schulman, Jacob Hilton, Fraser Kelton, Luke~E. Miller, Maddie Simens, Amanda Askell, Peter Welinder, Paul~Francis Christiano, Jan Leike, and Ryan~J. Lowe. 2022.
\newblock \href {https://api.semanticscholar.org/CorpusID:246426909} {Training language models to follow instructions with human feedback}.
\newblock \emph{ArXiv}, abs/2203.02155.

\bibitem[{Ozdayi et~al.(2023)Ozdayi, Peris, FitzGerald, Dupuy, Majmudar, Khan, Parikh, and Gupta}]{ozdayi2023controlling}
Mustafa~Safa Ozdayi, Charith Peris, Jack FitzGerald, Christophe Dupuy, Jimit Majmudar, Haidar Khan, Rahil Parikh, and Rahul Gupta. 2023.
\newblock Controlling the extraction of memorized data from large language models via prompt-tuning.
\newblock \emph{arXiv preprint arXiv:2305.11759}.

\bibitem[{Pan et~al.(2023{\natexlab{a}})Pan, Pan, Chen, Nakov, Kan, and Wang}]{MisinformationPan}
Yikang Pan, Liangming Pan, Wenhu Chen, Preslav Nakov, Min-Yen Kan, and William Wang. 2023{\natexlab{a}}.
\newblock On the risk of misinformation pollution with large language models.

\bibitem[{Pan et~al.(2023{\natexlab{b}})Pan, Pan, Chen, Nakov, Kan, and Wang}]{pan-etal-2023-risk}
Yikang Pan, Liangming Pan, Wenhu Chen, Preslav Nakov, Min-Yen Kan, and William Wang. 2023{\natexlab{b}}.
\newblock \href {https://aclanthology.org/2023.findings-emnlp.97} {On the risk of misinformation pollution with large language models}.
\newblock In \emph{Findings of the Association for Computational Linguistics: EMNLP 2023}, pages 1389--1403, Singapore. Association for Computational Linguistics.

\bibitem[{Pearce et~al.(2021)Pearce, Ahmad, Tan, Dolan-Gavitt, and Karri}]{pearce2021asleep}
Hammond Pearce, Baleegh Ahmad, Benjamin Tan, Brendan Dolan-Gavitt, and Ramesh Karri. 2021.
\newblock \href {http://arxiv.org/abs/2108.09293} {Asleep at the keyboard? assessing the security of github copilot's code contributions}.

\bibitem[{Pearce et~al.(2022)Pearce, Tan, Ahmad, Karri, and Dolan-Gavitt}]{pearce2022examining}
Hammond Pearce, Benjamin Tan, Baleegh Ahmad, Ramesh Karri, and Brendan Dolan-Gavitt. 2022.
\newblock \href {http://arxiv.org/abs/2112.02125} {Examining zero-shot vulnerability repair with large language models}.

\bibitem[{Pegoraro et~al.(2023)Pegoraro, Kumari, Fereidooni, and Sadeghi}]{Pegoraro2023ToCO}
Alessandro Pegoraro, Kavita Kumari, Hossein Fereidooni, and Ahmad-Reza Sadeghi. 2023.
\newblock To chatgpt, or not to chatgpt: That is the question!
\newblock \emph{ArXiv}, abs/2304.01487.

\bibitem[{Peng et~al.(2023)Peng, Yi, Wu, Wu, Zhu, Lyu, Jiao, Xu, Sun, and Xie}]{peng2023are}
Wenjun Peng, Jingwei Yi, Fangzhao Wu, Shangxi Wu, Bin~Benjamin Zhu, Lingjuan Lyu, Binxing Jiao, Tong Xu, Guangzhong Sun, and Xing Xie. 2023.
\newblock \href {https://www.microsoft.com/en-us/research/publication/are-you-copying-my-model-protecting-the-copyright-of-large-language-models-for-eaas-via-backdoor-watermark/} {Are you copying my model? protecting the copyright of large language models for eaas via backdoor watermark}.
\newblock In \emph{ACL 2023}.

\bibitem[{Perez et~al.(2022{\natexlab{a}})Perez, Huang, Song, Cai, Ring, Aslanides, Glaese, McAleese, and Irving}]{perez2022red}
Ethan Perez, Saffron Huang, Francis Song, Trevor Cai, Roman Ring, John Aslanides, Amelia Glaese, Nat McAleese, and Geoffrey Irving. 2022{\natexlab{a}}.
\newblock Red teaming language models with language models.
\newblock \emph{arXiv preprint arXiv:2202.03286}.

\bibitem[{Perez et~al.(2022{\natexlab{b}})Perez, Huang, Song, Cai, Ring, Aslanides, Glaese, McAleese, and Irving}]{Perez2022RedTL}
Ethan Perez, Saffron Huang, Francis Song, Trevor Cai, Roman Ring, John Aslanides, Amelia Glaese, Nathan McAleese, and Geoffrey Irving. 2022{\natexlab{b}}.
\newblock \href {https://api.semanticscholar.org/CorpusID:246634238} {Red teaming language models with language models}.
\newblock In \emph{Conference on Empirical Methods in Natural Language Processing}.

\bibitem[{Perez and Ribeiro(2022)}]{Perez2022IgnorePP}
F{\'a}bio Perez and Ian Ribeiro. 2022.
\newblock Ignore previous prompt: Attack techniques for language models.
\newblock \emph{ArXiv}, abs/2211.09527.

\bibitem[{Prabhumoye et~al.(2023)Prabhumoye, Patwary, Shoeybi, and Catanzaro}]{prabhumoye2023adding}
Shrimai Prabhumoye, Mostofa Patwary, Mohammad Shoeybi, and Bryan Catanzaro. 2023.
\newblock Adding instructions during pretraining: Effective way of controlling toxicity in language models.
\newblock \emph{arXiv preprint arXiv:2302.07388}.

\bibitem[{Qi et~al.(2020)Qi, Chen, Li, Yao, Liu, and Sun}]{qi2020onion}
Fanchao Qi, Yangyi Chen, Mukai Li, Yuan Yao, Zhiyuan Liu, and Maosong Sun. 2020.
\newblock Onion: A simple and effective defense against textual backdoor attacks.
\newblock \emph{arXiv preprint arXiv:2011.10369}.

\bibitem[{Qi et~al.(2021)Qi, Li, Chen, Zhang, Liu, Wang, and Sun}]{qi2021hidden}
Fanchao Qi, Mukai Li, Yangyi Chen, Zhengyan Zhang, Zhiyuan Liu, Yasheng Wang, and Maosong Sun. 2021.
\newblock Hidden killer: Invisible textual backdoor attacks with syntactic trigger.
\newblock \emph{arXiv preprint arXiv:2105.12400}.

\bibitem[{Quidwai et~al.(2023)Quidwai, Li, and Dube}]{Quidwai2023BeyondBB}
Mujahid~Ali Quidwai, Chun~Xing Li, and Parijat Dube. 2023.
\newblock Beyond black box ai-generated plagiarism detection: From sentence to document level.
\newblock \emph{ArXiv}, abs/2306.08122.

\bibitem[{Rajani et~al.(2019)Rajani, McCann, Xiong, and Socher}]{rajani-etal-2019-explain}
Nazneen~Fatema Rajani, Bryan McCann, Caiming Xiong, and Richard Socher. 2019.
\newblock \href {https://doi.org/10.18653/v1/P19-1487} {Explain yourself! leveraging language models for commonsense reasoning}.
\newblock In \emph{Proceedings of the 57th Annual Meeting of the Association for Computational Linguistics}, pages 4932--4942, Florence, Italy. Association for Computational Linguistics.

\bibitem[{Rao et~al.(2023)Rao, Vashistha, Naik, Aditya, and Choudhury}]{Rao2023TrickingLI}
Abhinav Rao, Sachin Vashistha, Atharva Naik, Somak Aditya, and Monojit Choudhury. 2023.
\newblock Tricking llms into disobedience: Understanding, analyzing, and preventing jailbreaks.
\newblock \emph{ArXiv}, abs/2305.14965.

\bibitem[{Ribeiro et~al.(2016)Ribeiro, Singh, and Guestrin}]{ribeiro2016should}
Marco~Tulio Ribeiro, Sameer Singh, and Carlos Guestrin. 2016.
\newblock " why should i trust you?" explaining the predictions of any classifier.
\newblock In \emph{Proceedings of the 22nd ACM SIGKDD international conference on knowledge discovery and data mining}, pages 1135--1144.

\bibitem[{Roy et~al.(2023)Roy, Naragam, and Nilizadeh}]{Roy2023GeneratingPA}
Sayak~Saha Roy, Krishna~Vamsi Naragam, and Shirin Nilizadeh. 2023.
\newblock Generating phishing attacks using chatgpt.
\newblock \emph{ArXiv}, abs/2305.05133.

\bibitem[{Russinovich et~al.(2024)Russinovich, Salem, and Eldan}]{russinovich2024great}
Mark Russinovich, Ahmed Salem, and Ronen Eldan. 2024.
\newblock Great, now write an article about that: The crescendo multi-turn llm jailbreak attack.
\newblock \emph{arXiv preprint arXiv:2404.01833}.

\bibitem[{Sadasivan et~al.(2023)Sadasivan, Kumar, Balasubramanian, Wang, and Feizi}]{Sadasivan2023CanAT}
Vinu~Sankar Sadasivan, Aounon Kumar, S.~Balasubramanian, Wenxiao Wang, and Soheil Feizi. 2023.
\newblock Can ai-generated text be reliably detected?
\newblock \emph{ArXiv}, abs/2303.11156.

\bibitem[{Sadik et~al.(2023)Sadik, Ceravola, Joublin, and Patra}]{Sadik2023AnalysisOC}
Ahmed~R. Sadik, Antonello Ceravola, Frank Joublin, and Jibesh Patra. 2023.
\newblock Analysis of chatgpt on source code.
\newblock \emph{ArXiv}, abs/2306.00597.

\bibitem[{Samvelyan et~al.(2024)Samvelyan, Raparthy, Lupu, Hambro, Markosyan, Bhatt, Mao, Jiang, Parker-Holder, Foerster, Rockt{\"a}schel, and Raileanu}]{Samvelyan2024RainbowTeam}
Mikayel Samvelyan, Sharath~Chandra Raparthy, Andrei Lupu, Eric Hambro, Aram~H. Markosyan, Manish Bhatt, Yuning Mao, Minqi Jiang, Jack Parker-Holder, Jakob Foerster, Tim Rockt{\"a}schel, and Roberta Raileanu. 2024.
\newblock \href {https://arxiv.org/abs/2402.16822} {Rainbow teaming: Open-ended generation of diverse adversarial prompts}.
\newblock \emph{arXiv preprint arXiv:2402.16822}.

\bibitem[{Sandoval et~al.(2022)Sandoval, Pearce, Nys, Karri, Garg, and Dolan-Gavitt}]{Sandoval2022LostAC}
Gustavo Sandoval, Hammond~A. Pearce, Teo Nys, Ramesh Karri, Siddharth Garg, and Brendan Dolan-Gavitt. 2022.
\newblock Lost at c: A user study on the security implications of large language model code assistants.

\bibitem[{Schwarzschild et~al.(2020)Schwarzschild, Goldblum, Gupta, Dickerson, and Goldstein}]{Schwarzschild2020JustHT}
Avi Schwarzschild, Micah Goldblum, Arjun Gupta, John~P. Dickerson, and Tom Goldstein. 2020.
\newblock \href {https://api.semanticscholar.org/CorpusID:219980448} {Just how toxic is data poisoning? a unified benchmark for backdoor and data poisoning attacks}.
\newblock \emph{ArXiv}, abs/2006.12557.

\bibitem[{Senadeera and Ive(2022)}]{Senadeera2022ControlledTG}
Damith~Chamalke Senadeera and Julia Ive. 2022.
\newblock \href {https://api.semanticscholar.org/CorpusID:254274934} {Controlled text generation using t5 based encoder-decoder soft prompt tuning and analysis of the utility of generated text in ai}.
\newblock \emph{ArXiv}, abs/2212.02924.

\bibitem[{Shao et~al.(2023)Shao, Huang, Zheng, and Chang}]{Shao2023QuantifyingAC}
Hanyin Shao, Jie Huang, Shen Zheng, and Kevin Chen-Chuan Chang. 2023.
\newblock \href {https://api.semanticscholar.org/CorpusID:258832523} {Quantifying association capabilities of large language models and its implications on privacy leakage}.
\newblock \emph{ArXiv}, abs/2305.12707.

\bibitem[{Shayegani et~al.(2024)Shayegani, Dong, and Abu-Ghazaleh}]{shayegani2024jailbreakpieces}
Erfan Shayegani, Yue Dong, and Nael Abu-Ghazaleh. 2024.
\newblock \href {https://openreview.net/forum?id=plmBsXHxgR} {Jailbreak in pieces: Compositional adversarial attacks on multi-modal language models}.
\newblock In \emph{The Twelfth International Conference on Learning Representations}.

\bibitem[{Shayegani et~al.(2023)Shayegani, Mamun, Fu, Zaree, Dong, and Abu-Ghazaleh}]{Shayegani2023SurveyOV}
Erfan Shayegani, Md. Abdullah~Al Mamun, Yu~Fu, Pedram Zaree, Yue Dong, and Nael~B. Abu-Ghazaleh. 2023.
\newblock \href {https://api.semanticscholar.org/CorpusID:264172191} {Survey of vulnerabilities in large language models revealed by adversarial attacks}.
\newblock \emph{ArXiv}, abs/2310.10844.

\bibitem[{Shayegani et~al.(2025)Shayegani, Shahariar, Abdali, Yu, Abu-Ghazaleh, and Dong}]{shayegani2025misalignedrolesmisplacedimages}
Erfan Shayegani, G~M Shahariar, Sara Abdali, Lei Yu, Nael Abu-Ghazaleh, and Yue Dong. 2025.
\newblock \href {http://arxiv.org/abs/2504.03735} {Misaligned roles, misplaced images: Structural input perturbations expose multimodal alignment blind spots}.

\bibitem[{Shen et~al.(2023)Shen, Chen, Backes, Shen, and Zhang}]{shen2023anything}
Xinyue Shen, Zeyuan Chen, Michael Backes, Yun Shen, and Yang Zhang. 2023.
\newblock " do anything now": Characterizing and evaluating in-the-wild jailbreak prompts on large language models.
\newblock \emph{arXiv preprint arXiv:2308.03825}.

\bibitem[{Shi et~al.(2023{\natexlab{a}})Shi, Tonolini, Aletras, Yilmaz, Kazai, and Jiao}]{Shi2023RethinkingSL}
Zhengxiang Shi, Francesco Tonolini, Nikolaos Aletras, Emine Yilmaz, Gabriella Kazai, and Yunlong Jiao. 2023{\natexlab{a}}.
\newblock \href {https://api.semanticscholar.org/CorpusID:258832439} {Rethinking semi-supervised learning with language models}.
\newblock \emph{ArXiv}, abs/2305.13002.

\bibitem[{Shi et~al.(2023{\natexlab{b}})Shi, Wang, Yin, Chen, Chang, and Hsieh}]{shi2023red}
Zhouxing Shi, Yihan Wang, Fan Yin, Xiangning Chen, Kai-Wei Chang, and Cho-Jui Hsieh. 2023{\natexlab{b}}.
\newblock Red teaming language model detectors with language models.
\newblock \emph{arXiv preprint arXiv:2305.19713}.

\bibitem[{Shin et~al.(2019)Shin, Allamanis, Brockschmidt, and Polozov}]{ShinProgramSynthesis2019}
Richard Shin, Miltiadis Allamanis, Marc Brockschmidt, and Oleksandr Polozov. 2019.
\newblock \emph{Program Synthesis and Semantic Parsing with Learned Code Idioms}. Curran Associates Inc., Red Hook, NY, USA.

\bibitem[{Shinn et~al.(2023)Shinn, Cassano, Labash, Gopinath, Narasimhan, and Yao}]{shinn2023reflexion}
Noah Shinn, Federico Cassano, Beck Labash, Ashwin Gopinath, Karthik Narasimhan, and Shunyu Yao. 2023.
\newblock \href {http://arxiv.org/abs/2303.11366} {Reflexion: Language agents with verbal reinforcement learning}.

\bibitem[{Si et~al.(2023)Si, Backes, and Zhang}]{si2023mondrian}
Wai~Man Si, Michael Backes, and Yang Zhang. 2023.
\newblock Mondrian: Prompt abstraction attack against large language models for cheaper api pricing.
\newblock \emph{arXiv preprint arXiv:2308.03558}.

\bibitem[{Siddiq et~al.(2022)Siddiq, Majumder, Mim, Jajodia, and Santos}]{Siddiq2022}
Mohammed~Latif Siddiq, Shafayat~H. Majumder, Maisha~R. Mim, Sourov Jajodia, and Joanna C.~S. Santos. 2022.
\newblock \href {https://doi.org/10.1109/SCAM55253.2022.00014} {An empirical study of code smells in transformer-based code generation techniques}.
\newblock In \emph{2022 IEEE 22nd International Working Conference on Source Code Analysis and Manipulation (SCAM)}, pages 71--82.

\bibitem[{Solaiman et~al.(2019)Solaiman, Brundage, Clark, Askell, Herbert-Voss, Wu, Radford, and Wang}]{Solaiman2019ReleaseSA}
Irene Solaiman, Miles Brundage, Jack Clark, Amanda Askell, Ariel Herbert-Voss, Jeff Wu, Alec Radford, and Jasmine Wang. 2019.
\newblock Release strategies and the social impacts of language models.
\newblock \emph{ArXiv}, abs/1908.09203.

\bibitem[{Spiess et~al.(2024)Spiess, Gros, Pai, Pradel, Rabin, Jha, Devanbu, and Ahmed}]{Spiess2024CalibrationAC}
Claudio Spiess, David Gros, Kunal~Suresh Pai, Michael Pradel, Md~Rafiqul~Islam Rabin, Susmit Jha, Prem Devanbu, and Toufique Ahmed. 2024.
\newblock \href {https://api.semanticscholar.org/CorpusID:267412346} {Calibration and correctness of language models for code}.

\bibitem[{Stapleton et~al.(2023)Stapleton, Taylor, Fox, Wu, and Zhu}]{stapleton2023seeing}
Logan Stapleton, Jordan Taylor, Sarah Fox, Tongshuang Wu, and Haiyi Zhu. 2023.
\newblock Seeing seeds beyond weeds: Green teaming generative ai for beneficial uses.
\newblock \emph{arXiv preprint arXiv:2306.03097}.

\bibitem[{Stoehr et~al.(2024)Stoehr, Gordon, Zhang, and Lewis}]{stoehr2024localizing}
Niklas Stoehr, Mitchell Gordon, Chiyuan Zhang, and Owen Lewis. 2024.
\newblock Localizing paragraph memorization in language models.
\newblock \emph{arXiv preprint arXiv:2403.19851}.

\bibitem[{Stokel-Walker(2022)}]{stokel2022ai}
Chris Stokel-Walker. 2022.
\newblock Ai bot chatgpt writes smart essays-should academics worry?
\newblock \emph{Nature}.

\bibitem[{Su et~al.(2023)Su, Zhuo, Wang, and Nakov}]{Su2023DetectLLMLL}
Jinyan Su, Terry~Yue Zhuo, Di~Wang, and Preslav Nakov. 2023.
\newblock Detectllm: Leveraging log rank information for zero-shot detection of machine-generated text.
\newblock \emph{ArXiv}, abs/2306.05540.

\bibitem[{Sun et~al.(2023)Sun, Zhang, Deng, Cheng, and Huang}]{Sun2023SafetyAO}
Hao Sun, Zhexin Zhang, Jiawen Deng, Jiale Cheng, and Minlie Huang. 2023.
\newblock Safety assessment of chinese large language models.
\newblock \emph{ArXiv}, abs/2304.10436.

\bibitem[{Tam et~al.(2022)Tam, Mascarenhas, Zhang, Kwan, Bansal, and Raffel}]{Tam2022EvaluatingTF}
Derek Tam, Anisha Mascarenhas, Shiyue Zhang, Sarah Kwan, Mohit Bansal, and Colin Raffel. 2022.
\newblock Evaluating the factual consistency of large language models through summarization.
\newblock \emph{ArXiv}, abs/2211.08412.

\bibitem[{Tam et~al.(2023)Tam, Mascarenhas, Zhang, Kwan, Bansal, and Raffel}]{tam-etal-2023-evaluating}
Derek Tam, Anisha Mascarenhas, Shiyue Zhang, Sarah Kwan, Mohit Bansal, and Colin Raffel. 2023.
\newblock \href {https://doi.org/10.18653/v1/2023.findings-acl.322} {Evaluating the factual consistency of large language models through news summarization}.
\newblock In \emph{Findings of the Association for Computational Linguistics: ACL 2023}, pages 5220--5255, Toronto, Canada. Association for Computational Linguistics.

\bibitem[{Tang et~al.(2023{\natexlab{a}})Tang, Uberti, and Shlomi}]{Tang2023BaselinesFI}
Leonard Tang, Gavin Uberti, and Tom Shlomi. 2023{\natexlab{a}}.
\newblock \href {https://api.semanticscholar.org/CorpusID:258967971} {Baselines for identifying watermarked large language models}.
\newblock \emph{ArXiv}, abs/2305.18456.

\bibitem[{Tang et~al.(2023{\natexlab{b}})Tang, Chuang, and Hu}]{Tang2023TheSO}
Ruixiang Tang, Yu-Neng Chuang, and Xia Hu. 2023{\natexlab{b}}.
\newblock The science of detecting llm-generated texts.
\newblock \emph{ArXiv}, abs/2303.07205.

\bibitem[{Tang et~al.(2023{\natexlab{c}})Tang, Kong, li~Huang, and Xue}]{Tang2023LargeLM}
Ruixiang Tang, Dehan Kong, Lo~li~Huang, and Hui Xue. 2023{\natexlab{c}}.
\newblock Large language models can be lazy learners: Analyze shortcuts in in-context learning.
\newblock \emph{ArXiv}, abs/2305.17256.

\bibitem[{Taori et~al.(2023)Taori, Gulrajani, Zhang, Dubois, Li, Guestrin, Liang, and Hashimoto}]{taori2023stanford}
Rohan Taori, Ishaan Gulrajani, Tianyi Zhang, Yann Dubois, Xuechen Li, Carlos Guestrin, Percy Liang, and Tatsunori~B Hashimoto. 2023.
\newblock Stanford alpaca: an instruction-following llama model (2023).
\newblock \emph{URL https://github. com/tatsu-lab/stanford\_alpaca}.

\bibitem[{Tian(2023)}]{GPTZero}
Edward Tian. 2023.
\newblock \href {https://gptzero.me/} {[link]}.

\bibitem[{Tirumala et~al.(2022)Tirumala, Markosyan, Zettlemoyer, and Aghajanyan}]{tirumala2022memorization}
Kushal Tirumala, Aram Markosyan, Luke Zettlemoyer, and Armen Aghajanyan. 2022.
\newblock Memorization without overfitting: Analyzing the training dynamics of large language models.
\newblock \emph{Advances in Neural Information Processing Systems}, 35:38274--38290.

\bibitem[{Tol and Sunar(2023)}]{Tol2023ZeroLeakUL}
M.~Caner Tol and Berk Sunar. 2023.
\newblock \href {https://api.semanticscholar.org/CorpusID:261214430} {Zeroleak: Using llms for scalable and cost effective side-channel patching}.
\newblock \emph{ArXiv}, abs/2308.13062.

\bibitem[{Vasilatos et~al.(2023)Vasilatos, Alam, Rahwan, Zaki, and Maniatakos}]{Vasilatos2023HowkGPTIT}
Christoforos Vasilatos, Manaar Alam, Talal Rahwan, Yasir Zaki, and Michail Maniatakos. 2023.
\newblock Howkgpt: Investigating the detection of chatgpt-generated university student homework through context-aware perplexity analysis.
\newblock \emph{ArXiv}, abs/2305.18226.

\bibitem[{Vaswani et~al.(2017)Vaswani, Shazeer, Parmar, Uszkoreit, Jones, Gomez, Kaiser, and Polosukhin}]{Vaswani2017AttentionIA}
Ashish Vaswani, Noam~M. Shazeer, Niki Parmar, Jakob Uszkoreit, Llion Jones, Aidan~N. Gomez, Lukasz Kaiser, and Illia Polosukhin. 2017.
\newblock \href {https://api.semanticscholar.org/CorpusID:13756489} {Attention is all you need}.
\newblock In \emph{Neural Information Processing Systems}.

\bibitem[{Vykopal et~al.(2023)Vykopal, Pikuliak, Srba, M{\'o}ro, Macko, and Bielikov{\'a}}]{Vykopal2023DisinformationCO}
Ivan Vykopal, Mat'uvs Pikuliak, Ivan Srba, R{\'o}bert M{\'o}ro, Dominik Macko, and M{\'a}ria Bielikov{\'a}. 2023.
\newblock \href {https://api.semanticscholar.org/CorpusID:265213085} {Disinformation capabilities of large language models}.
\newblock \emph{ArXiv}, abs/2311.08838.

\bibitem[{Wallace et~al.(2021)Wallace, Zhao, Feng, and Singh}]{wallace-etal-2021-concealed}
Eric Wallace, Tony Zhao, Shi Feng, and Sameer Singh. 2021.
\newblock \href {https://doi.org/10.18653/v1/2021.naacl-main.13} {Concealed data poisoning attacks on {NLP} models}.
\newblock In \emph{Proceedings of the 2021 Conference of the North American Chapter of the Association for Computational Linguistics: Human Language Technologies}, pages 139--150, Online. Association for Computational Linguistics.

\bibitem[{Wan et~al.(2023)Wan, Wallace, Shen, and Klein}]{wan2023poisoning}
Alexander Wan, Eric Wallace, Sheng Shen, and Dan Klein. 2023.
\newblock Poisoning language models during instruction tuning.
\newblock \emph{arXiv preprint arXiv:2305.00944}.

\bibitem[{Wang et~al.(2023{\natexlab{a}})Wang, Chen, Pei, Xie, Kang, Zhang, Xu, Xiong, Dutta, Schaeffer, Truong, Arora, Mazeika, Hendrycks, Lin, Cheng, Koyejo, Song, and Li}]{Wang2023DecodingTrustAC}
Boxin Wang, Weixin Chen, Hengzhi Pei, Chulin Xie, Mintong Kang, Chenhui Zhang, Chejian Xu, Zidi Xiong, Ritik Dutta, Rylan Schaeffer, Sang Truong, Simran Arora, Mantas Mazeika, Dan Hendrycks, Zi-Han Lin, Yu~Cheng, Sanmi Koyejo, Dawn~Xiaodong Song, and Bo~Li. 2023{\natexlab{a}}.
\newblock Decodingtrust: A comprehensive assessment of trustworthiness in gpt models.
\newblock \emph{ArXiv}, abs/2306.11698.

\bibitem[{Wang et~al.(2023{\natexlab{b}})Wang, Luo, Wang, and Yan}]{Wang2023BotOH}
Hong Wang, Xuan Luo, Weizhi Wang, and Xifeng Yan. 2023{\natexlab{b}}.
\newblock Bot or human? detecting chatgpt imposters with a single question.
\newblock \emph{ArXiv}, abs/2305.06424.

\bibitem[{Wang et~al.(2023{\natexlab{c}})Wang, Cao, Luo, Zhou, Xie, Jatowt, and Cai}]{Wang2023EnhancingLL}
Jiexin Wang, Liuwen Cao, Xitong Luo, Zhiping Zhou, Jiayuan Xie, Adam Jatowt, and Yi~Cai. 2023{\natexlab{c}}.
\newblock \href {https://api.semanticscholar.org/CorpusID:264487366} {Enhancing large language models for secure code generation: A dataset-driven study on vulnerability mitigation}.
\newblock \emph{ArXiv}, abs/2310.16263.

\bibitem[{Wang et~al.(2023{\natexlab{d}})Wang, yang Liu, Park, Chen, and Xiao}]{Wang2023AdversarialDA}
Jiong Wang, Zi~yang Liu, Keun~Hee Park, Muhao Chen, and Chaowei Xiao. 2023{\natexlab{d}}.
\newblock Adversarial demonstration attacks on large language models.
\newblock \emph{ArXiv}, abs/2305.14950.

\bibitem[{Wang et~al.(2023{\natexlab{e}})Wang, Zhang, Xie, Yao, Tian, Wang, Xi, Cheng, Liu, Zheng et~al.}]{wang2023easyedit}
Peng Wang, Ningyu Zhang, Xin Xie, Yunzhi Yao, Bozhong Tian, Mengru Wang, Zekun Xi, Siyuan Cheng, Kangwei Liu, Guozhou Zheng, et~al. 2023{\natexlab{e}}.
\newblock Easyedit: An easy-to-use knowledge editing framework for large language models.
\newblock \emph{arXiv preprint arXiv:2308.07269}.

\bibitem[{Wang et~al.(2023{\natexlab{f}})Wang, Li, Ren, Jiang, Zhang, and Qiu}]{Wang2023SeqXGPTSA}
Pengyu Wang, Linyang Li, Ke~Ren, Botian Jiang, Dong Zhang, and Xipeng Qiu. 2023{\natexlab{f}}.
\newblock \href {https://api.semanticscholar.org/CorpusID:264128397} {Seqxgpt: Sentence-level ai-generated text detection}.
\newblock \emph{ArXiv}, abs/2310.08903.

\bibitem[{Wang et~al.(2016)Wang, Liu, and Tan}]{wangDefectPrediction2016}
Song Wang, Taiyue Liu, and Lin Tan. 2016.
\newblock \href {https://doi.org/10.1145/2884781.2884804} {Automatically learning semantic features for defect prediction}.
\newblock In \emph{Proceedings of the 38th International Conference on Software Engineering}, ICSE '16, page 297–308, New York, NY, USA. Association for Computing Machinery.

\bibitem[{Wang et~al.(2023{\natexlab{g}})Wang, Wei, Schuurmans, Le, Chi, Narang, Chowdhery, and Zhou}]{wang2023selfconsistency}
Xuezhi Wang, Jason Wei, Dale Schuurmans, Quoc~V Le, Ed~H. Chi, Sharan Narang, Aakanksha Chowdhery, and Denny Zhou. 2023{\natexlab{g}}.
\newblock \href {https://openreview.net/forum?id=1PL1NIMMrw} {Self-consistency improves chain of thought reasoning in language models}.
\newblock In \emph{The Eleventh International Conference on Learning Representations}.

\bibitem[{Wang et~al.(2023{\natexlab{h}})Wang, Mansurov, Ivanov, Su, Shelmanov, Tsvigun, Whitehouse, Afzal, Mahmoud, Aji, and Nakov}]{Wang2023M4MM}
Yuxia Wang, Jonibek Mansurov, Petar Ivanov, Jinyan Su, Artem Shelmanov, Akim Tsvigun, Chenxi Whitehouse, Osama~Mohammed Afzal, Tarek Mahmoud, Alham~Fikri Aji, and Preslav Nakov. 2023{\natexlab{h}}.
\newblock M4: Multi-generator, multi-domain, and multi-lingual black-box machine-generated text detection.
\newblock \emph{ArXiv}, abs/2305.14902.

\bibitem[{Weber-Wulff et~al.(2023)Weber-Wulff, Anohina-Naumeca, Bjelobaba, Folt{\'y}nek, Guerrero-Dib, Popoola, Sigut, and Waddington}]{WeberWulff2023TestingOD}
Debora Weber-Wulff, Alla Anohina-Naumeca, Sonja Bjelobaba, Tom'{a}\v{s} Folt{\'y}nek, Jean~Gabriel Guerrero-Dib, Olumide Popoola, Petr Sigut, and Lorna Waddington. 2023.
\newblock \href {https://api.semanticscholar.org/CorpusID:259262442} {Testing of detection tools for ai-generated text}.
\newblock \emph{International Journal for Educational Integrity}, 19:1--39.

\bibitem[{Wei et~al.(2022)Wei, Wang, Schuurmans, Bosma, brian ichter, Xia, Chi, Le, and Zhou}]{wei2022chain}
Jason Wei, Xuezhi Wang, Dale Schuurmans, Maarten Bosma, brian ichter, Fei Xia, Ed~H. Chi, Quoc~V Le, and Denny Zhou. 2022.
\newblock \href {https://openreview.net/forum?id=_VjQlMeSB_J} {Chain of thought prompting elicits reasoning in large language models}.
\newblock In \emph{Advances in Neural Information Processing Systems}.

\bibitem[{Weidinger et~al.(2021)Weidinger, Mellor, Rauh, Griffin, Uesato, Huang, Cheng, Glaese, Balle, Kasirzadeh, Kenton, Brown, Hawkins, Stepleton, Biles, Birhane, Haas, Rimell, Hendricks, Isaac, Legassick, Irving, and Gabriel}]{Weidinger2021EthicalAS}
Laura Weidinger, John F.~J. Mellor, Maribeth Rauh, Conor Griffin, Jonathan Uesato, Po-Sen Huang, Myra Cheng, Mia Glaese, Borja Balle, Atoosa Kasirzadeh, Zachary Kenton, Sande~Minnich Brown, William~T. Hawkins, Tom Stepleton, Courtney Biles, Abeba Birhane, Julia Haas, Laura Rimell, Lisa~Anne Hendricks, William~S. Isaac, Sean Legassick, Geoffrey Irving, and Iason Gabriel. 2021.
\newblock Ethical and social risks of harm from language models.
\newblock \emph{ArXiv}, abs/2112.04359.

\bibitem[{Wen et~al.(2023)Wen, Ke, Sun, Zhang, Li, Bai, and Huang}]{Wen2023UnveilingTI}
Jiaxin Wen, Pei Ke, Hao Sun, Zhexin Zhang, Chengfei Li, Jinfeng Bai, and Minlie Huang. 2023.
\newblock \href {https://api.semanticscholar.org/CorpusID:265498356} {Unveiling the implicit toxicity in large language models}.
\newblock In \emph{Conference on Empirical Methods in Natural Language Processing}.

\bibitem[{Wolff(2020)}]{Wolff2020AttackingNT}
Max Wolff. 2020.
\newblock \href {https://api.semanticscholar.org/CorpusID:211532535} {Attacking neural text detectors}.
\newblock \emph{ArXiv}, abs/2002.11768.

\bibitem[{Wu et~al.(2021)Wu, Zhou, Zhu, Liu, Harandi, and Li}]{Wu2021PerformanceEO}
Jing Wu, Mingyi Zhou, Ce~Zhu, Yipeng Liu, Mehrtash Harandi, and Li~Li. 2021.
\newblock \href {https://api.semanticscholar.org/CorpusID:233346886} {Performance evaluation of adversarial attacks: Discrepancies and solutions}.
\newblock \emph{ArXiv}, abs/2104.11103.

\bibitem[{Wu et~al.(2025)Wu, Yu, Yogatama, Lu, and Kim}]{wu2025semantichubhypothesislanguage}
Zhaofeng Wu, Xinyan~Velocity Yu, Dani Yogatama, Jiasen Lu, and Yoon Kim. 2025.
\newblock \href {http://arxiv.org/abs/2411.04986} {The semantic hub hypothesis: Language models share semantic representations across languages and modalities}.

\bibitem[{de~Wynter et~al.(2023)de~Wynter, Wang, Sokolov, Gu, and Chen}]{de2023evaluation}
Adrian de~Wynter, Xun Wang, Alex Sokolov, Qilong Gu, and Si-Qing Chen. 2023.
\newblock An evaluation on large language model outputs: Discourse and memorization.
\newblock \emph{arXiv preprint arXiv:2304.08637}.

\bibitem[{Xu et~al.(2022{\natexlab{a}})Xu, Chen, Cui, Gao, and Liu}]{xu2022exploring}
Lei Xu, Yangyi Chen, Ganqu Cui, Hongcheng Gao, and Zhiyuan Liu. 2022{\natexlab{a}}.
\newblock Exploring the universal vulnerability of prompt-based learning paradigm.
\newblock \emph{arXiv preprint arXiv:2204.05239}.

\bibitem[{Xu et~al.(2022{\natexlab{b}})Xu, Chen, Cui, Gao, and Liu}]{Xu2022ExploringTU}
Lei Xu, Yangyi Chen, Ganqu Cui, Hongcheng Gao, and Zhiyuan Liu. 2022{\natexlab{b}}.
\newblock Exploring the universal vulnerability of prompt-based learning paradigm.
\newblock \emph{ArXiv}, abs/2204.05239.

\bibitem[{Yang et~al.(2023{\natexlab{a}})Yang, Xiang, Li, and Lu}]{Yang2023ACO}
Haomiao Yang, Kunlan Xiang, Hongwei Li, and Rongxing Lu. 2023{\natexlab{a}}.
\newblock \href {https://api.semanticscholar.org/CorpusID:261244059} {A comprehensive overview of backdoor attacks in large language models within communication networks}.
\newblock \emph{ArXiv}, abs/2308.14367.

\bibitem[{Yang et~al.(2024)Yang, Liu, Wu, Yang, Fung, Li, Huang, Cao, Wang, Wang, Ji, and Zhai}]{Yang2024IfLIA}
Ke~Yang, Jiateng Liu, John Wu, Chaoqi Yang, Y.~Fung, Sha Li, Zixuan Huang, Xu~Cao, Xingyao Wang, Yiquan Wang, Heng Ji, and Chengxiang Zhai. 2024.
\newblock \href {https://api.semanticscholar.org/CorpusId:266693465} {If llm is the wizard, then code is the wand: A survey on how code empowers large language models to serve as intelligent agents}.
\newblock \emph{ArXiv}, abs/2401.00812.

\bibitem[{Yang et~al.(2021)Yang, Li, Zhang, Ren, Sun, and He}]{Yang2021BeCA}
Wenkai Yang, Lei Li, Zhiyuan Zhang, Xuancheng Ren, Xu~Sun, and Bin He. 2021.
\newblock \href {https://api.semanticscholar.org/CorpusID:232404131} {Be careful about poisoned word embeddings: Exploring the vulnerability of the embedding layers in nlp models}.
\newblock \emph{ArXiv}, abs/2103.15543.

\bibitem[{Yang et~al.(2023{\natexlab{b}})Yang, Chen, Zhang, rui Liu, Qi, Zhang, Fang, and Yu}]{Yang2023WatermarkingTG}
Xi~Yang, Kejiang Chen, Weiming Zhang, Chang rui Liu, Yuang Qi, Jie Zhang, Han Fang, and Neng~H. Yu. 2023{\natexlab{b}}.
\newblock Watermarking text generated by black-box language models.
\newblock \emph{ArXiv}, abs/2305.08883.

\bibitem[{Yang et~al.(2023{\natexlab{c}})Yang, Cheng, Petzold, Wang, and Chen}]{Yang2023DNAGPTDN}
Xianjun Yang, Wei Cheng, Linda Petzold, William~Yang Wang, and Haifeng Chen. 2023{\natexlab{c}}.
\newblock Dna-gpt: Divergent n-gram analysis for training-free detection of gpt-generated text.
\newblock \emph{ArXiv}, abs/2305.17359.

\bibitem[{Yao et~al.(2023{\natexlab{a}})Yao, Lou, and Qin}]{Yao2023PoisonPromptBA}
Hongwei Yao, Jian Lou, and Zhan Qin. 2023{\natexlab{a}}.
\newblock \href {https://api.semanticscholar.org/CorpusID:264306255} {Poisonprompt: Backdoor attack on prompt-based large language models}.
\newblock \emph{ArXiv}, abs/2310.12439.

\bibitem[{Yao et~al.(2023{\natexlab{b}})Yao, Wang, Tian, Cheng, Li, Deng, Chen, and Zhang}]{yao2023editing}
Yunzhi Yao, Peng Wang, Bozhong Tian, Siyuan Cheng, Zhoubo Li, Shumin Deng, Huajun Chen, and Ningyu Zhang. 2023{\natexlab{b}}.
\newblock Editing large language models: Problems, methods, and opportunities.
\newblock \emph{arXiv preprint arXiv:2305.13172}.

\bibitem[{Yu et~al.(2023)Yu, Qi, Chen, Chen, Yang, Zhu, Zhang, and Yu}]{Yu2023GPTPT}
Xiao Yu, Yuang Qi, Kejiang Chen, Guoqiang Chen, Xi~Yang, Pengyuan Zhu, Weiming Zhang, and Neng~H. Yu. 2023.
\newblock Gpt paternity test: Gpt generated text detection with gpt genetic inheritance.
\newblock \emph{ArXiv}, abs/2305.12519.

\bibitem[{Zaib et~al.(2021)Zaib, Tran, Sagar, Mahmood, Zhang, and Sheng}]{Zaib2021BERTCoQACBC}
Munazza Zaib, Dai~Hoang Tran, Subhash Sagar, Adnan Mahmood, Wei~Emma Zhang, and Quan~Z. Sheng. 2021.
\newblock \href {https://api.semanticscholar.org/CorpusID:231930144} {Bert-coqac: Bert-based conversational question answering in context}.
\newblock In \emph{International Symposium on Parallel Architectures, Algorithms and Programming}.

\bibitem[{Zelikman et~al.(2022)Zelikman, Wu, Mu, and Goodman}]{zelikman2022star}
Eric Zelikman, Yuhuai Wu, Jesse Mu, and Noah Goodman. 2022.
\newblock \href {https://openreview.net/forum?id=_3ELRdg2sgI} {{ST}ar: Bootstrapping reasoning with reasoning}.
\newblock In \emph{Advances in Neural Information Processing Systems}.

\bibitem[{Zhan et~al.(2023)Zhan, He, Xu, Wu, and Stenetorp}]{Zhan2023G3DetectorGG}
Haolan Zhan, Xuanli He, Qiongkai Xu, Yuxiang Wu, and Pontus Stenetorp. 2023.
\newblock G3detector: General gpt-generated text detector.
\newblock \emph{ArXiv}, abs/2305.12680.

\bibitem[{Zhan et~al.(2024)Zhan, Liang, Ying, and Kang}]{Zhan2024InjecAgentBIA}
Qiusi Zhan, Zhixiang Liang, Zifan Ying, and Daniel Kang. 2024.
\newblock \href {https://arxiv.org/pdf/2403.02691.pdf} {Injecagent: Benchmarking indirect prompt injections in tool-integrated large language model agents}.
\newblock In \emph{Annual Meeting of the Association for Computational Linguistics}.

\bibitem[{Zhang et~al.(2023{\natexlab{a}})Zhang, Edelman, Francati, Venturi, Ateniese, and Barak}]{Zhang2023WatermarksIT}
Hanlin Zhang, Benjamin~L. Edelman, Danilo Francati, Daniele Venturi, Giuseppe Ateniese, and Boaz Barak. 2023{\natexlab{a}}.
\newblock \href {https://api.semanticscholar.org/CorpusID:265050535} {Watermarks in the sand: Impossibility of strong watermarking for generative models}.
\newblock \emph{ArXiv}, abs/2311.04378.

\bibitem[{Zhang et~al.(2021)Zhang, Li, Chen, Deng, Bi, Tan, Huang, and Chen}]{Zhang2021DifferentiablePM}
Ningyu Zhang, Luoqiu Li, Xiang Chen, Shumin Deng, Zhen Bi, Chuanqi Tan, Fei Huang, and Huajun Chen. 2021.
\newblock \href {https://api.semanticscholar.org/CorpusID:237353222} {Differentiable prompt makes pre-trained language models better few-shot learners}.
\newblock \emph{ArXiv}, abs/2108.13161.

\bibitem[{Zhang et~al.(2024{\natexlab{a}})Zhang, Chen, Ye, Yang, Chen, Wang, and Petzold}]{Zhang2024UnveilingTIA}
Xinlu Zhang, Z.~Chen, Xi~Ye, Xianjun Yang, Lichang Chen, William~Yang Wang, and Linda~R. Petzold. 2024{\natexlab{a}}.
\newblock \href {https://api.semanticscholar.org/CorpusId:270199509} {Unveiling the impact of coding data instruction fine-tuning on large language models reasoning}.
\newblock In \emph{AAAI Conference on Artificial Intelligence}.

\bibitem[{Zhang et~al.(2024{\natexlab{b}})Zhang, Chen, Jiang, Sun, Wang, and Wang}]{Zhang2024TowardsAHA}
Yuyang Zhang, Kangjie Chen, Xudong Jiang, Yuxiang Sun, Run Wang, and Lina Wang. 2024{\natexlab{b}}.
\newblock \href {https://api.semanticscholar.org/CorpusId:274776158} {Towards action hijacking of large language model-based agent}.
\newblock \emph{ArXiv}, abs/2412.10807.

\bibitem[{Zhang et~al.(2023{\natexlab{b}})Zhang, Yang, Ke, and Huang}]{Zhang2023DefendingLL}
Zhexin Zhang, Junxiao Yang, Pei Ke, and Minlie Huang. 2023{\natexlab{b}}.
\newblock \href {https://api.semanticscholar.org/CorpusID:265212812} {Defending large language models against jailbreaking attacks through goal prioritization}.
\newblock \emph{ArXiv}, abs/2311.09096.

\bibitem[{Zhao et~al.(2023{\natexlab{a}})Zhao, Wen, Luu, Zhao, and Fu}]{Zhao2023PromptAT}
Shuai Zhao, Jinming Wen, Anh~Tuan Luu, Junbo~Jake Zhao, and Jie Fu. 2023{\natexlab{a}}.
\newblock Prompt as triggers for backdoor attack: Examining the vulnerability in language models.
\newblock \emph{ArXiv}, abs/2305.01219.

\bibitem[{Zhao et~al.(2023{\natexlab{b}})Zhao, Zhou, Li, Tang, Wang, Hou, Min, Zhang, Zhang, Dong, Du, Yang, Chen, Chen, Jiang, Ren, Li, Tang, Liu, Liu, Nie, and rong Wen}]{Zhao2023ASO}
Wayne~Xin Zhao, Kun Zhou, Junyi Li, Tianyi Tang, Xiaolei Wang, Yupeng Hou, Yingqian Min, Beichen Zhang, Junjie Zhang, Zican Dong, Yifan Du, Chen Yang, Yushuo Chen, Z.~Chen, Jinhao Jiang, Ruiyang Ren, Yifan Li, Xinyu Tang, Zikang Liu, Peiyu Liu, Jianyun Nie, and Ji~rong Wen. 2023{\natexlab{b}}.
\newblock A survey of large language models.
\newblock \emph{ArXiv}, abs/2303.18223.

\bibitem[{Zhao et~al.(2025)Zhao, Zheng, Luo, Li, Ma, and Jiang}]{Zhao2025BlueSuffix}
Yunhan Zhao, Xiang Zheng, Lin Luo, Yige Li, Xingjun Ma, and Yu-Gang Jiang. 2025.
\newblock \href {https://openreview.net/forum?id=wwVGZRnAYG} {{BlueSuffix}: Reinforced blue teaming for vision-language models against jailbreak attacks}.
\newblock In \emph{Proceedings of the International Conference on Learning Representations (ICLR)}.
\newblock ArXiv:2410.20971.

\bibitem[{Zhong et~al.(2023)Zhong, Guo, Gao, Ye, and Wang}]{Zhong2023MemoryBankEL}
Wanjun Zhong, Lianghong Guo, Qi-Fei Gao, He~Ye, and Yanlin Wang. 2023.
\newblock \href {https://api.semanticscholar.org/CorpusID:258741194} {Memorybank: Enhancing large language models with long-term memory}.
\newblock \emph{ArXiv}, abs/2305.10250.

\bibitem[{Zhou et~al.(2024{\natexlab{a}})Zhou, Li, Li, Kang, Hu, Wu, and Meng}]{zhou2024purple}
Jingyan Zhou, Kun Li, Junan Li, Jiawen Kang, Minda Hu, Xixin Wu, and Helen Meng. 2024{\natexlab{a}}.
\newblock Purple-teaming llms with adversarial defender training.
\newblock \emph{arXiv preprint arXiv:2407.01850}.

\bibitem[{Zhou et~al.(2024{\natexlab{b}})Zhou, Hwang, Ren, and Sap}]{Zhou2024RelyingOT}
Kaitlyn Zhou, Jena~D. Hwang, Xiang Ren, and Maarten Sap. 2024{\natexlab{b}}.
\newblock \href {https://api.semanticscholar.org/CorpusID:266977353} {Relying on the unreliable: The impact of language models' reluctance to express uncertainty}.

\bibitem[{Zhu et~al.(2022)Zhu, Qin, Cui, Chen, Zhao, Fu, Deng, Liu, Wang, Wu, Sun, and Gu}]{zhu2022moderatefitting}
Biru Zhu, Yujia Qin, Ganqu Cui, Yangyi Chen, Weilin Zhao, Chong Fu, Yangdong Deng, Zhiyuan Liu, Jingang Wang, Wei Wu, Maosong Sun, and Ming Gu. 2022.
\newblock \href {https://openreview.net/forum?id=C7cv9fh8m-b} {Moderate-fitting as a natural backdoor defender for pre-trained language models}.
\newblock In \emph{Advances in Neural Information Processing Systems}.

\bibitem[{Zhuo et~al.(2023)Zhuo, Huang, Chen, and Xing}]{zhuo2023red}
Terry~Yue Zhuo, Yujin Huang, Chunyang Chen, and Zhenchang Xing. 2023.
\newblock Red teaming chatgpt via jailbreaking: Bias, robustness, reliability and toxicity.
\newblock \emph{arXiv preprint arXiv:2301.12867}.

\bibitem[{Ziegler et~al.(2019)Ziegler, Stiennon, Wu, Brown, Radford, Amodei, Christiano, and Irving}]{Ziegler2019FineTuningLM}
Daniel~M. Ziegler, Nisan Stiennon, Jeff Wu, Tom~B. Brown, Alec Radford, Dario Amodei, Paul Christiano, and Geoffrey Irving. 2019.
\newblock \href {https://api.semanticscholar.org/CorpusID:202660943} {Fine-tuning language models from human preferences}.
\newblock \emph{ArXiv}, abs/1909.08593.

\end{thebibliography}
\end{document}